\documentclass[12pt]{article}

\usepackage[margin=0.75in]{geometry}
\usepackage{color}
\usepackage{graphicx}
\usepackage{epsfig}
\graphicspath{{./eps/}}
\usepackage{epstopdf}
\usepackage{subcaption}
\usepackage[font=small,skip=0pt]{caption}

\usepackage{amsmath}
\usepackage{amssymb}
\usepackage{amsthm}
\usepackage{mathtools}

\usepackage{rotating}
\usepackage{verbatim}
\usepackage{hyperref}
\usepackage{bm}
\usepackage[normalem]{ulem}
\usepackage[authoryear]{natbib}
\usepackage{bbm}
\usepackage{array}
\usepackage{enumerate}
\usepackage{float}
\usepackage{authblk}

\usepackage{algorithm}
\usepackage[noend]{algpseudocode}
\renewcommand{\algorithmicensure}{\textbf{Input:}}

\usepackage[flushleft]{threeparttable}

\usepackage{footmisc}

\newtheorem{theorem}{Theorem}[section]

\newtheorem{proposition}[theorem]{Proposition}

\theoremstyle{definition}
\newtheorem{definition}{Definition}[section]

\usepackage{thmtools}

\declaretheorem[style=definition]{example}%

\renewcommand{\P}{\mathbb{P}}

\newcommand{\E}{\mathbb{E}}

\newcommand{\iid}{\stackrel{\emph{i.i.d.}}{\sim}}
\newcommand{\simind}{\stackrel{\mathrm{ind}}{\sim}}

\newcommand{\Perp}{\perp\!\!\!\perp}
\makeatletter
\newcommand*{\indep}{%
  \mathbin{%
    \mathpalette{\@indep}{}%
  }%
}
\newcommand*{\nindp}{%
  \mathbin{
    \mathpalette{\@indep}{\not}
  }%
}
\newcommand*{\@indep}[2]{
  \sbox0{$#1\perp\m@th$}
  \sbox2{$#1=$}
  \sbox4{$#1\vcenter{}$}
  \rlap{\copy0}
  \dimen@=\dimexpr\ht2-\ht4-.2pt\relax
  \kern\dimen@
  {#2}%
  \kern\dimen@
  \copy0 
} 
\makeatother

\newcommand{\bs}{\boldsymbol}

\newcommand{\var}{\ensuremath{\operatorname{Var}}}

\def\R{\mathbb{R}}
\def\Z{\mathbb{Z}}
\def\N{\mathbb{N}}

\def\I{\mathbb{I}}

\newcommand{\mcA}{\mathcal{A}}

\newcommand{\mcC}{\mathcal{C}}
\newcommand{\mcD}{\mathcal{D}}

\newcommand{\mcF}{\mathcal{F}}
\newcommand{\mcG}{\mathcal{G}}
\newcommand{\mcH}{\mathcal{H}}

\newcommand{\mcJ}{\mathcal{J}}
\newcommand{\mcK}{\mathcal{K}}
\newcommand{\mcL}{\mathcal{L}}
\newcommand{\mcM}{\mathcal{M}}
\newcommand{\mcN}{\mathcal{N}}

\newcommand{\mcR}{\mathcal{R}}

\newcommand{\Bern}{\mathrm{Bern}}
\newcommand{\Beta}{\mathrm{Beta}}
\newcommand{\Unif}{\mathrm{Unif}}
\newcommand{\Bin}{\mathrm{Bin}}

\newcommand{\invGamma}{\mathrm{invGamma}}
\newcommand{\truncBeta}{\mathrm{TruncBeta}}
\newcommand{\truncNorm}{\mathrm{truncNorm}}
\newcommand{\Multi}{\mathrm{Mult}}

\newcommand{\defeq}{\coloneqq}
\newcommand{\opt}{^\star}
\newcommand{\CS}{\mathrm{CS}}
\def\ident{\mathbf{I}}
\newcommand{\minp}{p_{\mathrm{min}}}
\newcommand{\negc}{\mathrm{-}}

\newcommand{\bX}{\bs{X}}

\newcommand{\bz}{\bs{z}}
\newcommand{\bd}{\bs{d}}

\newcommand{\by}{\bs{y}}
\newcommand{\br}{\bs{r}}
\newcommand{\bx}{\bs{x}}

\newcommand{\blip}{\text{BLiP}}
\newcommand{\blips}{\text{BLiP}\,}
\newcommand{\weight}{w}
\newcommand{\locations}{\mathcal{L}}
\newcommand{\mcLz}{\mathcal{L}_0}
\newcommand{\loc}{\ell}
\newcommand{\pyblip}{\texttt{pyblip}}
\newcommand{\pyblips}{\texttt{pyblip}\,}
\newcommand{\blipr}{\texttt{blipr}}
\newcommand{\bliprs}{\texttt{blipr}\,}

\newcommand{\Power}{\mathrm{Power}}
\newcommand{\FDR}{\mathrm{FDR}}
\newcommand{\FNR}{\mathrm{FNR}}

\usepackage[normalem]{ulem}
\def\incstrikeout{1pt} 
\ifdim\incstrikeout=1pt
	\newcommand{\strout}[1]{
    {\sout{#1}}
}
\else
	\newcommand{\strout}[1]{}
\fi

\graphicspath{{./fig/}}

\numberwithin{equation}{section}

\usepackage{xcolor}

\title{Controlled Discovery and Localization of Signals via Bayesian Linear Programming}
\author[1]{Asher Spector}
\author[2]{Lucas Janson}
\date{}
\affil[1]{Department of Statistics, Stanford University}
\affil[2]{Department of Statistics, Harvard University}

\begin{document}
\maketitle

\begin{abstract}
    Scientists often must simultaneously localize and discover signals. For instance, in genetic fine-mapping, high correlations between nearby genetic variants make it hard to identify the exact locations of causal variants. So the statistical task is to output as many disjoint regions containing a signal as possible, each as small as possible, while controlling false positives. Similar problems arise, e.g., when locating stars in astronomical surveys and in changepoint detection. Common Bayesian approaches to these problems involve computing a posterior distribution over signal locations. However, existing procedures to translate these posteriors into credible regions for the signals fail to capture all the information in the posterior, leading to lower power and (sometimes) inflated false discoveries. We introduce Bayesian Linear Programming (BLiP), which can efficiently convert any posterior distribution over signals into credible regions for signals. BLiP overcomes an extremely high-dimensional and nonconvex problem to verifiably nearly maximize expected power while controlling false positives. Applying BLiP to existing state-of-the-art analyses of UK Biobank data (for genetic fine-mapping) and the Sloan Digital Sky Survey (for astronomical point source detection) increased power by 30-120\% in just a few minutes of additional computation. BLiP is implemented in \pyblips (Python) and \blipr (R).
\end{abstract}

\section{Introduction}\label{sec::intro}

\subsection{Motivation}\label{subsec::motivation}

\textit{Localizing signals} is an important statistical task across disciplines. For example, consider the problem of controlled variable selection: given variables $X_1, \dots, X_p$, analysts seek to identify a few key variables which impact an outcome $Y$. Here, $X_1, \dots, X_p$ could represent genetic mutations or demographic data, and $Y$ could represent a disease status or economic outcome. However, when variables are highly correlated, it can be nearly impossible to certify that any individual variable is important. For example, if $X_1$ and $X_2$ are nearly perfectly correlated, analysts may not be able to distinguish between them even if the data make clear that $\{X_1, X_2\}$ contains at least one important variable. Yet selecting different variables may lead to qualitatively different scientific conclusions, such as in genetic studies, where intervening on a causal variant (e.g., $X_1$) might help cure a disease, whereas intervening on its highly correlated neighbor (e.g., $X_2$) could have no impact at all. As a result, analysts often must contend with significant uncertainty in which variables ought to be selected. Unfortunately, as observed by \cite{susie2020}, many modern variable selection methods cannot accomplish this task.

More generally, analysts in many settings may be able to tell that a signal exists without perfectly localizing it. Indeed, astronomers often can guarantee that a light source exists somewhere in a region of space without knowing its exact location. Similarly, economists may be able to tell that a time series has changed without knowing precisely when it did so. Lastly, this problem is particularly relevant in \textit{genetic fine-mapping}, where researchers attempt to identify genetic variants which cause outcomes such as cardiovascular disease. Indeed, genetic variants are highly locally correlated---for example, this paper analyzes a UK Biobank dataset where over $40\%$ of genetic variants are at least $99\%$ correlated with a nearby variant. In these settings, analysts often want to localize signals as precisely as possible, i.e., to make statements like ``there is a signal in this region, even if we do not know exactly where it is." These discovered regions should be as small as possible to yield precise scientific insights, all while controlling false positives to ensure findings are replicable, interpretable, and do not waste future resources.

However, localizing signals can be difficult because there are combinatorially many regions which might contain signals. As a result, existing frequentist methods can be computationally expensive, lose power, or lose Type I error control due to the high multiplicity and complex dependence structure of the underlying p-values (see Sections \ref{subsec::literature}, \ref{sec::sims}). In contrast, a Bayesian approach based on a single approximate posterior distribution of the signal locations may in principle be more computationally efficient and powerful than approaches which compute large numbers of p-values. That said, it can be very challenging to extract interpretable results from such high-dimensional posterior distributions \citep{susie2020}, and existing solutions fail to capture all of the information available in the solution, again leading to either power loss, inflated false positives, or both, as we shall see in two real data applications.

With this motivation, our paper introduces a novel procedure, called \textit{Bayesian Linear Programming} (\blip), which takes an approximate posterior as an input and uses it to localize signals as precisely as possible while controlling false positives. Before describing our contribution, however, we pause to give a formal problem statement.

\subsection{Problem Statement}\label{subsec::problemstatement}

In this section, we define the problem of \textit{resolution-adaptive signal detection}. 
Note this scientific problem is not inherently frequentist or Bayesian, although we will take a Bayesian approach in this paper.
To start, let $\locations$ denote a set of locations at which there might be signals. For example, $\locations$ might represent locations on a genome or positions in the night sky. For each location $\loc \in \locations$, let $I_\loc$ be the indicator of whether there is a signal at $\ell$. For example, in genetic fine-mapping, $I_{\loc} = 1$ if the genetic variant at location $\ell$ on the genome has a causal relationship with a trait of interest, e.g., heart disease. Intuitively, one can think of $I_{\loc}$ as either an unobserved latent variable or the indicator that the null hypothesis that there is no signal at $\loc$ is false. To aid intuition, we pause to give two concrete examples of this setup. (We give a third example, change point detection, in Appendix \ref{subsec::changepointsim}.)

\begin{example}[Variable selection in regression]\label{ex::sparsereg} Suppose we observe a response $Y$ and variables $X = (X_1,\dots,X_p)$, and we seek to discover ``important" variables. Here, the locations $\locations = \{1, \dots, p\}$ correspond to $X_1, \dots, X_p$, and there is a signal at location $\loc \in \locations$ if $X_\loc$ is ``important." For example, if $Y$ depends on $X$ through linear coefficients $\beta \in \R^p$, we might set $I_\loc = \I(\beta_\loc \ne 0)$. More generally, we could set $I_\loc = 1$ if the conditional distribution $Y \mid X$ depends on $X_{\loc}$, or equivalently, if $X_{\loc} \not \Perp Y \mid X_{-\ell}$. Note when $(X_1, \dots, X_p)$ are highly correlated, we may not be able to discover individual signal variables with confidence. However, we often do have power to discover that at least one variable in a group $G \subset \locations$ is important.
\end{example}

\begin{example}[Point source detection]\label{ex::pointsource} Astronomers often seek to locate point sources (e.g., stars) in the night sky. Here, we define the set of locations $\locations$ to be a region of the sky, so $\locations \subset \R^2$. For each coordinate $(x,y) \in \locations$, $I_{(x,y)}=1$ if and only if a source exists at location $(x,y)$. Our data consists of a set of photon counts $\{D_{ij}\}_{1 \le i \le d_1, 1 \le j \le d_2}$, where $D_{ij}$ counts the photons observed by a telescope in the $(i,j)$th pixel of an image. Most images have blur, making it difficult to distinguish whether a source is located in one pixel versus its neighbors. This motivates a resolution-adaptive approach, where instead of estimating signal locations $\{(x_i,y_i)\}_{i=1}^R$, we output regions $G_1, \dots, G_R$ which each contain a source with high confidence and are as small as possible.
\end{example}

Since it can be difficult to perfectly localize signals, we allow ourselves to discover any group or region $G \subset \locations$, which asserts that at least one signal exists in $G$. (We use the words ``group" and ``region" interchangeably.) This is a true detection if there actually is a signal in $G$, or formally, if $I_G \defeq \I\left(\exists \loc \in G : I_\loc = 1 \right) = 1$. Intuitively, detecting signals in larger regions corresponds to making discoveries at a coarser resolution, which is ``easier" statistically than detecting signals in smaller regions but yields less specific information. For example, in genetic fine-mapping, discovering a large region $G$ tells us that a causal genetic variant exists somewhere in $G$, whereas discovering $G = \{\loc\}$ precisely isolates the location of a single causal variant. Thus, we prefer to discover regions $G$ that are as small as possible.

With this motivation, the problem of \textit{resolution-adaptive signal detection} is to select a disjoint set of regions $G_1, \dots, G_R$ which maximizes expected power while controlling a false positive rate.\footnote{We use the word ``power" slightly loosely in this paper. Although ``power" often refers to an expected quantity, we use it to refer to the \textit{realized} (resolution-adjusted) number of true discoveries made using a single dataset.} We will precisely define statistical power in Section \ref{subsec::gap}, but for now, higher power intuitively corresponds to making more discoveries at a finer resolution. For example, we might take $\Power(G_1, \dots, G_R) = \sum_{r=1}^R \frac{1}{|G_r|}$ to count the number of discoveries but weight discoveries by the reciprocal of their size, which rewards discovering smaller groups \citep{mandozzi2016}.

\begin{definition}[Resolution-adaptive signal detection]\label{def::rasd} Suppose we seek to discover signals in a set of locations $\locations$. Let $R \ge 0$ be the number of discoveries and let $G_1, \dots, G_R$ denote the discovered regions. Given a notion of statistical power for a set of discovered regions (see Section \ref{subsec::gap}), denoted by $\Power(G_1,\dots,G_R)$, we seek to maximize
\begin{align}
    \max \,\,\,\,\,\,\,\,\,\,\,\, & \E\left[\Power(G_1, \dots, G_R)\right]  \label{eq::rasdpower} \\
    \mathrm{s.t.}\,\,\,\,\,\,\,\,\,\,\,\, & \FDR \defeq \E\left[\frac{\#\{1 \le r \le R : I_{G_r} = 0 \}}{\max(1, R)} \right] \le q,  \label{eq::rasdfdr} \\
    & G_1, \dots, G_R \subset \locations \text{ are disjoint.} \label{eq::rasddisj} 
\end{align}

Note that we constrain $G_1, \dots, G_R$ to be disjoint to improve interpretability and prevent double-counting. For example, discovering $\{\loc_1\} \subset \locations$ makes discovering $\{\loc_1, \loc_2\} \subset \locations$ logically redundant, so this should not count as two separate discoveries when calculating power or FDR. We focus on the FDR because it is a popular and appealing error rate, but this problem is still well-defined if we replace the FDR with another error rate, such as the family-wise error rate (FWER) or the local FDR. Indeed, \blips will be able to solve any of these problems.
\end{definition}

A simplification of this problem would be to fix a prespecified partition $G_1, \dots, G_m \subset \locations$ and test whether a signal exists in each of $G_1, \dots, G_m$. However, this non-adaptive approach will not optimally localize signals, because the best choice of partition $G_1, \dots, G_m$ depends on the unknown data-generating process. Informally, when the ``signal size" is large, we may be able to perfectly localize individual signals, whereas we may only be able to detect that a weak signal exists somewhere in a relatively large region. Indeed, in Appendix \ref{subsec::ftest}, we give a concrete example of a regression problem where the best partition depends on the unknown relationship between $Y$ and $X$. In contrast, resolution-adaptive methods can use the data to discover regions $G_1, \dots, G_R$ which are as small as possible. Of course, for computational reasons, it may not be possible to consider \textit{every} region $G \subset \locations$ as a potential discovery. However, we expect that methods which are more adaptive, meaning they can use the data to choose from among a larger set of candidate regions, to have higher power.

Although resolution-adaptive methods can be much more powerful than methods which fix a resolution in advance, any resolution-adaptive method must search over combinatorially many possible groupings of the locations. How can we construct such a method which is maximally powerful, computationally efficient, and still controls some notion of the false positive rate? The goal of our work is to answer this question.

\subsection{Contribution}

Our key contribution is to introduce \textit{Bayesian Linear Programming} (\blip), which is a method for performing resolution-adaptive signal detection. As an input, \blips takes an approximate posterior distribution over the location of the signals. For example, if $Y \mid X$ follows a generalized linear model (GLM) where nonzero coefficients are signals, one can use a Markov Chain Monte Carlo (MCMC) algorithm to sample from the posterior distribution of the model coefficients and use the MCMC samples as the input for \blips. Thus, \blips can accurately be described as a type of post-processing on the posterior (albeit with appealing statistical guarantees). As we shall see in Sections \ref{sec::gwas} and \ref{sec::astro}, this ``post-processing" can dramatically improve the power and calibration of applied analyses in settings ranging from GWAS to astronomical point-source detection. 

Given this input, \blips will output a set of disjoint regions, each containing a signal, which nearly maximizes expected power while controlling false positives. For example, in variable selection problems, \blips will return a set of disjoint groups of variables so that (1) nearly all groups contain at least one signal variable, (2) we discover as many groups as possible, and (3) the groups are as small as possible. Indeed, in genetic fine-mapping, this corresponds to identifying as many important genetic variants as possible and simultaneously localizing them as precisely as possible. 

At the outset, we highlight a few attractive features of \blip:

\textbf{Power}: Where competitors exist, \blips is often much more powerful than other methods, as we demonstrate in extensive simulations and two real data analyses. This is both because \blips can take advantage of arbitrary Bayesian methods to compute approximate posteriors, and also because \blips finds nearly the optimal set of discoveries, a claim we make precise in Section \ref{subsec::blipfdr}. For this reason, when competitors exist, \blips can wrap on top of those competitor methods to increase their power, often by wide margins. 

\textbf{Provable error control}: Given a correct posterior, \blips can provably control one of several error rates, including the FDR, FWER, and local FDR. 

\textbf{Flexibility}: Since \blips acts directly on an approximate posterior, it can be applied on top of any Bayesian model. Furthermore, analysts can use nearly any algorithm to compute the approximate posterior, allowing them to leveraging recent advances in Bayesian MCMC or variational inference. This allows analysts to flexibly trade-off computational efficiency and statistical power.

\textbf{Computational Efficiency}: Although computing an approximate posterior over signal locations may be expensive, \blips itself is extremely computationally efficient, allowing it to search over hundreds of millions of candidate regions to find a near-optimal set of discoveries. This allows \blips to make discoveries at resolutions which are as precise as possible even in very large-scale problems without obvious hierarchical structure. 

To quote \cite{susie2020}, ``the output from Bayesian Variable Selection methods is typically a complex posterior distribution, and this can be difficult to distill into results that are easily interpretable." \blips is designed to solve exactly this problem (in the more general signal detection setting), and thus it offers a highly general framework to extract interpretable and replicable information from complex high-dimensional posterior distributions.

\subsection{Related Literature}\label{subsec::literature}

One natural approach when searching for signals is to partition the locations into groups in an unsupervised manner and then test whether there is a signal in each group. Here, note that in the frequentist literature, detecting a signal in a group $G \subset \locations$ is often framed as rejecting a group null hypothesis of the form $H_G = \bigcap_{\ell \in G} H_{\ell}$, where $H_{\ell}$ is the null hypothesis that no signal exists at location $\ell$. For example, in variable selection problems, many methods use $X$ to cluster the variables into groups and then test if there exists a signal (e.g., a nonzero coefficient) in each group using both $X$ and $Y$ \citep{esl2001}. The key difference between \blips and these methods is that \blips simultaneously groups the locations \textit{and} discovers signals at the same time. For example, in variable selection, if $X_1$ and $X_2$ are moderately correlated, unsupervised clustering methods have to guess whether or not to group $X_1$ and $X_2$ together without looking at $Y$: in contrast, \blips can use the full data to determine if there is enough signal to distinguish between $X_1$ and $X_2$ (see Appendix \ref{subsec::ftest} for a concrete example where the best grouping is discernible from $Y \mid X$ but not $X$ alone). As a result, \blips can use the full data to localize signals as precisely as possible: this is the advantage of taking a resolution-adaptive approach. 

A few recent works have developed frequentist methods to search over hierarchical sets of groups to detect signals at as fine a resolution as possible \citep{yekutieli2008, mandozzi2016, benjamini2007, hierinf2018, multilayerknock2019, knockoffzoom2019, fbh2021}. In the language of this paper, these methods perform resolution-adaptive signal detection or something quite similar to it, and indeed, these can be powerful methods in many settings. That said, they have a few disadvantages. First, each of these methods can only search over a somewhat limited set of candidate groups, often corresponding to a hierarchical tree. As a result, their adaptivity is limited. Although \blips also can only make discoveries from a candidate set of groupings, \blips places no restrictions on this candidate set and is efficient enough to search over, e.g., thousands of possible hierarchical groupings of the variables. As \cite{susie2020} have observed, this is important in practical applications like genetic fine-mapping where the variables do not follow a single hierarchical structure. Second, these methods either assume complete independence of p-values \citep{yekutieli2008} or make conservative assumptions to control the FDR under dependence \citep{fbh2021}. As a result, we show in Section \ref{sec::sims} that each of these methods is either unable to reliably control the FDR or loses statistical power. \blip, in contrast, provably controls the Bayesian FDR at nearly the nominal level and has high power, since it is highly adaptive. Lastly, there is also a much larger literature on hierarchical testing more broadly, including methods which control, e.g., the FWER \citep{meinshausen2008, goeman2012, mandozzi2016, hierinf2018} or weighted versions of the FDR \citep{benjamini1997, bogomolov2020}. (We note that \blips uses a linear program to maximize power subject to FDR constraints; similar optimization techniques have been used to design hypothesis tests before in a different setting, e.g., design of adaptive clinical trials \citep{heller2022}.) 
Furthermore, there are resolution-adaptive methods in change point detection which control the FWER (see, e.g., \cite{frick2014, fang2016, fang2020, fryzlewicz2020}) or a slightly modified FDR \citep{limunk2016}. It would be impossible to exhaustively compare to each method from this vast literature, however, and thus we only compare \blips to methods which return a disjoint set of discoveries and target the FDR as defined in Definition \ref{def::rasd}, which is sometimes known as the ``outer nodes FDR." This choice is also motivated by our real data applications, where this FDR is a standard choice of error rate.

Our work is perhaps closest in spirit to that of \cite{susie2020}, who introduced ``SuSiE," a method for sparse Bayesian linear regression.\footnote{As is common in the literature, we will use the name ``SuSiE" to refer to both the SuSiE model and the Iterative Bayesian stepwise selection (IBSS) algorithm.} SuSiE uses a novel variational approximation (which is accurate when the number of signals is small) to approximate the posterior and then processes that approximate posterior to perform resolution-adaptive inference. In contrast, \blips performs only the latter task, but it can do so on any posterior. For example, \blips can apply directly to the approximate posterior from SuSiE, as we show in Section \ref{sec::sims}, where \blips uniformly improves SuSiE's power. Even when the SuSiE variational approximation is inaccurate, \blips can often partially correct this issue, leading to improved power and FDR control. Note that \blips can also apply on top of the refined SuSiE procedure suggested in \cite{zou2021}, although the refined procedure was too computationally expensive for us to use in our simulations. To our knowledge, the only other comparable Bayesian method is DAP-G \citep{dapg2018}, which also requires a specific approximation to the posterior to perform resolution-adaptive inference. In principle, \blips can also wrap on top of DAP-G to improve its power, although we focused only on combining SuSiE with \blips in our simulations because SuSiE outperformed DAP-G. Alternatively, \blips can be combined with any other method to approximate the posterior, e.g., MCMC, and thus \blips offers an attractive alternative to perform resolution-adaptive inference without (necessarily) making any variational approximation.

Since \blips takes a posterior distribution over the signal locations as an input, our work leverages (but does not contribute to) the existing literature on efficiently approximating such posteriors. Indeed, \blips can leverage arbitrary advances in MCMC and variational inference to compute this approximate posterior (see \cite{mcmcbook2011} and \cite{blei2017}, respectively, for canonical reviews of these subjects). In the context of variable selection, our work draws on methods for sparse Bayesian linear regression \citep{mitchell1988, mcculloch1997, nprior2018}, Bayesian probit and logistic regression \citep{albertchib1993, polygamma2013}, and variational methods for Bayesian regression \citep{susie2020}. These types of methods are also commonly used in genetic fine-mapping problems (e.g., \cite{guanstephens2011, carbonettostephens2012, benner2015, dapg2018, polyfun2019}), change point detection (see \cite{cappello2021} for a recent review), and point-source detection (see, e.g., \cite{starnet2021}). Overall, this literature is vast, so our review is necessarily incomplete, but the main point is this: this literature is in fact a core motivation for our work. Indeed, previous Bayesian methods for resolution-adaptive signal detection require using specific variational approximations to the posterior distribution. In contrast, \blips can wrap on top of any posterior approximation algorithm, so further advances in posterior approximation can immediately be translated to better inference in resolution-adaptive signal detection problems.

\subsection{Notation}

For any $k \in \N$, let $[k]$ denote the set $\{1, \dots, k\}$. For any subset $J \subset [k]$ and vector $v \in \R^k$, let $v_J$ denote the entries of $v$ corresponding to the set $J$. Similarly, for any matrix $M \in \R^{m \times k}$, let $M_J$ denote the columns of $M$ corresponding to $J$, and let $M_j$ denote the $j$th column for $j \in [k]$. To ease notation, let $M_{-j}$ denote the set of all columns of $M$ except column $j$, and similarly $v_{-j}$ denotes all but the $j$th element of $v$. Much of this paper focuses on regression problems, and in these settings, we let $\by = (y_1, \dots, y_n) \in \R^n$ denote $n$ independent samples of an outcome of interest, and we let $\bX \in \R^{n \times p}$ denote the design matrix for $p$ variables. For convenience, we will let the non-bold notation $Y \in \R$ and $X = (X_1, \dots, X_p) \in \R^p$ refer to an arbitrary single data-point of the outcome and covariates. We let $\locations$ denote the set of locations, and for any $G \subset \locations$, $I_G$ is the indicator of whether a signal exists in $G$. Note that we use the bold notation $\ident_k$ to specify the $k \times k$ identity matrix. In settings where we obtain samples from a posterior distribution, we let $N$ denote the number of posterior samples. For any point $x \in \R^d$, let $S_r(x) \subset \R^d$ denote the closed $d$-sphere of radius $r$ around $x$, and let $C_r(x) \subset \R^d$ denote the closed $d$-cube of side-length $2r$ around $x$. In all problems, we let $\mcD$ denote all of the observed data.

\subsection{Outline}\label{subsec::outline}

The outline of the rest of the paper is as follows. In Section \ref{sec::blip}, we formally define our notion of resolution-adjusted power and then introduce \blip. In Section \ref{sec::sims}, we demonstrate the advantages of \blips over competitors via simulations. In Sections \ref{sec::gwas} and \ref{sec::astro}, we apply \blips to a genetic fine-mapping analysis of UK Biobank data and to detect point sources in the Sloan Digital Sky Survey, respectively. Section \ref{sec::discussion} concludes with a discussion of future directions. For convenience, Section \ref{sec::urls} includes a list of links to aid replication of the analyses in this paper.

\section{Bayesian Linear Programming for Resolution-Adaptive Signal Detection}\label{sec::blip}

\subsection{Resolution-Adjusted Power}\label{subsec::gap}

In this section, we formally define \textit{resolution-adjusted power}. Given locations $\locations$, intuitively, we would like to make as many true discoveries as possible while controlling, e.g., the false discovery rate. However, as discussed in Section \ref{subsec::problemstatement}, there is some ambiguity in how to count the number of discoveries, since discovering a smaller group gives us more specific information than discovering a larger group. For example, it is not obvious whether it is preferable to discover one region of size one (thereby perfectly localizing one signal) or to instead discover two regions, each of size two.

To resolve this ambiguity, we suggest weighting discoveries to prioritize discovering smaller groups, so that, e.g., discovering a group of size $3$ counts as ``fewer" discoveries than discovering a group of size $2$. Indeed, one proposal from \cite{mandozzi2016} is to assign a group of size $m$ weight $\frac{1}{m}$, so that discovering a group of size $1$ and a group of size $2$ would count as $1.5$ discoveries total. Intuitively, it makes sense that if we can only say that a signal exists in one of two locations, this gives us (by some measure) half as much information as discovering the true location of the signal. However, this definition may not be appropriate for every problem: for example, in our astronomical application in Section \ref{sec::astro}, we consider two other weighting schemes which are more natural choices for astronomy. Similarly, in variable selection problems, some analysts may be more willing to discover the group $\{1,2\}$ if $X_1$ and $X_2$ are highly correlated. For this reason, the rest of this paper allows for analysts to use an arbitrary \emph{weight function} $\weight$ which maps each possible group $G$ to a weight. After discovering a set of groups $G_1, \dots, G_R$, we define power to be the sum of the weights associated with each true discovery.

\begin{definition}[Resolution-Adjusted Power]\label{def::power}  Given a set of locations $\locations$ and data $\mcD$, suppose a method discovers a disjoint set of groups $G_1, \dots, G_R \subset \locations$. Let $\weight : 2^{\mcL} \to \R$ be a \textit{weighting function} on $2^{\mcL}$, the set of all subsets of $\locations$. Define the resolution-adjusted power as follows:
\begin{equation}\label{eq::power}
    \Power(G_1,\dots, G_R) = \sum_{r=1}^R I_{G_r} \weight(G_r),
\end{equation}
where it may be helpful to recall that $I_{G_r}$ is the indicator of whether a signal truly exists in $G_r \subset \locations$.
\end{definition}

\blips can optimize for \textit{any} possible weight function; however, by default, we suggest choosing $\weight(G) = |G|^{-1}$, because this choice is simple and interpretable, and it reflects the intuition that discovering a group of size $m$ gives roughly $m$ times less information than discovering an individual signal. Indeed, this choice has been used in several recent papers \citep{mandozzi2016, buzdugan2016, hierinf2018, guo2016, renaux2021}. Of course, in variable selection problems, one might wonder if $\weight(G)$ should account for the correlation structure of $X_G$. For example, if $X_1$ and $X_2$ are perfectly correlated, does it still make sense to count $G = \{X_1, X_2\}$ as only half a discovery? By default, we argue that the answer is yes when $X_1$ and $X_2$ represent distinct scientific hypotheses. For example, in genetic fine-mapping, if $X_1$ is causal and $X_2$ is not, then discovering $X_1$ might help develop a drug, whereas discovering $X_2$ is a false positive, no matter the correlation between $X_1$ and $X_2$. Lastly, it is important to note that reasonable modifications of $\weight(G)$ tend to give similar results; for example, we show empirically in Appendix \ref{appendix::weightfn} that applying increasing functions to our default yields a similar set of discoveries. Thus, as long as the user prefers to discover smaller groups, $\weight(G) = |G|^{-1}$ is a reasonable default. That said, there are settings where this default choice may not be ideal. For example, in Section \ref{sec::astro}, we use two different choices of $\weight$ to optimize for two different scientific objectives. Thus, for the rest this section, we work with an arbitrary weight function $\weight$.

\subsection{Bayesian Linear Programming for FDR Control}\label{subsec::blipfdr}

Having formally defined resolution-adjusted power, we now introduce a Bayesian Linear Program (\blip) to perform resolution-adaptive signal detection. In this paper, we take a Bayesian approach, meaning that we aim to maximize expected power and control the false positive rate conditional on the data $\mcD$. Although \blips can control other error rates as well, we start by discussing how to control the FDR, as it is the de facto error rate of choice in many applications including genetic fine-mapping.

This problem initially seems intractable, but it turns out that extremely high quality solutions can be found via a linear programming relaxation. Before we discuss this relaxation, however, note that it still may be too challenging computationally to search over all possible groupings, since there are an enormous (often infinite) number of possible subsets of $\locations$.
To narrow down the search space, we require that $G_1, \dots, G_R$ are members of a set of \textit{candidate groupings}, $\mcG$. This is not a particularly restrictive requirement, since one can make $\mcG$ as large as is computationally feasible, and our algorithm can handle hundreds of millions of candidate groupings. For example, in a variable selection problem, one could cluster the variables $X_1, \dots, X_p$ using (literally) a thousand different clustering algorithms and then let $\mcG$ denote the union of the groups created by the clustering algorithms. We offer more suggestions for constructing $\mcG$ in Section \ref{subsec::guidelines}.

Given candidate groupings $\mcG$, the first key observation is that maximizing power corresponds to maximizing a linear function. To see this, let $p_G = \E[I_G \mid \mcD]$ be the posterior probability that there is a signal in group $G$, also known as a posterior inclusion probability (PIP). Then, let $x_G \in \{0,1\}$ be the indicator of whether our procedure discovers group $G$: here, $\{x_G\}_{G \in \mcG}$ are our optimization variables which completely determine our discoveries $G_1, \dots, G_R$. Using this notation plus the definition of resolution-adjusted power, we obtain that

\begin{equation}\label{eq::linearpower}
    \E[\Power(G_1, \dots, G_R) \mid \mcD] \,\,\,\defeq\,\,\, \E\left[\sum_{r=1}^R I_{G_r} \weight(G_r) \mid \mcD \right] \,\,\,=\,\,\, \sum_{G \in \mcG} p_G \weight(G) x_G,
\end{equation}

where the right-most equality sums over $\mcG$ because $x_G = 1$ if and only if $G \in \{G_1, \dots, G_R\}$. Note that for simplicity, Equation (\ref{eq::linearpower}) assumes that the weights $w(G)$ are constants which do not depend on any unknown parameters. However, one can also allow $w(G)$ to depend on unknown parameters by simply replacing $p_G w(G)$ with $\E\left[I_G w(G)\mid \mcD\right]$ (which will be no harder to approximate than $p_G w(G)$ given a set of samples from the posterior distribution) in Equation~\eqref{eq::linearpower}. Then, the rest of the analysis in the paper will carry through without further modification.

Equation (\ref{eq::linearpower}) tells us that if we have access to $\{p_G\}_{G \in \mcG}$, the objective function is a linear function of  $\{x_G\}_{G \in \mcG}$. While computing the PIPs is often challenging, this problem is well studied (see Section \ref{subsec::literature}), and thus at the least, analysts have many tools available to them to try to compute the PIPs. Therefore, for now, we assume that we have access to the PIPs---later, in Sections \ref{sec::sims} and Appendix \ref{sec::mcmc}, we will suggest practical ways to approximate the PIPs and study whether \blips is robust PIPs which are only approximately correct.

Notably, the FDR constraint can also be formulated as a linear constraint. In particular, let $V = R - \sum_{r=1}^R I_{G_r}$ be the number of false discoveries. Controlling the FDR at level $q$ requires that
\begin{equation}\label{eq::fdrv1}
    \FDR \defeq \E\left[\frac{V}{R} \mid \mcD \right] = \frac{\sum_{G \in \mcG} (1 - p_G) x_G}{\sum_{G \in \mcG} x_G} \le q,
\end{equation}
where in the above equation we use the convention that $0 / 0 = 0$. Multiplying by $\sum_{G \in \mcG} x_G$ on both sides yields the linear constraint $\sum_{G \in \mcG} (1 - p_G - q) x_G \le 0$. Thus, as the following proposition makes clear, the resolution-adaptive signal detection problem can be formulated as a mixed-integer linear program (LP). A proof is given in Appendix \ref{subsec::proofs}.

\begin{proposition}\label{prop::blipequiv} The solution to the resolution-adaptive signal detection problem in Definition \ref{def::rasd} is the same as the solution to the following mixed-integer LP:
\begin{align}
    \max_{\{x_G\}_{G \in \mcG} : x_G \in [0,1]} \,\,\,\,\,\,\,\,\,\,\,\, & \sum_{G \in \mcG} p_G \weight(G) x_G \label{eq::power_reduction} \\
    \mathrm{s.t.}\,\,\,\,\,\,\,\,\,\,\,\,\,\,\,\,\,\,\,\, & \sum_{G \in \mcG} (1-p_G-q) x_G \le 0, \label{eq::fdr_red} \\
    & \sum_{G \in \mcG : \ell \in G} x_G \le 1 \,\,\,\,\,\, \forall \ell \in \locations, \label{eq::disj_red} \\
    & x_G \in \{0,1\} \,\,\,\,\,\, \forall G \in \mcG. \label{eq::intconstraint}
\end{align}
\end{proposition}
Note that when $\mcG$ or $\mcL$ are infinite sets, it is possible to reduce this integer LP to an equivalent finite-dimensional problem assuming (i) the expected number of signals is finite and (ii) a very mild regularity condition on $\mcG$. In Appendix \ref{appendix::infloc}, we introduce an efficient algorithm based on edge clique covers \citep{erdos1966} to perform this reduction.

Mixed-integer LPs are NP complete but nonetheless are very well studied \citep{intlpbook2008}, so it is often possible to get approximate or even exact solutions fairly efficiently. That said, we would like to be able to search over millions or even billions of candidate groupings, so we would still like to improve efficiency. To this end, in large problems, we instead suggest solving the relaxed problem (\ref{eq::power_reduction})-(\ref{eq::disj_red}), which drops the integer constraint that $x_G \in \{0,1\}$ and only requires that $x_G \in [0,1]$. The relaxed problem is a vanilla linear program with highly sparse constraints, so it can be solved extremely efficiently (see, e.g., \cite{leeaaron2015, highs2015}). Relaxations of this form are a common technique in optimization problems which often yield good results in practice \citep{boyd2004}. Furthermore, it turns out that this problem is sufficiently similar to a \textit{knapsack} problem that the relaxed version returns solutions which are almost entirely composed of integers. Usually, any non-integer values roughly correspond to the need to randomize the discoveries to achieve an FDR of exactly level $q$. Indeed, in Appendix \ref{subsec::knapsack}, we show that the solution to the relaxed LP typically has a single-digit number of non-integer values when searching over tens of thousands of groups. Although we cannot prove that this will happen in general, Appendix \ref{subsec::knapsack} gives intuition to explain this phenomenon, and we will present further empirical evidence for this claim shortly. 

This motivates Algorithm \ref{alg::blip}, which defines \blips for FDR control. To start, \blips solves the relaxed LP corresponding to (\ref{eq::power_reduction})-(\ref{eq::disj_red}), which typically yields a solution $\{x_G\opt\}_{G \in \mcG}$ with only a few non-integer values of $x_G\opt$, denoted $\mcH = \{G \in \mcG : x_G\opt \not \in \{0,1\}\}$. To obtain a deterministic rejection set, \blips then runs the integer linear program (\ref{eq::power_reduction})-(\ref{eq::intconstraint}) while holding the values of $\{x_G\opt : G \in \mcG \setminus \mcH\}$ constant and only optimizing over $\{x_G : G \in \mcH\}$. Typically, this integer LP has only a few variables, usually in the single digits, so it runs practically instantly, yielding the output of \blip. Note that technically speaking, up to two things could go wrong in this step. First, if the integer LP is very large, it may be challenging to solve it efficiently, although in this case the user will know this in advance and can use polynomial-time heuristic methods instead (see Appendix \ref{subsec::randomization}). Second, it is technically possible for the integer LP to be infeasible, in which case we propose a backtracking algorithm that iteratively finds the group $G_{\min}$ in $\{G : x_G\opt =1\}$ with the smallest PIP and adds $G_{\min}$ to $\mcH$. This guarantees that \blips finds a feasible solution in a finite number of backtracking steps. That said, neither of these two phenomenon ever occurred in any of the simulations or real data applications in this paper, despite our applying \blips tens of thousands of times in very large scale settings. This suggests that these modifications are only required in pathological examples. In all cases, however, the output of \blips is a feasible solution to (\ref{eq::power_reduction})-(\ref{eq::intconstraint}) and thus provably controls the FDR by Proposition \ref{prop::blipequiv}, even in pathological examples.

\begin{algorithm}[h!]
\caption{\blips for FDR control.}\label{alg::blip}
\algorithmicensure\, Candidate groups $\mcG$, PIPs $\{p_G\}_{G \in \mcG}$, a weighting function $\weight$, a nominal level $q$.
\begin{algorithmic}[1]
        \State Perform adaptive preprocessing on $\mcG$ using the PIPs (Appendix \ref{subsec::prefilter}).
        \smallskip
        \State Solve the relaxed LP corresponding to (\ref{eq::power_reduction})-(\ref{eq::disj_red}) to obtain a solution set $\{x_G\opt \}_{G \in \mcG}$. Let $\mcH = \{G \in \mcG : x_G\opt \not \in \{0,1\}\}$ denote the non-integer solutions. 
        \smallskip 
        \State Fix $x_G=x^\star_G$ for $G\in\mcG\setminus\mcH$ and run the integer LP (\ref{eq::power_reduction})-(\ref{eq::intconstraint}) on $\{x_G : G \in \mcH\}$ and $v$, yielding (integer) solutions $\{x_G^{\star\star} : G \in \mcG\}$.
        \smallskip 
        \State If the integer LP in Step 3 is feasible, detect signals in $\{G : x_G^{\star\star} = 1\}$ and terminate. \newline Else, define $G_{\min} \defeq \arg\min_{\{G : x_G\opt = 1\}} p_G$, set $\mcH = \mcH \cup G_{\min}$, and return to Step 3.
	\end{algorithmic}  
\end{algorithm}

Lastly, we can further speed up computation by \textit{adaptively preprocessing} the candidate groups $\mcG$. The main idea is that after looking at the data, we can tell that certain groups and locations have a very low chance of containing a signal, so we exclude these groups from consideration before running the LP. We develop three algorithms to accomplish this, although for brevity, we defer details to Appendix \ref{subsec::prefilter}. Often, these heuristics improve computation by multiple orders of magnitude, since when the signals are sparse, most groups and locations have a low posterior probability of containing a signal and can be discarded. That said, it is important to note that this step in no way compromises \blip's provable Type I error guarantees, and furthermore, these common sense heuristics have almost no impact on power empirically. Intuitively, this makes sense: if a group $G$ has only a $0.001\%$ chance of containing a signal, it is extremely unlikely we would have discovered it anyway, so there is no need to consider it in the LP. Note that unlike methods which by default restrict $\mcG$ to be very small, adaptive preprocessing does not sacrifice the adaptability of \blip, since it uses the full data to prune the candidate groups $\mcG$. Overall, these heuristics substantially improve computation and we have never observed them to meaningfully reduce power, so we apply them by default for the rest of the paper.

The overall runtime of \blips is dominated by a single large sparse linear program, whose computational complexity is at most $O(|\mcG_0|^2 |\mcL_0|)$ \citep{boyd2004}, where $\mcG_0 \subset \mcG, \mcL_0 \subset \mcL$ are the candidate groups and locations (respectively) that were not eliminated by the preprocessing (recall that in sparse problems, $|\mcG_0| << |\mcG|$ and $|\mcL_0| << |\mcL|$). Additionally, LP solvers may in practice be much faster than their worst case performance; usually, it is possible to solve LPs with millions of variables and constraints \citep{boyd2004}. Further speedups can be obtained by exploiting the sparsity of the LP \citep{leeaaron2015}. Overall, in Sections \ref{sec::sims}-\ref{sec::astro}, we find that running \blips is always less expensive than computing its input PIPs.

\begin{figure}[!ht]
\centering
\includegraphics[width=1\linewidth]{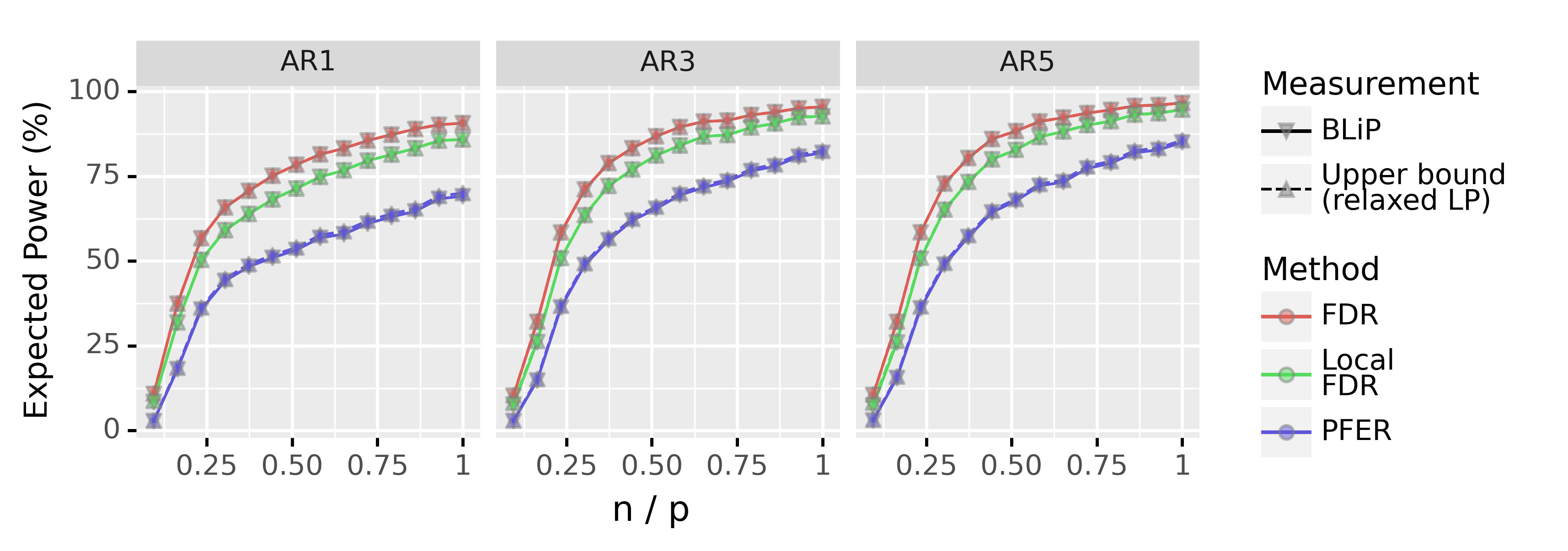}
\label{fig::objgap} 

\caption{We sample $X \sim \mcN(0, \Sigma)$ for various $\Sigma$ and $Y \mid X \sim \mcN(X \beta, 1)$,  where $\beta$ has $50$ nonzero components sampled from $\mcN(0,1)$ and $p = 1000$. Here, the facets correspond to different $\Sigma$: see Appendix \ref{subsec::glmdetails} for precise details on the setting. We run Algorithm \ref{alg::blip} on top of a generic spike-and-slab MCMC algorithm with greater than $50,000$ candidate groups, and we plot the objective function (expected power) achieved by \blips and the relaxed LP. We also run \blips to control the PFER and local FDR as discussed in Section \ref{subsec::blipother}. In all cases, the power of the relaxed LP is an upper bound on the expected power of any valid resolution-adaptive method. For all error rates, the upper bound completely overlaps with \blip's expected power, which suggests \blips is effectively optimal in this setting. Note that we can create an analogous plot for the FWER, but its interpretation is slightly more complicated, so we defer it to Appendix \ref{subsec::knapsack}.
Error bars denote two standard errors, but they are so small they are often not visible.
}\label{fig::blipintsol}
\end{figure}

Before discussing other error rates, we briefly discuss the heuristic claim that \blips finds ``nearly" the optimal grouping of the locations in the candidate set $\mcG$. As discussed in Appendix \ref{subsec::knapsack}, approximate optimality follows from the fact that the relaxed LP usually admits a solution which is almost entirely composed of integers. As a result, the expected power (objective function value) achieved by \blips is extremely close to the expected power obtained by the non-integer LP solution, which is an upper-bound on the maximum achievable power. Of course, even when this not true, one can compute and compare the expected power achieved by the relaxed LP and \blip, so \blips also comes with ``warning lights" which signal when it is not optimal. However, Figure \ref{fig::blipintsol} confirms empirically that in three challenging high-dimensional variable selection problems with tens of thousands of candidate groups, the nominal expected power achieved by \blips is indistinguishable from the upper bound provided by the relaxed LP. This suggests \blips is finding effectively the optimal solution. Lastly, even if there are rare cases when \blips is not optimal, our simulations and data applications suggest that \blips is more powerful than its competitors in practical settings.

\subsection{Other error rates}\label{subsec::blipother}

In this section, we introduce variants of \blips that control the local FDR and FWER. We start with the local FDR, which requires that the PIP for each discovered group must be at least $1 - q$. To control the local FDR, we can start by letting $\mcJ = \{G \in \mcG : p_G \ge 1 - q\}$ denote the set of groups which have a PIP of at least $1-q$. Then, we can solve the following relaxed linear program to optimize for power while controlling the local FDR:
\begin{align}
    \max_{\{x_G\}_{G \in \mcJ} : x_G \in [0,1]} \,\,\,\,\,\,\,\, & \sum_{G \in \mcJ} p_G \weight(G) x_G \nonumber \\
    \mathrm{s.t.} \,\,\,\,\,\,\,\,\,\,\,\,\,\,\,\,\,\,\, & \sum_{G \in \mcJ : \ell \in G} x_G \le 1 \,\,\,\,\,\, \forall \ell \in \locations. \nonumber
\end{align}
As before, once we solve this LP, most of the solutions will be integers, and we can apply an integer LP or polynomial-time heuristic to the non-integer values to obtain discoveries $G_1, \dots, G_R$, just as in Algorithm \ref{alg::blip}. These discoveries trivially control the local FDR, since we have already filtered out groups $G$ which have a PIP below $1 - q$.

The last error rate we will explicitly address is the family-wise error rate (FWER), which requires that there is at most probability $q$ that we make any false discoveries, or equivalently, that $\P(V > 0 \mid \mcD) \le q$. Since $\P(V > 0 \mid \mcD) \le \E[V \mid \mcD]$, one valid way to control the FWER at level $q$ is to ensure $\E[V \mid \mcD] \le q$, which is also known as controlling the per-family error rate (PFER)---indeed, this is the approximation used by the standard Bonferroni correction. This is slightly conservative but works well empirically when $q$ is fairly small, e.g., if $q \lessapprox 0.2$. The advantage of making this approximation is that $\E[V \mid \mcD] = \sum_{G \in \mcG} (1 - p_G) x_G$ is linear in $\{x_G\}$. As a result, to control the FWER, we can run Algorithm \ref{alg::blip} but replace the FDR constraint (\ref{eq::fdr_red}) with the linear constraint $\E[V \mid \mcD] \le q$. Another alternative is to control the PFER at level $v$ and perform a binary search over $v$ to find the largest value $v\opt$ such that $\P(V > 0 \mid \mcD) \le q$. Either method is guaranteed to control the FWER, though the latter will in general have slightly increased power at the cost of some extra computation.

By adding extra constraints or modifying the weights in existing constraints, one can easily modify \blips to control various other error rates, such as a weighted FDR \citep{benjamini1997, bogomolov2020}, or even to control the FDR across multiple hierarchies simultaneously \citep{pfilter2017, multilayerknock2019}. However, we leave this possibility for future work. 

\subsection{Choosing the candidate groups and computing the PIPs}\label{subsec::guidelines}

In this section, we discuss how to compute the primary inputs to \blip: the candidate groups $\mcG$ and the PIPs $\{p_G\}_{G \in \mcG}$. 
To begin with, how can one obtain good candidate groups $\mcG$? Our overall advice is as follows: up to computational limits, we suggest adding every conceivably useful group to $\mcG$, since adding more groups to \blips should only improve its power without affecting validity. Furthermore, \blips is usually efficient enough to handle hundreds of millions of candidate groups. That said, we recommend two general approaches to find potentially useful groups.

First, we can let $\mcG$ include the set of all \textit{contiguous} regions below some maximum size $m$. This approach makes sense when the locations have spatial or temporal structure that we can exploit. For example, suppose $\locations = \{\loc_1, \dots, \loc_p\}$ consists of $p$ ordered locations of genetic variants on the genome. In this context, a contiguous group is of the form $\{\ell_i, \ell_{i+1}, \dots, \ell_{i+k}\}$ for some $i, k \in \N$. Since genetic variants exhibit local biological similarities and mostly local correlations, considering contiguous groups is arguably more interpretable and likely more useful than considering groups of far-flung genetic variants. This option is also attractive in change point detection (Appendix \ref{subsec::changepointsim}), where the set of locations $\locations = \{1, \dots, T\}$ is a set of ordered times. Notably, there are roughly $m \cdot p$ contiguous groups of length $m$ or less when considering $p$ locations, so the number of candidate groups scales linearly with $p$. Lastly, when there are more dimensions of space and/or time, we suggest using spherical subsets of $\locations$ as candidate regions. For example, in our application to astronomical point source detection in Section \ref{sec::astro}, $\locations = [0,1]^2$, and we let $\mcG$ include the set of circles of radius $\epsilon$ centered at one of a few million equidistant lattice points in $[0,1]^2$, for many possible values of $\epsilon$. We introduce efficient algorithms for this setting in Appendix \ref{appendix::infloc}.

The second main approach we recommend is tailored to regression problems, where we seek to discover important variables among $X = (X_1, \dots, X_p)$. In this setting, we recommend applying many different clustering algorithms to $X$ and letting $\mcG$ denote the union of those groups. For example, one could generate candidate groupings by hierarchically clustering $X$ based off its correlation matrix, or its partial correlation matrix \citep{pcacluster2012}, or any combination thereof. Furthermore, we re-run these algorithms many times using different tuning parameters to add more candidate groups to $\mcG$. Finally, there is no need to choose between multiple approaches: when $(X_1, \dots, X_p)$ exhibit spatial structure, we can combine this approach with that of the previous paragraph.

The second important input to \blips is the set of PIPs $\{p_G\}_{G \in \mcG}$. Since one can compute the PIPs based on the output of nearly any Bayesian MCMC algorithm or variational method, the question of how to compute the PIPs is essentially orthogonal to our contribution (\blip), so we largely defer discussion of computing $\{p_G\}_{G \in \mcG}$ to Appendix \ref{sec::mcmc}, where we give detailed suggestions. However, in this section, we will touch upon one important point, which is that of robustness. By construction, \blips provably controls various Bayesian error rates, such as the FDR, under the assumption that the PIPs $\{p_G\}_{G \in \mcG}$ are correct. However, in many practical settings of interest, (a) Bayesian MCMC algorithms may not fully converge or (b) one may not have an accurate Bayesian prior. This is indeed an important concern: \blips is only as good as its inputs, and like most Bayesian selective inference procedures, it will violate error control if the PIPs are inaccurate.

With this motivation, in Appendix \ref{sec::mcmc}, we offer guidelines on how to compute PIPs in a way that is robust to these concerns. In brief, our advice is as follows. To address (a), we generally recommend sampling from multiple MCMC chains (or variational algorithms, when appropriate) with uniformly random initialization. Even when each individual chain fails to converge, we argue in Appendix \ref{sec::mcmc} that this approach will typically overstate the uncertainty in the location of a signal, leading to conservative but valid inference. To address (b), we recommend using \textit{hierarchical priors} with conservative or estimated hyper-parameters to account for uncertainty in the Bayesian model, such as modeling the sparsity level in regression problems. Since these are fairly standard techniques in Bayesian analysis, we defer details to Appendix \ref{sec::mcmc}. Furthermore, in case there is non-negligible uncertainty in the PIPs, we note the possibility of replacing $p_G$ in Equation (\ref{eq::fdr_red}) with $p_G^{\mathrm{lower}}$, a high-confidence (or uniform) lower bound on $p_G$. For example, $p_G^{\mathrm{lower}}$ could account for uncertainty in the choice of prior or uncertainty due to MCMC error, although empirically this did not seem to be necessary in our experiments. This is equivalent to running a robust linear program \citep{robustopt2009}, and it is easy to see that if $p_G \ge p_G^{\mathrm{lower}}$ does hold for all $G$, then \blips will still provably control the FDR at level $q$. Overall, in Section \ref{sec::sims}, we often use \blips in combination with fairly uninformative priors for fairer comparison with frequentist methods. Despite this, \blips is very powerful and consistently controls the FDR, even in high-dimensional settings where we have no reason to believe MCMC algorithms will fully converge.

\section{Simulations}\label{sec::sims}

In this section, we demonstrate that \blips is powerful, robust, and efficient compared to its competitors. We focus on the case of variable selection in Gaussian linear models and probit regression, although Appendix \ref{subsec::changepointsim} also includes simulations which show that \blips holds promise as a method for change point detection. Note that Appendices \ref{subsec::pipguide} and \ref{subsec::gwassims} contain further simulations related to the robustness of \blips and genetic fine-mapping, respectively. All simulations were run on cores with 4 gigabytes of memory.

\subsection{Variable selection for Gaussian linear models}\label{subsec::linregsim}

We begin by applying \blips to perform variable selection in (sparse) Gaussian linear models. We simulate $Y \mid X \sim \mcN(X \beta, \sigma^2)$ where $\beta$ has $\lceil s p \rceil$ randomly chosen nonzero coefficients, for sparsity $s \in (0,1)$. The nonzero coefficients are i.i.d. $\mcN(0, \tau^2)$ random variables. The locations $\locations = [p]$ correspond to $(X_1, \dots,X_p)$, and we say $I_j = 1$ if $\beta_j \ne 0$. To capture a challenging setting where the variables are correlated, we sample $\bX$ from a nonstationary AR(k) model, meaning $\bX$ exhibits high local correlations with a complex structure---see Appendix \ref{appendix::simdetails} for details. Unless otherwise specified, we set $k = 5$ and $p = 1000$, and we control the FDR at level $q = 0.1$. We plot resolution-adjusted power as a percent, defined as the resolution-adjusted power divided by the total number of true signals. Note that given correct PIPs, \blips provably controls the FDR conditional on the data. Thus, it also should control the FDR in our plots, which average over the randomness in both the data and the data-generating parameters, e.g. $\beta$. (This is common practice even in frequentist papers.) Of course, in practice, we cannot perfectly compute the PIPs due to misspecification of the prior or other error in approximating the posterior. Thus, these simulations also assess \blip's robustness to (somewhat) inaccurate PIPs.

We run \blips on top of two types of models. First, we apply \blips on top of a standard Gibbs sampler for linear spike-and-slab (LSS) models, and we consider both a nearly well-specified case, where the sampler uses the true values of $s, \tau^2$ and $\sigma^2$, and a misspecified case, where the sampler uses uninformative priors on $s$, $\tau^2$, and $\sigma^2$; see Appendix \ref{subsec::lss} for detailed model descriptions.\footnote{The ``nearly well-specified" case is not perfectly well-specified because the prior assumes there are $\Bin(p, s)$ signals, whereas in our simulations there are always exactly $\lceil sp \rceil$ signals.} Second, we apply \blips on top of SuSiE \citep{susie2020}. We consider four main competitors: the focused Benjamini Hochberg (FBH) procedure \citep{fbh2021}, the ``Yekutieli" procedure \citep{yekutieli2008}, SuSiE, and DAP-G \citep{dapg2018}. The FBH and Yekutieli act on p-values, which we obtain using a lasso-based distilled conditional randomization test (dCRT) \citep{mxknockoffs2018, dcrt2020}. We used the dCRT because it was powerful in our simulations and because it allows us to obtain frequentist p-values in high-dimensional settings. We allow SuSiE/DAP-G to use the true sparsity level $s$. See Appendix \ref{appendix::simdetails} for more details, including details on the method we used to generate candidate groups.

\begin{figure}[!h]
    \centering
    \includegraphics[width=\linewidth]{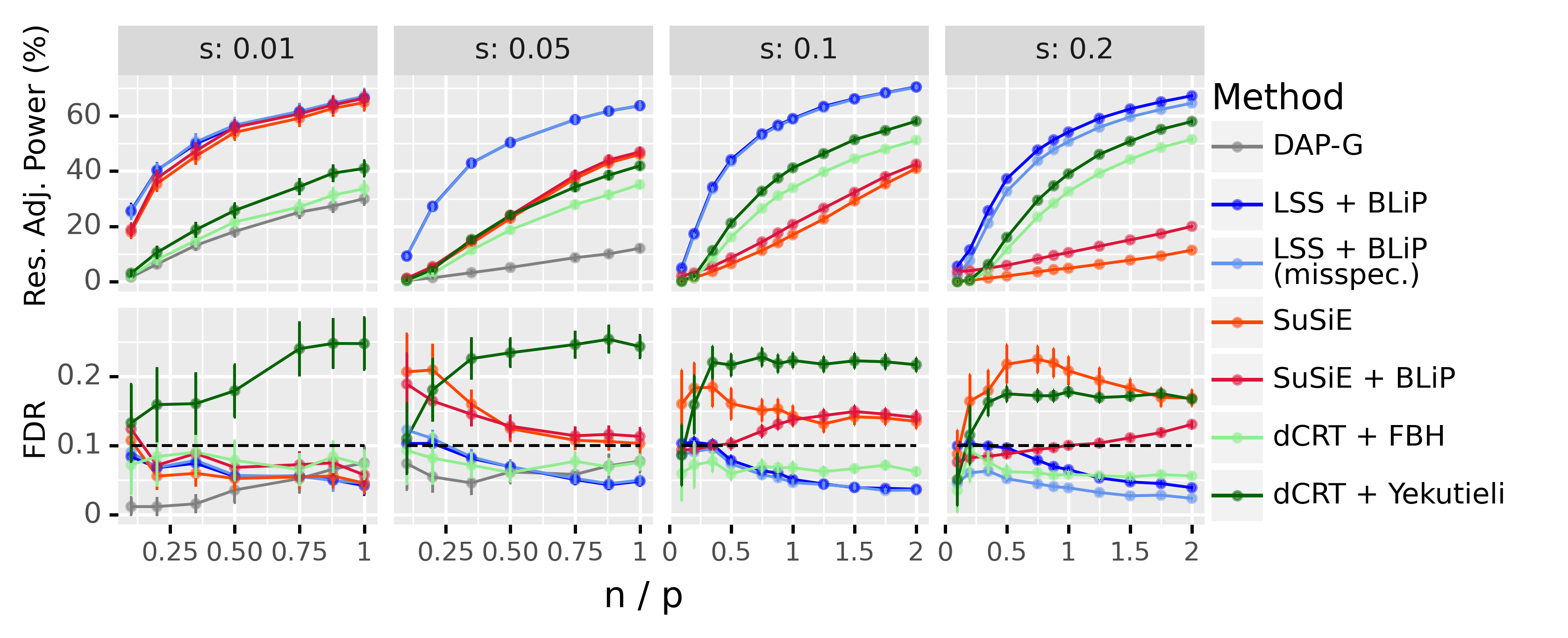}
    \caption{Resolution-adaptive variable selection for Gaussian linear models. Note that DAP-G is prohibitively expensive for $s > 0.05$, so we do not run DAP-G except in the left-most panels.}
    \label{fig::linear_hdim}
\end{figure}

Figure \ref{fig::linear_hdim} shows the (resolution-adjusted) power and FDR for each method while varying the number of data-points $n$ and the sparsity $s$. It shows that LSS + \blips is uniformly the most powerful method and reliably controls the FDR, even in high-dimensional settings where $p = 10n$ and the prior is misspecified. Remarkably, LSS + \blips (misspec.) always achieves the same power as the (nearly) well-specified LSS model with oracle knowledge of the hyperparameters. Besides LSS + \blip, only FBH and DAP-G reliably control the FDR, and LSS + \blips often has double their power (or more). Similarly, LSS + \blips is often many times more powerful than SuSiE and the Yekutieli procedure, even though these methods violate FDR control. Furthermore, SuSiE + \blips has uniformly higher power than SuSiE and often simultaneously improves upon SuSiE's FDR control. Overall, Figure \ref{fig::linear_hdim} confirms the intuition from Section \ref{subsec::literature} about why \blips outperforms its competitors, which we discuss in the next two paragraphs.

First, SuSiE behaves as suggested in Section \ref{sec::intro}. Recall that (roughly speaking) each iteration of SuSiE makes a variational approximation assuming that there is exactly one additional signal. When $s = 0.01$ and there are $10$ signals total, this approximation is closest to the truth and SuSiE nearly matches the power of LSS + \blip. Yet as $s$ increases, the approximation becomes less accurate. As a result, SuSiE has much lower power than LSS + \blips and violates FDR control in each of the other panels. Note that SuSiE's variational approximation is inaccurate when the absolute number of signals is large, so SuSiE may perform poorly even in very sparse problems with large $p$ (see Appendix \ref{subsec::glmdetails} for a simulation confirming this). However, applying \blips on top of SuSiE can remedy this problem to some extent. Indeed, any disjoint output from SuSiE is a feasible output for \blip, so we expect \blips to uniformly improve SuSiE's power, which is supported by all of our simulations. Furthermore, \blips can simultaneously improve FDR control, as shown in the panels where $s \ge 0.1$. See Appendix \ref{appendix::susie} for details on how SuSiE's approximation breaks down and how \blips can partially correct this problem.

LSS + \blips also has uniformly higher power than the FBH/Yekutieli procedures, often by a factor of two or more. We suspect that \blips gains power because \blips can search over hundreds of times more candidate groups than the FBH/Yekutieli, which are restricted to search over a single hierarchical tree, and also because \blips explicitly maximizes power when searching over candidate groups. In contrast, the FBH/Yekutieli only search heuristically over their input p-values and may not find a maximally powerful rejection set. (Note that without a hierarchical structure, the FBH is not guaranteed to return a disjoint rejection set while provably controlling the FDR.) That said, there are many differences between \blips and the FBH/Yekutieli, so we cannot perfectly account for this power differential. For example, we implemented the most powerful versions of the FBH/Yekutieli we could subject to computational constraints, but perhaps there are more powerful p-values that could be used with the frequentist methods, although we were unable to find such p-values. Overall, we hope that the results speak for themselves, since either way it is not clear how one could apply the FBH/Yekutieli procedures in a way that is more powerful than LSS + \blip. Note lastly that the Yekutieli procedure fails to control the FDR in any setting, presumably because it only has FDR guarantees when its input p-values are fully independent. The p-values also do not provably satisfy the conditions under which the FBH controls the FDR, but the FBH seems to be more robust.

\begin{figure}
    \centering
    \includegraphics[width=0.65\textwidth]{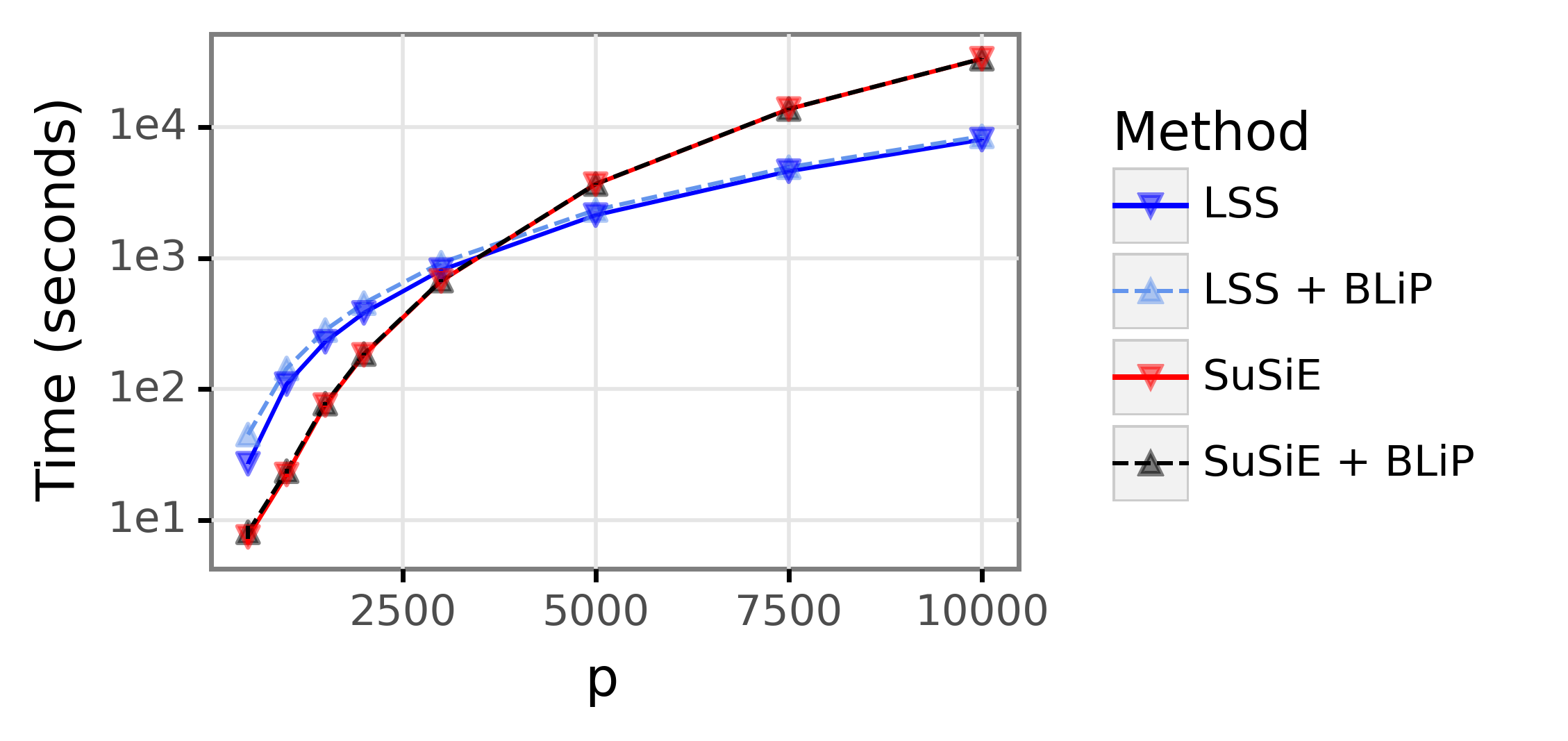}
    \caption{This figure shows computation time while varying $p$ with $n=p/2$ and $s=0.05$. It distinguishes between the computation required to fit the underlying model and the total time to both fit the model and run \blip. It shows that the cost of running \blips is effectively negligible compared to fitting a variational method (SuSiE) or Gibbs sampling in a spike-and-slab model. All methods controlled the FDR (see Appendix \ref{subsec::glmdetails}).}
    \label{fig::linear_vp}
\end{figure}

Lastly, Figure \ref{fig::linear_vp} analyzes the runtime of \blips in large-scale settings, with $s = 0.05, n= 0.5p$ and $p$ varied from $500$ to $10,000$. It shows that the cost of applying \blips is trivial compared to the cost of running SuSiE or the LSS sampler: when $p = 10,000$, \blips runs in a few minutes, whereas fitting LSS and SuSiE requires $2$ and $9$ hours, respectively. 

\subsection{Variable selection for probit regression}\label{subsec::binregsim}

We now apply \blips to variable selection for probit regression. In particular, 
consider the same setting as Section \ref{subsec::linregsim}, except instead of observing $Y$, we observe $Z = \I(Y \ge 0)$. We apply \blips on top of a Gibbs sampler for this probit spike-and-slab (PSS) model,  using the data augmentation scheme from \cite{albertchib1993}.  See Appendix \ref{subsec::pss} for more details on the Gibbs sampler. We compare to the FBH/Yekutieli procedures, which we apply to p-values from a logistic-lasso-based dCRT. Although the PSS sampler has the same asymptotic runtime as the LSS sampler $(O(n_{\mathrm{iter}} np))$, in practice it is roughly an order of magnitude slower due to the data augmentation step. Indeed, inference is generally slower and more challenging for binary data---for this reason, many applied analyses in, e.g., genetic fine-mapping simply apply methods designed for linear regression to binary outcomes \citep{polyfun2019}. For this reason, we also compare to SuSiE and LSS + \blip, as in Section \ref{subsec::linregsim}. 

\begin{figure}
    \centering
    \includegraphics[width=\linewidth]{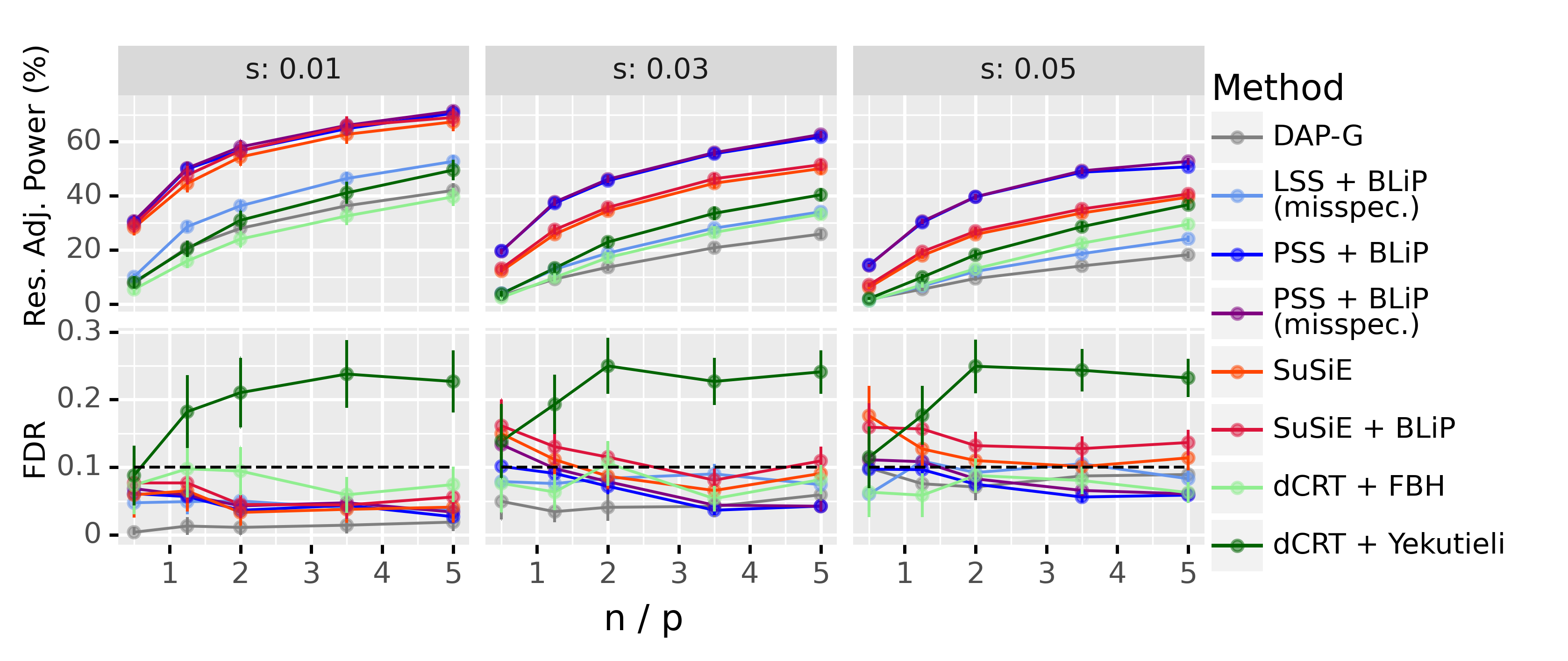}
    \caption{Power and FDR for resolution-adaptive variable selection in the probit model. Note that both the well-specified and misspecified versions of PSS + \blips are present in all panels, although usually their power curves entirely overlap. See Appendix \ref{subsec::glmdetails} for more details.}
    \label{fig::binreg_power_fdr}
\end{figure}

Figure \ref{fig::binreg_power_fdr} shows the results in three settings with $p = 400$, with varied $s$ and $n$. The results show that PSS + \blips is uniformly the most powerful, often by a wide margin. SuSiE + \blips has (uniformly) slightly higher power than SuSiE alone, with a slightly higher realized FDR (although occasionally the realized FDR is lower). This suggests that \blips can enhance power in binary regression problems as well as linear regression.

\section{Application to genetic fine-mapping}\label{sec::gwas}

As discussed in Section \ref{sec::intro}, resolution-adaptive methods are particularly attractive in fine-mapping problems, where genetic variants are often very highly correlated with each other. As a result, it can be very challenging to detect individually important genetic variants. Resolution-adaptive methods instead allow the analyst to localize causal variants as precisely as possible given the data at hand, and for this reason, a few recent works \citep{polyfun2019, susie2020, wallace2021} have used resolution-adaptive methods in fine-mapping problems. Furthermore, Bayesian variable selection methods are already commonly used in genetic fine-mapping problems \citep{guanstephens2011, carbonettostephens2012, benner2015, dapg2018, polyfun2019}. All this suggests that \blips can help solve an important problem in the domain of fine-mapping.

\begin{figure}
    \includegraphics[width=\linewidth]{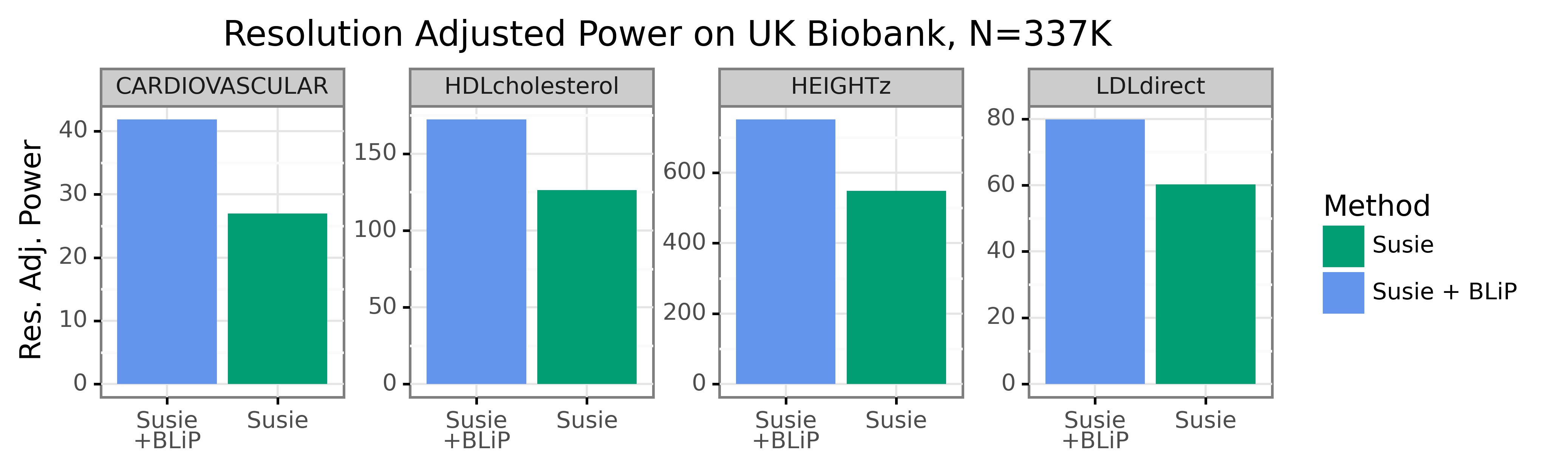}
    \caption{This figure shows the resolution-adjusted number of discoveries made by each method in our application to UK Biobank data. It shows that \blips increases SuSiE's power by $30$-$50\%$.}
    \label{fig::gwas_power}
\end{figure}

\begin{figure}
    \centering
    \includegraphics[width=\linewidth]{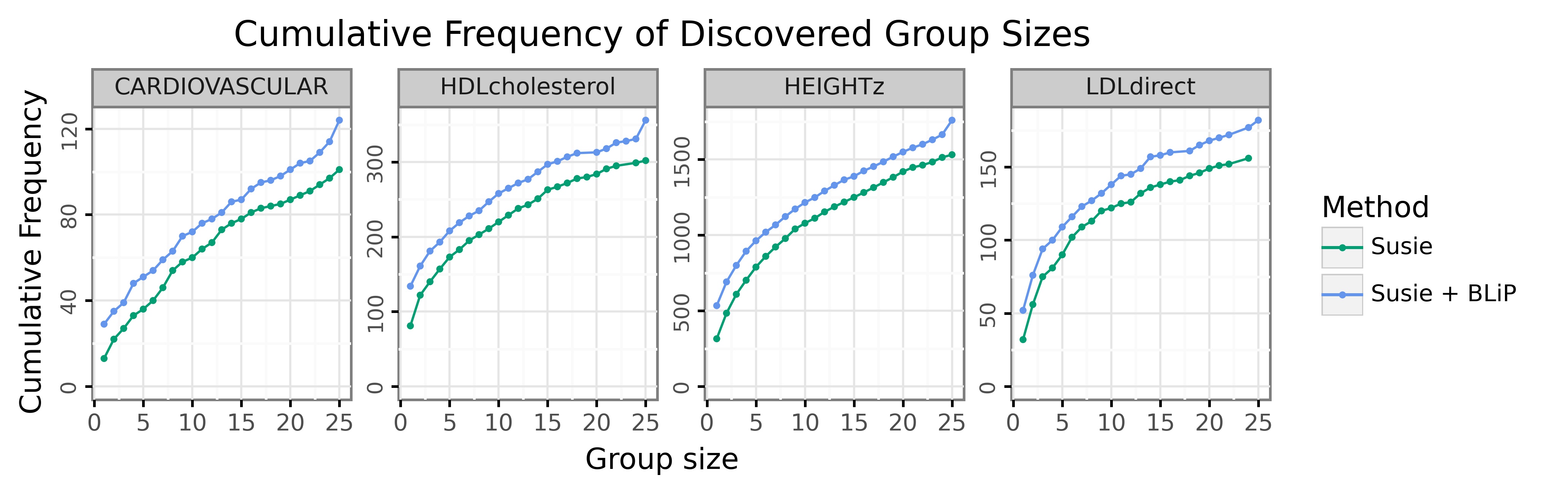}
    \caption{This figure plots the cumulative frequency of the discovered group sizes. To be precise, the point with x-value $k$ on the green curve (resp. blue curve) counts the number of groups of size $k$ or less discovered by SuSiE (resp. SuSiE + \blip). This figure shows that SuSiE + \blips makes more discoveries at finer resolutions than SuSiE alone, i.e., \blips detects more causal variants and localizes them more precisely.}
    \label{fig::gwas_groupsize}
\end{figure}

To test \blip's effectiveness in fine-mapping problems, we apply \blips to a dataset of $n \approx 337,000$ individuals from the UK Biobank with $p \approx 19,000,000$ genetic variants. In particular, we seek to identify causal genetic variants for four traits of interest: cardiovascular disease, height, low-density lipoprotein (LDL) cholesterol, and high-density lipoprotein (HDL) cholsterol. This dataset was previously analyzed by \cite{polyfun2019}, and indeed, our work explicitly builds upon theirs. SuSiE is an attractive model in this setting because we expect a priori that each genetic locus has a small number of causal variants, and our simulations suggest that SuSiE performs nearly as well as full Bayesian inference when the number of signals is small. Thus, we run \blips directly on top of the SuSiE model that \cite{polyfun2019} fit on this dataset to identify causal genetic variants. Running \blips requires less than $1$ minute of computation per trait, but as shown by Figure \ref{fig::gwas_power}, SuSiE + \blips is consistently 30--50\% more powerful than SuSiE alone. Crucially, this result is not at all sensitive to our definition of resolution-adjusted power: as shown by Figure \ref{fig::gwas_groupsize}, for every $k$, SuSiE + \blips discovers more groups of size $k$ or less than SuSiE alone, meaning that SuSiE + \blips is making more discoveries at finer resolutions by nearly any metric. Indeed, for every group $G$ discovered by SuSiE, SuSiE + \blips discovers a group $G'$ which overlaps with $G$. This suggests that \blips is successfully optimally localizing signals based on the information available in the SuSiE model---see Appendix \ref{appendix::susie} for more intuition on why SuSiE + \blips can substantially outperform SuSiE alone. Notably, SuSiE + \blips makes more singleton discoveries than SuSiE alone, in part because SuSiE + \blips can use PIPs which are provably more powerful than the default SuSiE algorithm---however, we caution that discovering singleton groups is not the primary purpose of \blips and the interpretation of this result is subtle, so we discuss this further in Appendix \ref{subsec::singletons}. Lastly, note that each group we discover is roughly (but not perfectly) contiguous, meaning that each group only contains nearby genetic variants. As a result, detecting a signal in a group of size $k$ is fairly interpretable: it tells us that one of $k$ neighboring genetic variants has a causal effect on a trait. See Appendix \ref{subsec::gwasmethods} for methodological details.

To validate our findings, we confirm in simulations using the real genotype data that SuSiE + \blips controls the FDR in this context, as shown in Appendix \ref{subsec::gwassims}. Furthermore, we compare our findings to those of previous work. To start, as a sanity check, we compare the discoveries from SuSiE + \blips with those of the SuSiE model from \cite{polyfun2019} (this is the model represented by the green bars in Figure \ref{fig::gwas_power}).\footnote{Note that this is not a replication analysis, since both analyses use the same dataset and model.} Appendix \ref{subsec::gwascomp} shows that SuSiE + \blips replicates every finding from the SuSiE model but makes roughly $15$--$20\%$ more discoveries (note this number is not resolution-adjusted). Since SuSiE + \blips has $30$--$50\%$ higher resolution-adjusted power than SuSiE, this indicates that the power gain comes both from more precisely localizing existing discoveries and from making some entirely new discoveries. Crucially, of the new discoveries made by SuSiE + BLiP, we found that $45$--$65$\% could be corroborated by a separate study in the NHGRI-EBI GWAS Catalog \citep{gwascatalog2019}, which is comparable to the corroboration rate of the initial analysis from \cite{polyfun2019}. 
This is arguably a remarkable (and positive) result, since a priori one might expect that any novel discoveries would informally be ``harder to discover" and thus corroborate, precisely because the initial model did not discover them. Nonetheless, the \textit{additional} discoveries from SuSiE + BLiP were corroborated at a similar rate to the original discoveries.
See Appendix \ref{subsec::gwascomp} for more details, as well as an additional comparison to \cite{farh2015}. Overall, these results give further evidence that \blips meaningfully enhanced SuSiE’s power to find real causal variants, and they give no indication that the increased resolution-adjusted power of SuSiE + \blips can be explained away as false discoveries. All code, data, and results are publicly available at \url{https://github.com/amspector100/ukbb_blip}.

Lastly, we remind the reader that \blips can be applied on top of any Bayesian model, yielding higher power with very little additional computational cost. For example, in this section, SuSiE uses a relatively uninformative prior for simplicity, but several recent works have incorporated information from other sources into the prior, including functional annotations, other complex traits, and prior knowledge about the distribution of genetic effect sizes \citep{polyfun2019, oconnor2020, weeks2019, trippe2021}. Similarly, the fine-mapping literature contains many inferential algorithms besides SuSiE  \citep{carbonettostephens2012, caviar2014, benner2015, fastpaintor2016}, and some works have even used full Bayesian MCMC \citep{zhao2019}. \blips can wrap on top of any of these methods, and, we hope, enhance their power to make meaningful scientific discoveries.

\section{Application to astronomical point source detection}\label{sec::astro}

In this section, we apply \blips to the problem of astronomical point source detection. To formalize this problem, suppose that $\locations = [0,1]^2$ corresponds to some patch of the night sky, and we want to detect point sources (e.g., stars) in $\locations$. We assume we observe data $\{D_{i,j}^{(b)}\}_{i, j \in [d], b \in [B]}$. Here, $b$ denotes the \textit{band} of an image (i.e., color), and $D_{i,j}^{(b)}$ (informally) counts the photons observed by a telescope in the $(i,j)$th pixel of an image, i.e., the subregion $\left[\frac{i-1}{d}, \frac{i}{d}\right] \times \left[\frac{j-1}{d}, \frac{j}{d}\right]$. If a source exists at $(x,y) \in \locations$, we expect the pixels close to $(x,y)$ to be brighter, allowing us to detect the star. However, the exact location of the source is always uncertain, both because of the discrete nature of the pixels and also because there are often multiple sources in the same pixel---indeed, due to the increasing sensitivity of telescopes, \cite{feder2020} observed that astronomical images have become increasingly crowded over time. This phenomenon makes it difficult to discover as many point sources as possible, and it also causes ambiguity when deciding whether a bright pixel represents a single point source or two nearby point sources. The latter problem is particularly important since many downstream applications will yield misleading results unless sources are properly separated \citep{portillo2017}, such as analyses which attempt to classify point sources based on, e.g., their intensity. These challenges have motivated the development of Bayesian models which can simultaneously capture uncertainty about both the number and locations of sources, including approaches which can scale to catalogue massive astronomical surveys \citep{regier2019, starnet2021}. However, extracting interpretable results from these models can be challenging \citep{feder2020}.  We now show that \blips can solve exactly these types of problems. 

In this section, we apply \blips to a $100$ x $100$ pixel sub-image of the Messier 2 (M2) star cluster, taken by the Sloan Digital Sky Survey (SDSS). Several previous works \citep{portillo2017, feder2020, starnet2021} used this sub-image to validate point source detection methods, and indeed, our analysis directly builds on that of \cite{starnet2021}, who introduced StarNet, a Bayesian model for point source detection. The M2 region is a good testing ground because it was also imaged by the Hubble Space Telescope (HST) during the ACS Globular Cluster Survey \citep{hubble2007}. Since the HST has a much greater resolution than the Sloan telescope, previous works used the HST detections as the ground truth. In particular, \cite{starnet2021} compared StarNet's MAP estimates of source locations to the HST locations. To account for both measurement differences between telescopes and also the statistical uncertainty in the positions of sources, each of these previous works classified an estimated source position as a ``true source" if it was within $0.5$ pixels of an HST source. Equivalently, let $\mcM$ be the set of MAP estimates: for $(x,y) \in \mcM$, we can think of StarNet as detecting a signal in $C_{\epsilon}(x,y)$, where $C_{\epsilon}(x,y)$ is the square of radius $\epsilon$ centered at $(x,y)$. This procedure chooses $\epsilon$ before seeing the data, so it performs inference at a fixed resolution: we refer to it as the ``MAP baseline at resolution $\epsilon$," where \cite{starnet2021} picked $\epsilon = \frac{0.5}{100}$. Since \blips automatically accounts for statistical uncertainty in the location of sources, we use a more stringent standard: to account only for measurement differences, we classify each detected region from \blips as a true detection if an HST source lies within $0.01$ pixels of the region. (For fair comparison, we also give the baseline an additional $0.01$ pixels of slack.)  Despite this more stringent standard, we will see later that \blips outperforms the MAP baseline for any $\epsilon$.

\begin{figure}
    \centering
    \includegraphics[width=\linewidth]{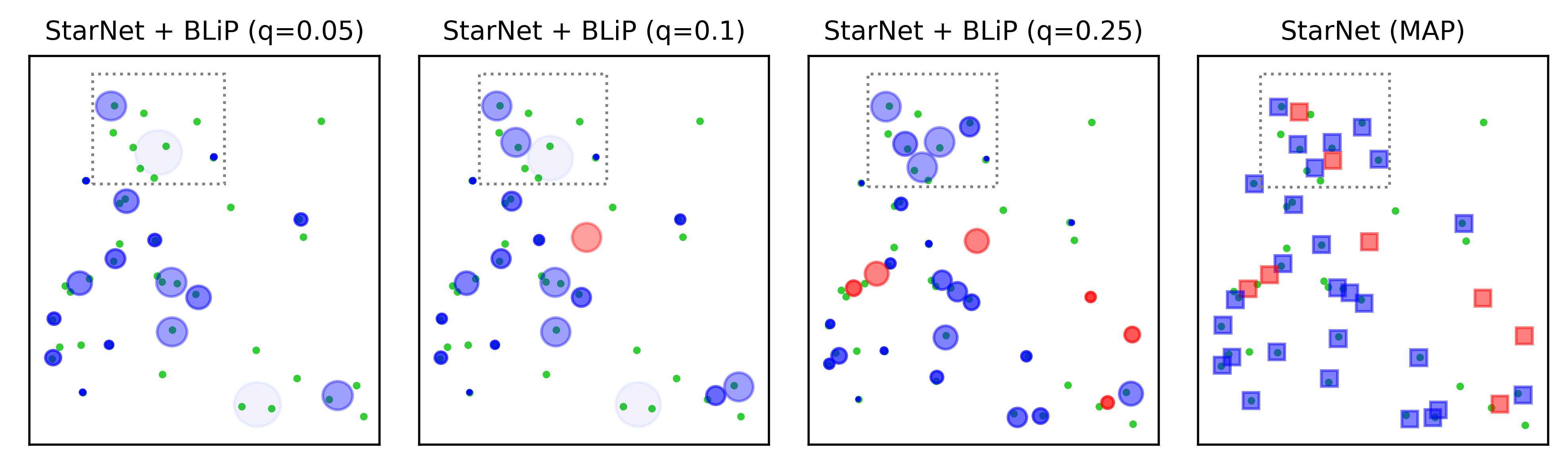}
    \caption{This figure shows the detections made by \blips for a $20$ x $20$ pixel  sub-image of the $100 \times 100$ M2 image using the ``inverse radius" weight function. True and false positives are shown in blue and red, respectively, and the opacity of each outputted region is proportional to the inverse radius. Sources detected by the HST are treated as the ground truth and shown in green. Note that \blips outputs circles of varying sizes, which reflects the varying degrees of uncertainty we have about the position of different sources. Furthermore, the right-most plot shows the output from \cite{starnet2021}, using their criterion for whether these are false positives. These plots show that at low nominal levels, \blips successfully filters out false positives from the set of MAP estimators. Furthermore, as the nominal level increases, \blips can better separate nearby sources while maintaining FDR control, as shown in the region highlighted by the dashed black box.}
    \label{fig::m2_picture}
\end{figure}

In particular, we apply \blips directly on top of posterior samples from \cite{starnet2021}'s pretrained StarNet model. Although in principle \blips can output regions of arbitrary shapes, for simplicity we use a set of roughly $280$ million overlapping circles of various radii as candidate groups. We applied \blips with two weight functions to proxy two possible astronomical motivations. First, we used an ``inverse radius" weight function, which assigns the weight $\frac{1}{\epsilon}$ to any candidate group with radius $\epsilon$. This reflects a preference for discovering smaller regions. However, some practitioners may care less about perfectly localizing signals as long as the signals are correctly \textit{separated}. In our second analysis, \blips outputs both a region $G \subset \locations$ and a contiguous credible interval $J \subset \N$ for the number of point sources in the region. For example, outputting $J = \{1, 2\}$ asserts that there are either one or two point sources in $G$. We use the ``separation-based" weight function $\weight(G, J) = \frac{1}{|J|}$, which gives \blips the incentive to precisely separate point sources whenever possible, since, e.g., discovering two regions containing exactly one star is four times as valuable as discovering a single region containing either one or two stars. For this second analysis, we count $(G, J)$ as a true discovery if and only if the number of Hubble stars within $0.01$ pixels of the region $G$ is an element of $J$. Of course, there are myriad other potential ways to apply \blip, including using different weight functions or even applying \blips in higher dimensions to simultaneously capture uncertainty about the flux of discovered sources. That said, we hope our two analyses demonstrate the flexibility of \blip, since it shows that \blips can be directly applied to pretrained models and also can accommodate varying scientific objectives. See Appendix \ref{appendix::astro} for further methodological details, and see Section \ref{sec::discussion} for a discussion of more extensions of \blip.

\begin{figure}
    \centering
    \includegraphics[width=\linewidth]{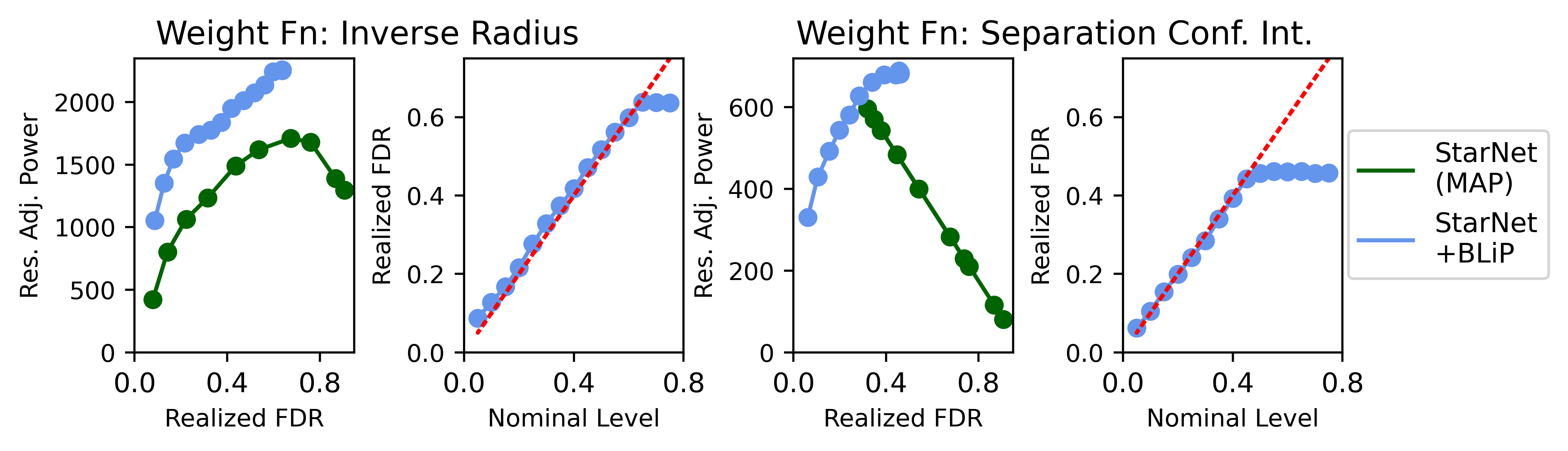}
    \caption{This figure shows the power and FDR of StarNet + \blips at various nominal levels compared to the power of the fixed-resolution baseline based on MAP estimation. It shows that StarNet + \blips successfully controls the FDR and simultaneously outperforms the baseline. The different dots on the green curves correspond to varying the fixed-resolution $\epsilon$. Note that for the right-hand power plot, the StarNet (MAP) points are co-linear because according to the weight-function, each region discovered by the baseline gets the (maximal) weight of one, no matter the resolution. As a result, the realized power is a linear function of the realized FDR. Furthermore, the green line does not extend any further (the power of the baseline never goes higher for any fixed resolution).}
    \label{fig::m2_power_fdr}
\end{figure}

In Figure \ref{fig::m2_picture}, we plot the detections from \blips at various nominal levels against the true HST sources and the MAP estimates from \cite{starnet2021}.  The figure shows that \blips is able to adaptively quantify uncertainty in the locations of the sources and simultaneously filter out clear false positives from the set of MAP estimates. To formally test error control, we run \blips to control the FDR at level $q$ for various values of $q$ and compare its discoveries to the HST sources. As depicted in Figure \ref{fig::m2_power_fdr}, \blips controls the FDR close to the exact nominal level for both weight functions, although for large $q$ it becomes conservative due to the adaptive preprocessing steps (Appendix \ref{subsec::prefilter}). We interpret the remarkable alignment between realized and nominal FDR as a testament both to \blip's capacity to control the FDR and to the pretrained StarNet model, which appears to be very well calibrated. 

Lastly, to analyze the power of \blip, we compare StarNet + \blips to the MAP baseline from \cite{starnet2021}, applied at various fixed resolutions $\epsilon$. Figure \ref{fig::m2_power_fdr} shows that \blips achieves a much better power-FDR trade-off than the baseline. For example, when controlling the FDR at $q=0.05$, StarNet + \blips is more than twice as powerful than the MAP-based approach using the inverse-radius weight function. (Note that in addition to providing higher power for any realized FDR, \blip's output also comes with an FDR guarantee, unlike the MAP approach.) Similarly, for both weight functions, StarNet + \blips run with $q = 0.30$ is more powerful than the baseline for \textit{any} fixed resolution, even when the realized FDR for the baseline is as high as $90\%$. This is the benefit of resolution-adaptivity: whereas the baseline must discover each source at some fixed resolution $\epsilon$, \blips can use the data to account for the fact that we may be more uncertain of the location of some sources than others. And for the same reason, \blips is better able to quantify uncertainty about whether two potential point sources have been properly separated for the ``separation-based" weight function. In contrast, StarNet (MAP) always has a realized FDR above $30\%$. Overall, \blips simultaneously controls the FDR and achieves much higher power than the MAP-based approach for both weight functions. Taken together, these results suggest that \blips holds promise as a method for uncertainty quantification in point source detection. All code and data are publicly available at \url{https://github.com/amspector100/DeblendingStarfields}. Note that the whole analysis ran in under ten minutes on a laptop and required 11 gigabytes of memory.

\section{Discussion}\label{sec::discussion}

This paper introduces \blip, a powerful method for performing resolution-adaptive signal detection. Our simulations and two applications show that \blips is computationally efficient and very powerful while simultaneously providing provable error control. Another notable benefit of \blip, however, is that it is a highly flexible method, and it also solves a very general problem. As a result, there are many possible extensions of \blips which may be of interest.

First, in this paper, we use \blips to maximize expected power while controlling the FDR or other error rates, but in principle, \blip-like algorithms could optimize for different objective functions under different constraints. For example, it is fairly straightforward to tweak \blips to maximize the harmonic mean of the expected true positive rate and expected power, also known as the F1 score (see Appendix \ref{subsec::balance} for details). This objective function balances maximizing power against minimizing FDR, which may be useful in settings where strict FDR control is not necessary. Indeed, some previous work in astronomical point source detection attempted to maximize the F1 score \citep{starnet2021}, suggesting this may be of interest. Similarly, as mentioned in Section \ref{subsec::blipfdr}, one can add or modify the linear constraints in the LP formulation of \blips to control different notions of the error rate, such as controlling a weighted FDR \citep{benjamini1997, bogomolov2020} or achieving multilayer FDR control \citep{pfilter2017, multilayerknock2019}.

Second, in Sections \ref{sec::sims}-\ref{sec::gwas}, we apply \blips on top of Bayesian models with fairly uninformative priors. However, using more sophisticated models or algorithms could improve \blip's performance. To start, recent advances in Bayesian inference for spike-and-slab models or change point detection could boost power and computational efficiency \citep{cappello2021, biswas2022}, and generic advances in MCMC could ensure that the inputs to \blips are at least unbiased \citep{jacob2020}. More specifically, recent works in fine-mapping have developed informative priors based on, e.g., functional annotation data \citep{polyfun2019, weeks2019, trippe2021}. Indeed, \blips can apply directly on top of nearly any Bayesian fine-mapping model or algorithm, e.g., those developed in \cite{caviar2014, benner2015, fastpaintor2016}. Of course, in some cases, it might be worthwhile to develop sensible heuristics to optimize performance. For example, methods like FINEMAP \citep{benner2015} partially enumerate over configurations of causal variants to compute approximate PIPs, and it might make sense to account for this approximation before applying \blip. Such questions may point to promising directions for future research.

Lastly, we hope \blips may prove useful in many domains not considered in this paper. For example, \blips might apply in problems with spatially structured signals, even if the signals are not isolated. One such application is the analysis of fMRI data, where researchers often study questions like ``in which regions of the brain did an intervention change brain activity?" In this case, for some regions $G$, it might make sense to define the PIP $p_G$ as the posterior probability that \emph{every} point in $G$ is a signal point \citep{fmri2022}. However, having made this adjustment to $\{p_G\}$, one could otherwise run \blips exactly as written to perform resolution-adaptive signal detection. More generally, it can be challenging to extract interpretable results from complex posterior distributions in high-dimensional problems, such as in the Bayesian analysis of phylogenetic trees \citep{huelsenbeck2001, phylodist2020} or the Bayesian estimation of graphical models \citep{spikeslabbggm2018}. By carefully defining ``signals" of interest, could \blips help extract intelligible results in these settings? We did not consider this question in this paper, but perhaps future work will address it.

\section{URLs}\label{sec::urls}

For convenience, below are a collection of links which include all the code and data necessary to replicate our analyses.

\pyblip, a Python package implementing \blip: \url{https://github.com/amspector100/pyblip}.

\blipr, an R package implementing \blip: \url{https://github.com/amspector100/blipr}.

The code for the paper's simulations: \url{https://github.com/amspector100/blip_sims}.

The code for the astronomical application: \url{https://github.com/amspector100/DeblendingStarfields}.

The code for the fine-mapping analysis: \url{https://github.com/amspector100/ukbb_blip}.

Open-access data for the fine-mapping analysis: \url{https://data.broadinstitute.org/alkesgroup/UKBB_LD/} and \url{https://data.broadinstitute.org/alkesgroup/polyfun_results/}.

\section*{Acknowledgements}

The authors would like to thank Niloy Biswas, Jun Liu, Minsuk Shin, and Benjamin Spector for suggestions related to Bayesian regression; Vinay Kashyap and Douglas Finkbeiner for their insight on the astronomical application; and Hilary Finucane and Luke O'Connor for valuable comments on the fine-mapping application. A.S. was partially supported by the Two Sigma Graduate Fellowship Fund and a Graduate Research Fellowship from the National Science Foundation. L.J. was partially supported by the William F. Milton Fund and a CAREER grant from the National Science Foundation (Grant \#DMS2045981).

\bibliography{references}
\bibliographystyle{apalike}

\appendix

\section{A toy example motivating resolution-adaptivity}\label{subsec::ftest}

In this section, we give a toy example of a variable selection problem where the optimal way to group the variables depends on the unknown relationship between $X$ and $Y$.

\begin{example}\label{ex::ftest} Suppose we observe regression data $\bX \in \R^{n \times 2}, \by \in \R^n$ such that $\bX^T \bX = \begin{bmatrix} 1 & 0.99 \\ 0.99 & 1 \end{bmatrix}$ and $\by \sim \mcN(\bX \beta, \ident_n)$. $X_{\loc}$ is a signal variable if $\beta_{\loc} \ne 0$, for $\loc \in \{1,2\}$. Here, we could either try to detect individual signals (the ``finer resolution"), or we could try to detect whether \textit{either} $X_1$ or $X_2$ is a signal variable (the ``coarser resolution"). A non-adaptive method must choose one option \textit{before} observing $\by$. Which is the correct choice?
\end{example}

\begin{figure}[!ht]
    \centering
    \includegraphics[width=0.7\linewidth]{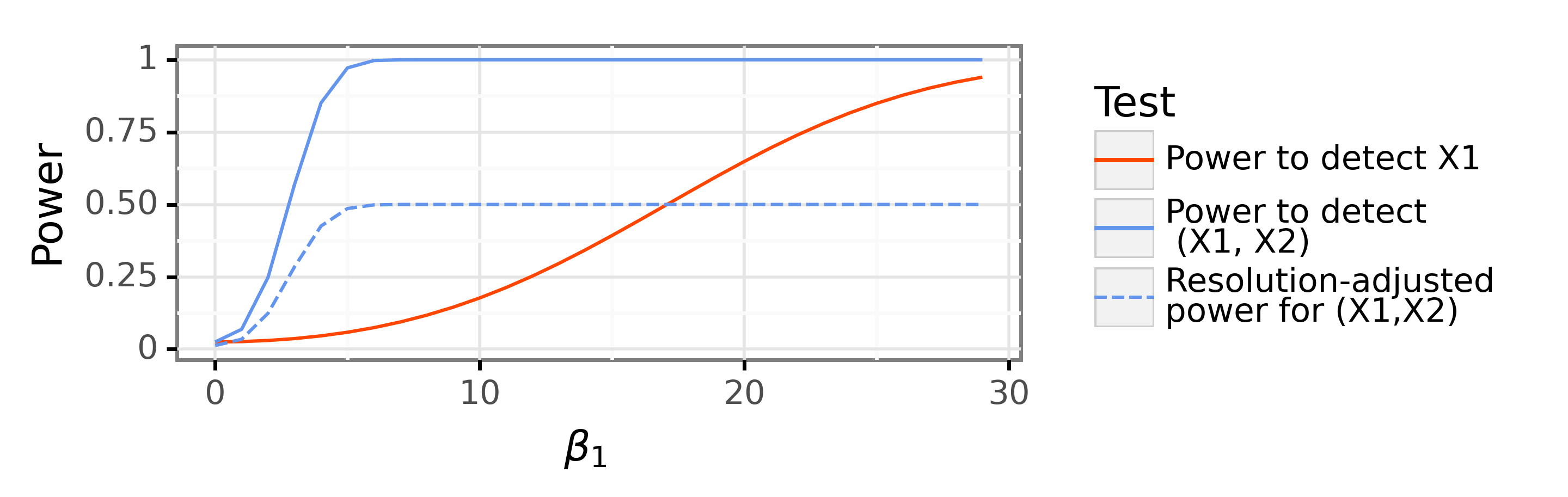}
    \caption{Power of F-tests described in Example \ref{ex::ftest} while varying $\beta_1$. This figure shows that the correct choice of resolution in Example \ref{ex::ftest} depends on the unknown coefficient $\beta_1$. Note resolution-adjusted power is formally defined in Section \ref{subsec::gap}: the main idea is that discoveries at coarser resolutions count as fewer discoveries since they yield less specific information.}
    \label{fig::ftest}
\end{figure}

As we will prove in a moment, the answer to this question depends on the unknown coefficients $\beta \in \R^2$. Suppose in truth that $\beta_1 \ne 0, \beta_2 = 0$, although the analyst does not know this. In Figure \ref{fig::ftest}, we plot the power of the standard F-tests corresponding to the two previous resolutions, with appropriate multiplicity corrections. As shown by Figure \ref{fig::ftest}, detecting $X_1$ individually requires a much larger signal size than detecting a signal in $(X_1, X_2)$, simply because $X_1$ and $X_2$ are correlated. Indeed, for $\beta_1 \approx 5$, we have nearly a $100\%$ chance of detecting a signal in $(X_1, X_2)$ but less than a $5\%$ chance of detecting $X_1$. However, when $\beta_1$ is large, we can detect $X_1$ individually with high probability, which yields more specific information. For this reason, the resolution-adjusted power of the finer-resolution method is higher for large $\beta_1$ and lower for small $\beta_1$ in Figure \ref{fig::ftest}. This means that every non-adaptive method will make the wrong choice in some scenarios, whereas adaptive methods like \blips can use the data to choose a resolution.

To derive the power curves in Figure \ref{fig::ftest}, we review classical results for the power of the F-tests testing (1) whether $\beta_1 = 0$ and (2) whether $(\beta_1, \beta_2) = 0$. As notation, let $H = \bX (\bX^T \bX)^{-1} \bX^T$ be the projection matrix onto the column space of $\bX$ and let $H_2$ be the projection matrix onto the column space of $\bX_2$. Let $\chi_{k, \mu}^2$ denote a non-central $\chi^2$ distribution with non-centrality parameter $\mu$. Then Cochran's theorem (see \cite{agresti2015} Chapter $3$ for a review) tells us that for $n \ge 3$,
\begin{equation*}
    \by^T (H - H_2) \by \sim \chi_{1,\mu_1}^2, \,\,\,\, 
    \by^T H \by \sim \chi_{2, \mu_2}^2, \,\,\,\,
    \by^T (\ident_n - H) \by \sim \chi_{n-2}^2 ,
\end{equation*}
where in our setting each of the first two quantities is marginally independent of the third quantity, and furthermore,
\begin{equation*}
    \mu_1 = \beta^T \bX^T (H - H_2) \bX \beta = (1 - \rho)^2 \beta_1^2 \\ 
\end{equation*}
\begin{equation*}
    \mu_2 = \beta^T \bX^T H \bX \beta = \beta^T \bX^T \bX \beta = \beta_1^2.
\end{equation*}
Note that in the last step of the previous two equations, we use the fact that $\beta_2 = 0$ by assumption as well as the form of $\bX^T \bX$. As a result, we can calculate the distributions of the two standard $F$-statistics:
\begin{equation*}
    T_1 = \frac{\by^T (H - H_2) \by}{\by^T (\ident_n - H) \by / (n-2)} \sim F_{1, n-2, (1-\rho^2)\beta_1^2}, 
\end{equation*}
\begin{equation*}
    T_2 = \frac{\by^T H \by  / 2}{\by^T (\ident_n - H) \by / (n-2)} \sim F_{2, n-2, \beta_1^2},
\end{equation*}
where $F_{n_1, n_2, \mu}$ denotes a noncentral $F$ distribution with $n_1, n_2$ degrees of freedom and non-centrality parameter $\mu$. At a significance level of $5\%$, the standard F-tests for the two resolutions reject when $T_1 \ge t_1$ and $T_2 \ge t_2$, respectively, where $t_1, t_2$, are the $0.95$th quantiles of the $F_{1, n-2}, F_{2,n-2}$ distributions. For the finer resolution, the analyst without oracle knowledge would have to test two hypotheses (one for each variable): thus we apply a Bonferroni correction and let $t_1$ denote the $0.975$th quantile of $F_{1,n-2}$. Then, the power of these tests can be calculated as the CDF of the relevant non-central $F$-distributions evaluated at $t_1, t_2$. 


\section{Further details on BLiP}\label{appendix::blip}

\subsection{Proof of Proposition \ref{prop::blipequiv}}\label{subsec::proofs}

In this section, we prove Proposition \ref{prop::blipequiv}.

\begingroup
\def\thetheorem{\ref{prop::blipequiv}}
\begin{proposition} The solution to the resolution-adaptive signal detection problem in Definition \ref{def::rasd} is the same as the solution to the following mixed-integer LP:
\begin{align}
    \max_{\{x_G\}_{G \in \mcG} : x_G \in [0,1]} \,\,\,\,\,\,\,\,\,\,\,\, & \sum_{G \in \mcG} p_G \weight(G) x_G \tag{\ref{eq::power_reduction}} \\
    \mathrm{s.t.}\,\,\,\,\,\,\,\,\,\,\,\,\,\,\,\,\,\,\,\, & \sum_{G \in \mcG} (1-p_G-q) x_G \le 0, \tag{\ref{eq::fdr_red}} \\
    & \sum_{G \in \mcG : \ell \in G} x_G \le 1 \,\,\,\,\,\, \forall \ell \in \locations, \tag{\ref{eq::disj_red}} \\
    & x_G \in \{0,1\} \,\,\,\,\,\, \forall G \in \mcG. \tag{\ref{eq::intconstraint}}
\end{align}
\begin{proof} Note that throughout this proof, we use the convention that $\frac{0}{0} = 0$. Recall that Problem \ref{def::rasd} maximizes expected power subject to the constraint that the selected groups $G_1, \dots, G_R$ are disjoint and control the FDR at level $q$. As argued in Section \ref{subsec::blipfdr}, the first line (\ref{eq::power_reduction}) is equivalent to maximizing expected power. Additionally, (\ref{eq::disj_red}) is equivalent to the disjointness constraint. Indeed, consider two candidate groups $G, G'$ which are not disjoint. This means $G, G'$ contain a common location $\ell \in \locations$, and therefore (\ref{eq::disj_red}) requires that $x_G + x_{G'} \le 1$. Thus, (\ref{eq::disj_red}) permits at most one of $G, G'$ to be discovered. Conversely, if $G_1, \dots, G_R$ are all disjoint, $x_{G_1}=1, \dots, x_{G_R}=1$ does not violate (\ref{eq::disj_red}) since no two groups in $G_1, \dots, G_R$ share a location $\ell$.

It remains to show that (\ref{eq::fdr_red}) is equivalent to controlling the FDR. For the first direction, suppose $G_1, \dots, G_R$ control the FDR at level $q$. By Equation (\ref{eq::fdrv1}), this implies $\frac{\sum_{G \in \mcG} (1 - p_G) x_G}{\sum_{G \in \mcG} x_G} \le q$, and multiplying by $\sum_{G \in \mcG} x_G$ on both sides implies (\ref{eq::fdr_red}) holds. For the other direction, note that if $\{x_G\}_{G \in \mcG}$ is feasible, Equation (\ref{eq::fdr_red}) implies that $\sum_{G \in \mcG} (1-p_G) x_G \le q \sum_{G \in \mcG} x_G$. Dividing both sides of this inequality by $\sum_{G \in \mcG} x_G$ implies that we control the FDR: the only subtlety is that $\sum_{G \in \mcG} x_G$ could be zero. However, $\sum_{G \in \mcG} (1-p_G) x_G=0$ whenever $\sum_{G \in \mcG} x_G = 0$. Since we use the convention that $0/0 = 0$, we can divide by $\sum_{G \in \mcG} x_G$ in all cases to yield the result.
\end{proof}
\end{proposition}
\endgroup

\subsection{Integer solutions to the relaxed LP from Section \ref{sec::blip}}\label{subsec::knapsack}

In this section, we give some intuition as to why the relaxed LPs described in Section \ref{sec::blip} tend to return (largely) integer solutions. To see this, we briefly review a canonical relaxed knapsack problem. Note that for simplicity, in this section we consider the version of \blips which controls the PFER, but this intuition extends to other error rates as well. At the end of this section, we also present the corresponding plot of Figure \ref{fig::blipintsol} for the FWER.

Suppose we have $m$ liquids with weights $w = (w_1, \dots, w_{m}) \in \R^{m}_{\ge 0}$ and volumes $v = (v_1, \dots, v_{m}) \in \R^{m}_{\ge 0}$. The ``knapsack problem" is to select a proportion of each liquid $x_1, \dots, x_{m} \in [0,1]$ which maximizes the total weight $w^T x$ subject to the volume constraint $v^T x \le v^*$ for $v^* \ge 0$---here, the volume constraint corresponds to ensuring the liquids fit in a ``knapsack." This is a simple LP, and it is well known that optimal solution has at most one non-integer value, in particular because the greedy solution is optimal. In other words, the optimal solution is to order the liquids in decreasing order of their ``density" $w_j / v_j$ and select all of the first $k$ liquids where $k$ is the largest number such that $\sum_{j=1}^{k} v_j \le v^*$. Then, we select some fractional proportion of the $k+1$th liquid to use up the rest of our volume budget. In other words, the solution is of the form $x_1, \dots, x_k = 1$, $x_{k+1} \in [0,1], x_{k+2}, \dots, x_{\ell} = 0$ assuming that $\frac{w_1}{v_1} \ge \frac{w_2}{v_2} \ge \dots \ge \frac{w_m}{v_m}$.

When controlling the PFER, the relaxed version of the resolution-adaptive signal detection problem is simply a knapsack problem with extra linear constraints that prevent some variables (or ``liquids") from being selected simultaneously. In many cases, despite the extra constraints, the optimal solution to the relaxed LP in \blips consists of only one non-integer value, just like in a vanilla knapsack problem. In this case, let $G\opt$ be the unique group such that $x_{G\opt} \not \in \{0,1\}$. Given such a solution, to exactly achieve PFER control and maximize expected power, one could simply detect signals in all the groups $G$ such that $x_{G} = 1$ and detect a signal in group $G\opt$ with probability $x_{G\opt}$. Alternatively, to avoid a randomized procedure, one could only discover $\{G : x_G = 1\}$ at the cost of making up to one fewer discovery and being slightly conservative.

Of course, the extra constraints do often affect the solution to the LP. To build intuition for this, consider a simple example where there are only two groups, $G = \{1\}$ and $G' = \{1,2\}$, with $\weight(G) = 1, \weight(G') = 0.5$ and PIPs $p_G = 0.8$, $p_{G'} = 0.9$. Thinking in terms of a knapsack problem, $G$ and $G'$ have ``weights" $p_G \weight(G) = 0.9$, $p_{G'} \weight(G') = 0.45$ and ``volumes" $1 - p_G = 0.2, 1 - p_{G'} = 0.1$, respectively. Thus, the relaxed signal detection problem is of the form
\begin{align}
    \max \,\,\,\,\,\,\,\,\,\,\,\,& 0.8 x_G + 0.45 x_{G'} \label{eq::2groupex} \\
    \mathrm{s.t.} \,\,\,\,\,\,\,\,\,\,\,\,& 0.2 x_G + 0.1 x_{G'}\le q, \label{eq::pfer2groupex} \\
    & x_G + x_{G'} \le 1, \label{eq::disjointness2groupex} \\
    & x_G, x_{G'} \in [0,1]. \nonumber
\end{align}
As usual, the objective corresponds to maximizing power, (\ref{eq::pfer2groupex}) corresponds to controlling the PFER, and (\ref{eq::disjointness2groupex}) is the disjointness constraint. Without the disjointness constraint (\ref{eq::disjointness2groupex}), the optimal (greedy) solution would be to first make $x_{G'}$ as large as possible under the PFER constraint (\ref{eq::pfer2groupex}). Then, if it is possible to set $x_{G'} = 1$ and there is additional slack in the PFER constraint, we would make $x_G$ as large as possible while controlling the PFER. However, this behavior changes when we once again consider the disjointness constraint. For example, when $q = 0.15$, after setting $x_{G'} = 1$, there is extra slack in the PFER budget, but one cannot naively set $x_{G'} = 1, x_G > 0$ without violating the disjointness constraint (\ref{eq::disjointness2groupex}). So instead, the optimal solution is to \textit{simultaneously} make $x_{G'}$ smaller and $x_G$ larger until there is no slack in the PFER constraint. In this case, the optimal solution is to set $x_G = 0.5$ and $x_{G'} = 0.5$. Intuitively, what is happening is that we would like to make discoveries at a finer resolution and discover a signal in the group $\{1\}$: however, we cannot do that and control the PFER at level $q = 0.15$. Instead, the most powerful procedure is a randomized one, which discovers $\{1\}$ or $\{1,2\}$ each with probability $0.5$. And the most powerful non-randomized procedure is the slightly conservative one, which discovers $\{1,2\}$ with probability $1$. In the general formulation of the LP, we say that $G, G'$ are a \textit{randomized pair} if $x_G + x_{G'} = 1$ with $x_G, x_{G'}$ non-integers. Notably, this same intuition extends to each of the other error rates, where to achieve a Type I error of exactly $q$ in expectation, we would have to randomly choose between two or more discovery sets. Of course, \blips yields a deterministic solution, so the output of \blips will usually be (very slightly) conservative for this reason.

\begin{figure}
    \centering
    \includegraphics[width=\linewidth]{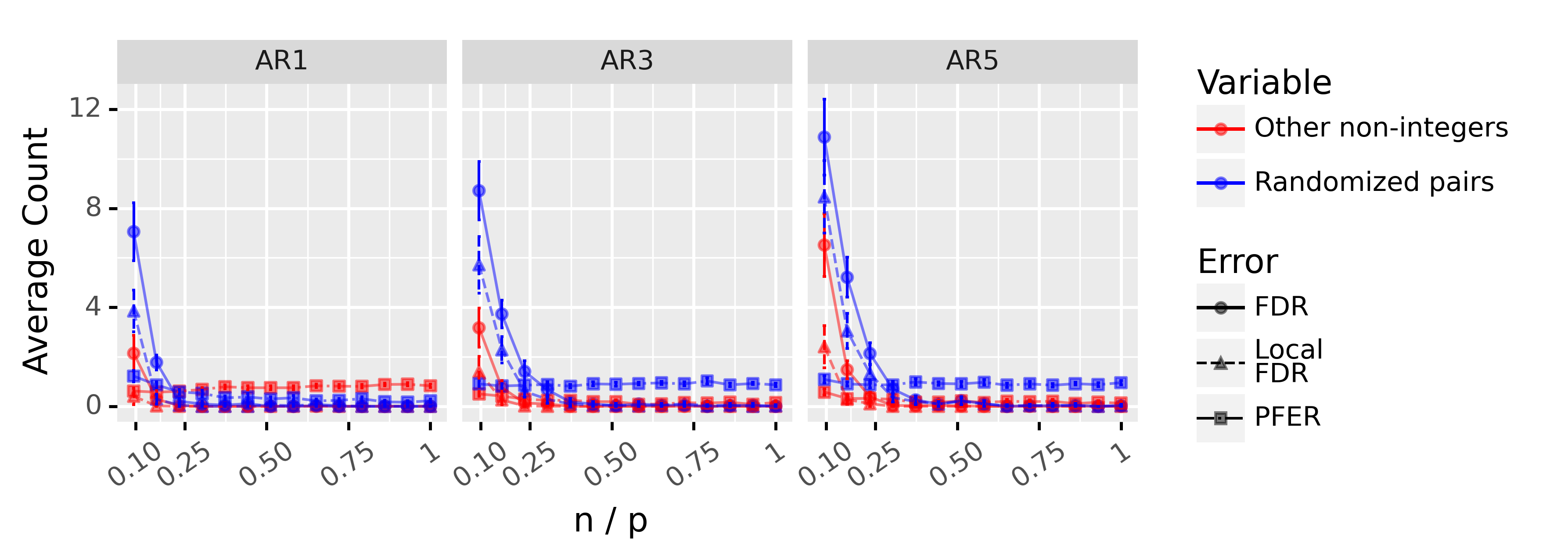}
    \caption{In the same setting as Figure \ref{fig::blipintsol}, this figure shows the average number of randomized pairs and non-integer solutions to the relaxed LP (\ref{eq::power_reduction})-(\ref{eq::disj_red}), and the corresponding relaxed LPs for other error rates. The total number of candidate groups exceeds $50,000$ for all methods, so this plot shows that only a tiny proportion of variables in the LP have nonzero values. See Appendix \ref{subsec::glmdetails} for the precise simulation setting.
    }
    \label{fig::randcounts}
\end{figure}

The main takeaway from all this is as follows: in general, we should expect the solution to the relaxed LP to be nearly all integers, with a few randomized pairs and a few other non-integer values to help us achieve a PFER equal to $q$. Indeed, Figure \ref{fig::randcounts} confirms that in the same setting as Figure \ref{fig::blipintsol}, solving the relaxed LP yields integer solutions for all but a few decision variables out of many thousands. Notably, as $n/p$ gets smaller, the number of non-integer values increases, which corresponds to the fact that in low-power settings, more randomization may be necessary to achieve exact optimal power.

\begin{figure}
    \centering
    \includegraphics[width=\linewidth]{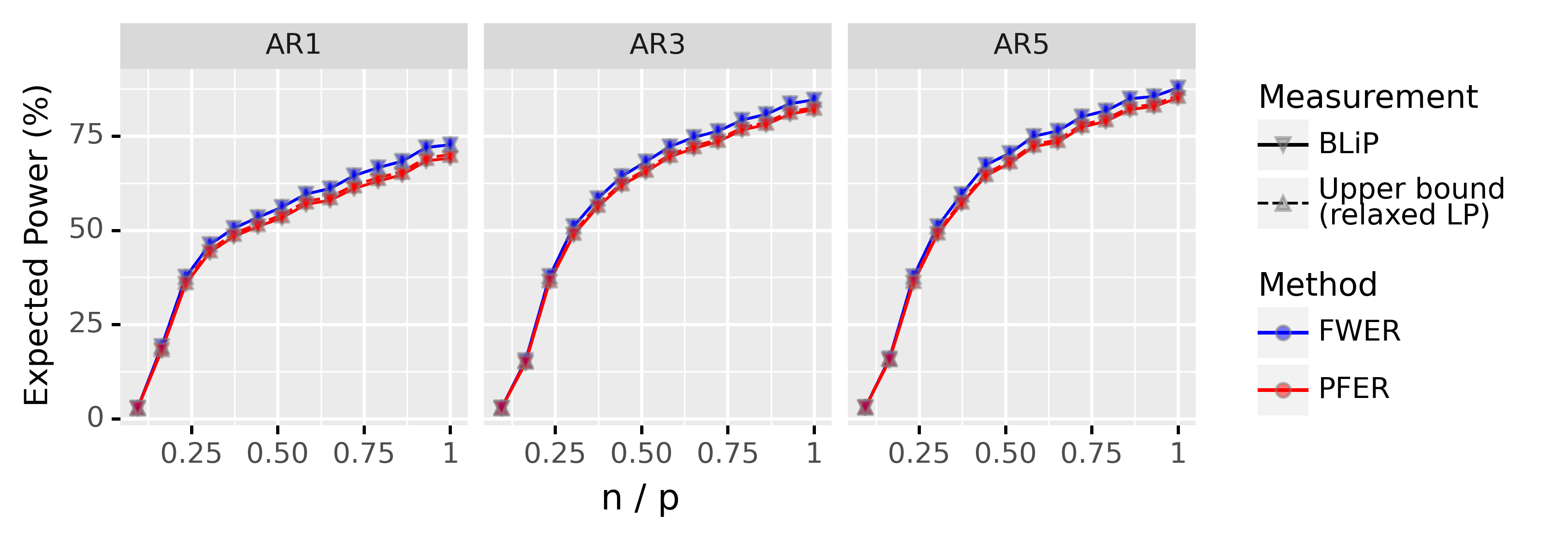}
    \caption{This plot is identical to Figure \ref{fig::blipintsol} except we apply \blips to control the FWER. Note that for the FWER, the upper bound from the relaxed LP is not a provable upper bound on the power of any method controlling the FWER at level $q$. Instead, as discussed in Appendix \ref{subsec::knapsack}, it upper bounds the power of any method controlling the PFER at some level $q\opt$, where $q\opt \ge q$ is chosen to be as large as possible while still controlling the FWER at level $q$. The plot also shows that choosing $q\opt \ge q$ yields a slight boost in power for the FWER method compared to the PFER method. See Appendix \ref{subsec::glmdetails} for simulation details.} 
    \label{fig::fwerintsol}
\end{figure}

Lastly, we now present Figure \ref{fig::fwerintsol}, the corresponding plot of Figure \ref{fig::blipintsol} for the FWER. Recall that to control the FWER, we use a more heuristic method which controls the PFER at the largest nominal level $q\opt$ such that the FWER is controlled. In particular, $q\opt \ge q$ because the PFER is a stricter notion of error control than the FWER, although usually $q\opt$ is only slightly larger than $q$. Figure \ref{fig::blipintsol} shows that this binary search yields very slightly higher power than choosing $q\opt = q$. Since the FWER constraint cannot be represented as a linear constraint on the optimization variables $\{x_G\opt\}_{G \in \mcG}$, unlike in Figure \ref{fig::blipintsol}, the LP bound in Figure \ref{fig::fwerintsol} is not an upper bound on the power of any method which controls the FWER. Instead, it is an upper bound on any method that controls the PFER at level $q\opt$, with $q\opt$ identified via binary search. This is not a principled upper bound, but at least it shows that our relaxation of the integer LP (\ref{eq::power_reduction})-(\ref{eq::intconstraint}) yields a nearly optimal solution to the integer LP. And of course, the method still provably controls the FWER---we just cannot give any provable guarantees about its optimality.

\subsection{Adaptive preprocessing}\label{subsec::prefilter}

In this section, we detail the adaptive preprocessing steps discussed in Section \ref{subsec::blipfdr}.

First, after we have computed PIPs, we can \textit{pre-filter the candidate groups} $\mcG$ before adding them into the linear program in \blip. Indeed, whenever we seek to control the FWER, local FDR, or PFER below some level $q$, we cannot detect a signal in a group $G \in \mcG$ unless $p_G \ge 1 - q$: otherwise we would definitionally violate error control. Thus, we can throw out all groups $G$ where $p_G < 1 - q$. We also suggest performing a similar procedure when controlling the FDR, where we throw out all groups such that $p_G < \kappa$ for some $\kappa \le 1 - q$. Although this does not provably preserve optimality, in practice we do not expect this to reduce power for an appropriate $\kappa$. For example, suppose $q = 0.05$ and $p_G \le 0.1$, which means that group $G$ has at most a $10\%$ chance of containing a signal. It is unlikely that a FDR controlling procedure would select $G$. Even if selecting $G$ is optimal for power, it is arguably desirable to \textit{not} select $G$ anyway, since $G$ would be ``free-riding" on the other selected groups with high PIPs. In practice, we suggest setting $\kappa = 0.5$.

Second, when $\locations$ is discrete, we suggest \textit{pre-filtering the locations}. Suppose for example that for some location $\loc \in \locations$, the posterior probability that there is a signal exactly at $\loc$ is extremely low, so $p_{\{\loc\}} \approx 0$. In this case, if $G_1 \subset \locations$ contains $\loc$, we could just as easily discover $G_2 = G_1 \setminus \{\loc\}$, which yields more precise information. Formally, a union bound yields that $p_{G_1} \le p_{G_2} + p_{\{\loc\}} \approx p_{G_2}$, so $G_1$ has roughly the same chance of containing a signal as $G_2$, even though $G_2 \subset G_1$. As a result, there is effectively no reason to include $\loc$ in any group in the candidate groups. This suggests the following procedure: after computing marginal PIPs $\{p_{\{\loc\}} : \ell \in \locations\}$, we only construct candidate groups based on locations with a sufficiently high chance of harboring a signal, i.e., $\mcL_{\kappa} = \{\ell : p_{\{\loc\}} \ge \kappa\}$. Typically, we choose $\kappa$ to be quite small, e.g., $\kappa \approx 0.001$. This approach often actually increases power in practice, since it prunes out irrelevant locations from our candidate groups, yielding a better set of candidate groups and allowing us to make discoveries at finer resolutions. Note that this approach is not appropriate when the set of locations is continuous and $p_{\{\loc\}} = 0$ for all $\ell \in \locations$.

Lastly, when controlling the FDR or local FDR, we suggest \textit{pre-narrowing} $\mcG$. To define this, suppose $G_1, G_2 \in \mcG$ such that $G_1 \subset G_2$, which means detecting a signal in $G_1$ yields more precise information than detecting a signal in $G_2$. As a result, it is usually a safe assumption that $\weight(G_1) > \weight(G_2)$ (this is true in all of our applications). Suppose also that $p_{G_1} \weight(G_1) \ge p_{G_2} \weight(G_2)$, meaning that detecting a signal in $G_1$ leads to higher expected power than detecting a signal in $G_2$. If this is true and $p_{G_1} \le \alpha$ for some $\alpha < q$, this means that we can already discover $G_1$ at below the nominal level, so there is not much reason to discover $G_2$: as a result, we exclude $G_2$ from consideration. Like the previous approach, this is a heuristic method, although it provably preserves validity. We suggest setting $\alpha = q/2$.

\subsection{Polynomial-time alternatives to the integer LP}\label{subsec::randomization}

Our algorithm for \blips (Algorithm \ref{alg::blip}) involves solving a relaxed LP and then running an integer LP on the non-integer solutions from the relaxed LP. In practice, the number of non-integer solutions from the relaxed LP is typically in the single-digits, as in Figure \ref{fig::randcounts}. As a result, the integer LP runs extremely quickly. That said, it is mathematically possible for the relaxed LP to admit solutions which include a very large number of non-integer values, in which case the integer LP may be too expensive. We emphasize that this never occurred in any of our simulations or real data applications, including in the fine-mapping application in Section \ref{sec::gwas} with $19$ million covariates and hundreds of millions of candidate groups. For the sake of completeness, however, we will detail a heuristic polynomial time alternative to the integer LP. Note that this section follows the notation from Section \ref{subsec::blipfdr}.

Given optimal solutions $\{x_G\opt\}_{G \in \mcG}$ to the relaxed LP (\ref{eq::power_reduction})-(\ref{eq::disj_red}) with $x_G\opt \in [0,1]$, we seek to obtain integer solutions $\{x_G^{\star\star}\}_{G \in \mcG}$ which are feasible for the integer LP (\ref{eq::power_reduction})-(\ref{eq::intconstraint}) and have as high expected power as possible. The key idea is as follows: we will use $\{x_G\opt\}$ as guidelines and sample random feasible integer solutions $\{z_G^{(i)}\}_{G \in \mcG}$ for $1 \le i \le n_{\mathrm{sample}}$, and then pick the solution $\{z_G^{(i)}\}_{G \in \mcG}$ which maximizes expected power. Intuitively, since $\{x_G\opt\}_{G \in \mcG}$ are the optimal solution to the relaxed problem, we might expect that there is an integer solution ``close to" $\{x_G\opt\}$ which has similar expected power. One simple way to convert $\{x_G\opt\}$ into a reasonably good solution would be to sample $z_G^{(i)}$ such that $\P(z_G^{(i)} = 1) = x_G\opt$. Indeed, one can think of this as a directed search algorithm, where we search over solutions ``close to" $x_G\opt$ by sampling $z_G^{(i)} \sim \Bern(x_G\opt)$. Although it is not always possible to ensure $\P(z_G^{(i)} = 1) = x_G\opt$ while simultaneously satisfying the disjointness constraint (\ref{eq::disj_red}), Algorithm \ref{alg::randpoly} iteratively samples $\{z_G^{(i)}\}$ in a way that guarantees $\{z_G^{(i)}\}$ are a feasible solution to the integer LP while ensuring $\P(z_G^{(i)} = 1) \approx x_G\opt$ at each step. 

\begin{algorithm}[h!]
\caption{Polynomial time alternative to the integer LP in \blip.}\label{alg::randpoly}
\algorithmicensure\, Locations $\mcL$, PIPs $\{p_G\}_{G \in \mcG}$ and solutions $\{x_G\opt\}$ to the relaxed LP (\ref{eq::power_reduction})-(\ref{eq::disj_red}). 
\begin{algorithmic}[1]
        \State For each $\ell \in \mcL$, compute $s_{\ell} = \sum_{G :\ell \in G} x_G\opt$, the target probability of discovering a group containing $\ell$.
        \State Sort $\mcL$ in decreasing order of $\{s_{\ell}\}$.
        \For {$i=1,2,\dots,n_{\mathrm{sample}}$}
            \State Initialize $\mcD = \{\}$ (the discovered groups). Later, we will set $z_G^{(i)} = 1$ iff $G \in \mcD$.
            \State Initialize $\mcF = \mcG$. Here, $\mcF$ represents the set of ``feasible" groups we can add to $\mcD$ without violating the disjointness constraint.
            \For {$\ell \in \mcL$}
                \State Let $\mcF_{\ell} = \{G \in \mcF : \ell \in G\}$ denote the feasible groups containing location $\ell$.
                \State Sample $U \sim \Unif(0,1)$.
                \If {$U \le s_{\ell}$} 
                    \State Sample a group $G_{\ell}$ from $\mcF_{\ell}$, with $\P(G_{\ell} = G) \propto x_G\opt$ for $G \in \mcF_{\ell}$.
                    \State Set $\mcD = \mcD \cup \{G_{\ell}\}$ and $\mcF = \mcF \setminus \{G' : G_{\ell} \cap G' \ne \emptyset\}$.
                \Else {}
                    \State Set $\mcF = \mcF \setminus \mcF_{\ell}$ and continue to the next location.
                \EndIf
    		\EndFor
    		\State Sort $\mcD$ in increasing order of $\{p_G\}_{G \in \mcD}$. 
    		\State Iteratively remove elements of $\mcD$ until the discoveries $\mcD$ satisfy the FDR constraint (\ref{eq::fdrv1}).
    		\State For $G \in \mcG$, set $z_G^{(i)} = \I(G \in \mcD)$. 
		\EndFor
		\State Return $\{z_G^{(i\star)}\}_{G \in \mcG}$ for $i\star = \arg \max_i \sum_{G \in \mcG} (1 - p_G) \weight(G) z_G^{(i)}$.
	\end{algorithmic} 
\end{algorithm}

Since each  $\{z_G^{(i)}\}_{G \in \mcG}$ is feasible by construction, Algorithm \ref{alg::randpoly} is guaranteed to return a feasible solution to the integer LP (\ref{eq::power_reduction})-(\ref{eq::intconstraint}), and furthermore, its runtime is polynomial in $|\mcG| |\mcL|$. In our simulations, however, we found that in practice Algorithm \ref{alg::randpoly} does not run faster than the integer LP in the default version of \blip, and both algorithms achieve similar power. Since Algorithm \ref{alg::randpoly} returns a randomized solution, for the rest of the paper, we always use the default version of \blips specified in Algorithm \ref{alg::blip} (which is deterministic).

\subsection{Balancing the false positive and false negative rates}\label{subsec::balance}

In this section, we discuss extending \blips to balance the expected false positive and false negative rates. Throughout this section, we make one very mild assumption, which is that perfectly localizing a signal attains the maximum weight of $1$. Formally, this means $\weight(G) \le 1$ for all $G \in \mcG$, and $\weight(\{\ell\}) = 1$ for all $\ell \in \mcL$. We also assume that $\mcL = \{1, \dots, p\}$ is discrete and finite, although we assume this mainly for convenience (the analysis extends to the continuous case). We use the notation defined in Section \ref{sec::blip}.

For any set of detections defined by $\{x_G\}_{G \in \mcG}$, we define the Bayesian expected (resolution-adjusted) false negative rate (FNR) as  
\begin{equation*}
    \E[\FNR \mid \mcD] \defeq 1 - \frac{\sum_{G \in \mcG} \weight(G) p_G x_G}{\sum_{\ell \in \mcL} p_{\{\ell\}}}.
\end{equation*}
In the equation above, note that the numerator $\sum_{G \in \mcG} \weight(G) p_G x_G$ is simply the expected resolution-adjusted power, and the denominator is the expected number of signals, which is also the maximum expected  resolution-adjusted power by the aforementioned assumptions. Thus, the expected FNR is zero if we discover signals at every location (which corresponds to maximal power), and it is one if we make no discoveries. It is important to note that $\E[\FNR \mid \mcD]$ is a linear function of the optimization variables $\{x_G\}_{G \in \mcG}$, since the PIPs $\{p_G\}_{G \in \mcG}$ are constants.

There are many possible objective functions which balance the FDR against the expected FNR. For brevity, we will focus on the F1 score, which is defined as the harmonic mean of the precision and recall. In particular, define the resolution-adjusted F1 score as
\begin{equation*}
    \mathrm{F1} \defeq \frac{2}{(1-\FDR)^{-1} + (1-\E[\FNR \mid \mcD])^{-1}}.
\end{equation*}
We are interested in optimizing $\{x_G\}_{G \in \mcG}$ to maximize the F1 score subject to the disjointness constraint (\ref{eq::disj_red}), without necessarily controlling the FDR at some fixed level $q \in (0,1)$. However, note that the F1 score is decreasing in both the FDR and expected FNR. Since \blips minimizes the expected FNR given a constraint on the FDR, this implies that the solution $\{x_G\opt\}_{G \in \mcG}$ which maximizes the F1 score is \textit{also} a solution to the \blips integer LP (\ref{eq::power_reduction})-(\ref{eq::intconstraint}) for some FDR level $q\opt$. To see this, suppose that $\{x_G\opt\}_{G \in \mcG}$ maximizes the F1 score. Then, $\{x_G\opt\}_{G \in \mcG}$ must be an optimal solution for the integer LP (\ref{eq::power_reduction})-(\ref{eq::intconstraint}) with $q\opt$ equal to the FDR level of $\{x_G\opt\}$, e.g., $q \opt = \frac{\sum_{G \in \mcG} (1-p_G) x_G}{\sum_{G\in \mcG} x_G\opt}$. Otherwise, it would be possible to strictly improve the F1 score by decreasing the FNR without increasing the FDR. Thus, finding the F1-maximal solution simply corresponds to finding $q\opt$ and solving \blips to control the FDR at level $q\opt$.

This observation suggests that to maximize the F1 score, we can first run \blips while controlling the FDR at various levels $q \in (0,1)$. By performing a line search or binary search over $q$, we can pick the value $q\opt \in (0,1)$ which attains the highest value of the F1 score. This algorithm is  heuristic, but we suspect it will yield a reasonably good solution, since finding $q\opt$ only requires solving a univariate optimization over $q$. Overall, this shows that by slightly changing the objective function for \blip, analysts could optimize for many different objectives, including balancing power with false positive control.

\section{Guidelines for computing the PIPs}\label{sec::mcmc}

In this section, we detail the methods we used to compute PIPs as an input to \blip. We start by offering a few general guidelines in Section \ref{subsec::pipguide}. Next, Sections \ref{subsec::lss} and \ref{subsec::pss} detail the Gibbs samplers we used in variable selection problems for sparse linear regression and sparse probit regression, respectively. We picked these samplers because some other existing software yielded inaccurate PIPs which caused \blips to substantially violate error control. In contrast, these (very simple) samplers performed well empirically and are quite efficient. Finally, Section \ref{appendix::susie} describes how to compute PIPs based on SuSiE \citep{susie2020} and explains why SuSiE + \blips can outperform SuSiE alone by wide margins.

\subsection{General guidelines}\label{subsec::pipguide}

In this section, we offer a few general guidelines on computing the PIPs $\{p_G\}_{G \in \mcG}$. To begin with, it is straightforward to estimate $p_G$ if we can sample from the posterior distribution of the locations of true signals. Indeed, let $\gamma_1, \dots, \gamma_N \subset \locations$ be such samples, which are straightforward to obtain from standard Bayesian methods. For example, in sparse linear regression problems where $Y \mid X \sim \mcN(X \beta, \sigma^2)$ and the signals correspond to nonzero coefficients in $\beta$, one can use any standard method to obtain samples $\tilde{\beta}^{(1)}, \dots, \tilde{\beta}^{(N)} \in \R^p$ from the posterior of $\beta$ and then set $\gamma_i = \{j : \tilde{\beta}_j^{(i)} \ne 0\}$. Having obtained $\gamma_1, \dots, \gamma_N$, we can estimate $p_G$ as the average number of posterior samples $\gamma_i$ which include a signal in $G$, i.e., $p_G \approx \frac{1}{N} \sum_{i=1}^N \I(G \cap \gamma_i \ne \emptyset)$. That said, obtaining good posterior samples can be challenging in high dimensional problems where (a) the prior may be misspecified and (b) standard MCMC algorithms may not converge quickly. Below, we describe a few heuristics to ensure the posterior samples and \blips are robust to these issues.

To address (a), we recommend using \textit{hierarchical priors} which account for some uncertainty in the Bayesian model. For example, when performing variable selection, many Bayesian models require some knowledge of $p_1$, the proportion of signals. Unless one has very precise prior knowledge of these parameters, we suggest picking a fairly uninformative prior for them, such as letting $p_1 \sim \Unif(0, p_{\max})$ for some $p_{\max} \le 1$. To determine the hyper-parameters for these priors (e.g., $p_{\max}$), we suggest either using a conservative choice or taking an empirical Bayesian approach. (Here, ``conservative" choices mean choices that may yield an error rate slightly below the nominal level.) The simulations in Section \ref{sec::sims} show that even fairly conservative choices, such as when $p_{\max} = \frac{1}{2} p_1$, do not lose much power and reliably control the FDR even when the hyperprior is quite different from the true sparsity level.

To address (b), when using MCMC algorithms to sample from the posterior, we generally recommend sampling from \textit{multiple MCMC chains with random initialization}. Even if each chain does not converge, we expect that aggregating results across chains will usually overestimate the uncertainty in the location of a signal. This allows \blips to control the error rate, even if it is somewhat conservative. For example, suppose that in a bivariate regression problem,  $\{X_1, X_2\}$ clearly contains a signal variable, but $X_1$ and $X_2$ are highly correlated, so it is not clear which one is the signal variable. Consider a worst-case scenario where the MCMC algorithm randomly initializes (e.g.) $X_1$ to be a signal variable, but then keeps $X_1$ as a signal variable at every iteration. If we compute $p_{\{1\}}$ just using this chain, we will falsely conclude that $p_{\{1\}} \approx 1$. However, if we run $5-10$ MCMC chains which each have a $50\%$ chance of initializing $X_1$ or $X_2$ as a signal variable, then we will conclude $p_{\{1\}}, p_{\{2\}} \approx 50\%$, or equivalently, that we know $X_1$ or $X_2$ is a signal but we are maximally uncertain about which one is the signal. This will yield (conservative) error rate control in this toy example, even though the MCMC algorithm did not converge whatsoever.
This intuition also extends to variational approaches which use random initialization or MCMC sampling.

\begin{figure}[!htb]
    \centering
    \subfloat{{\includegraphics[width=0.46\textwidth]{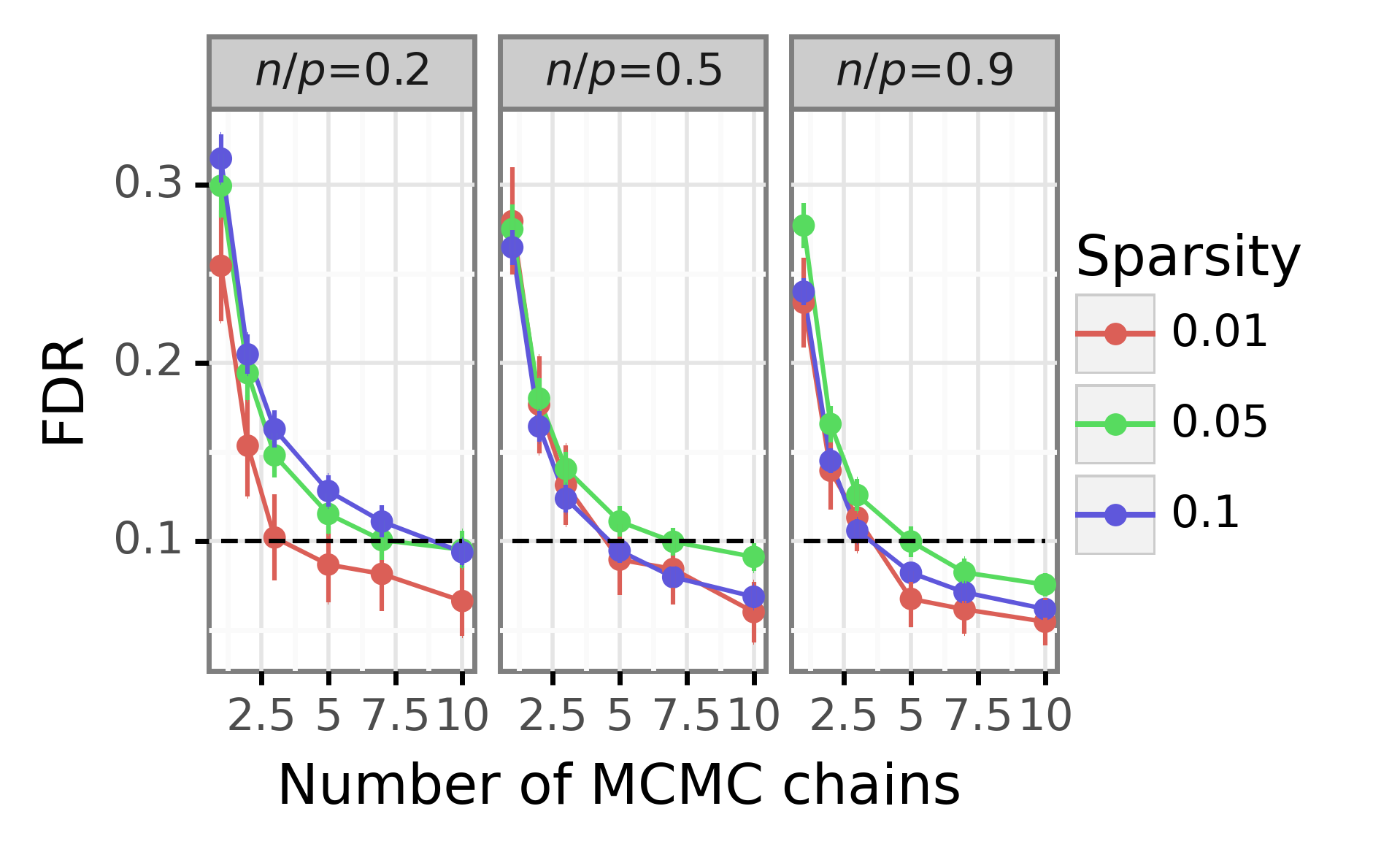}}}\label{subfig::manychainsfdr}
    \subfloat{{\includegraphics[width=0.46\textwidth]{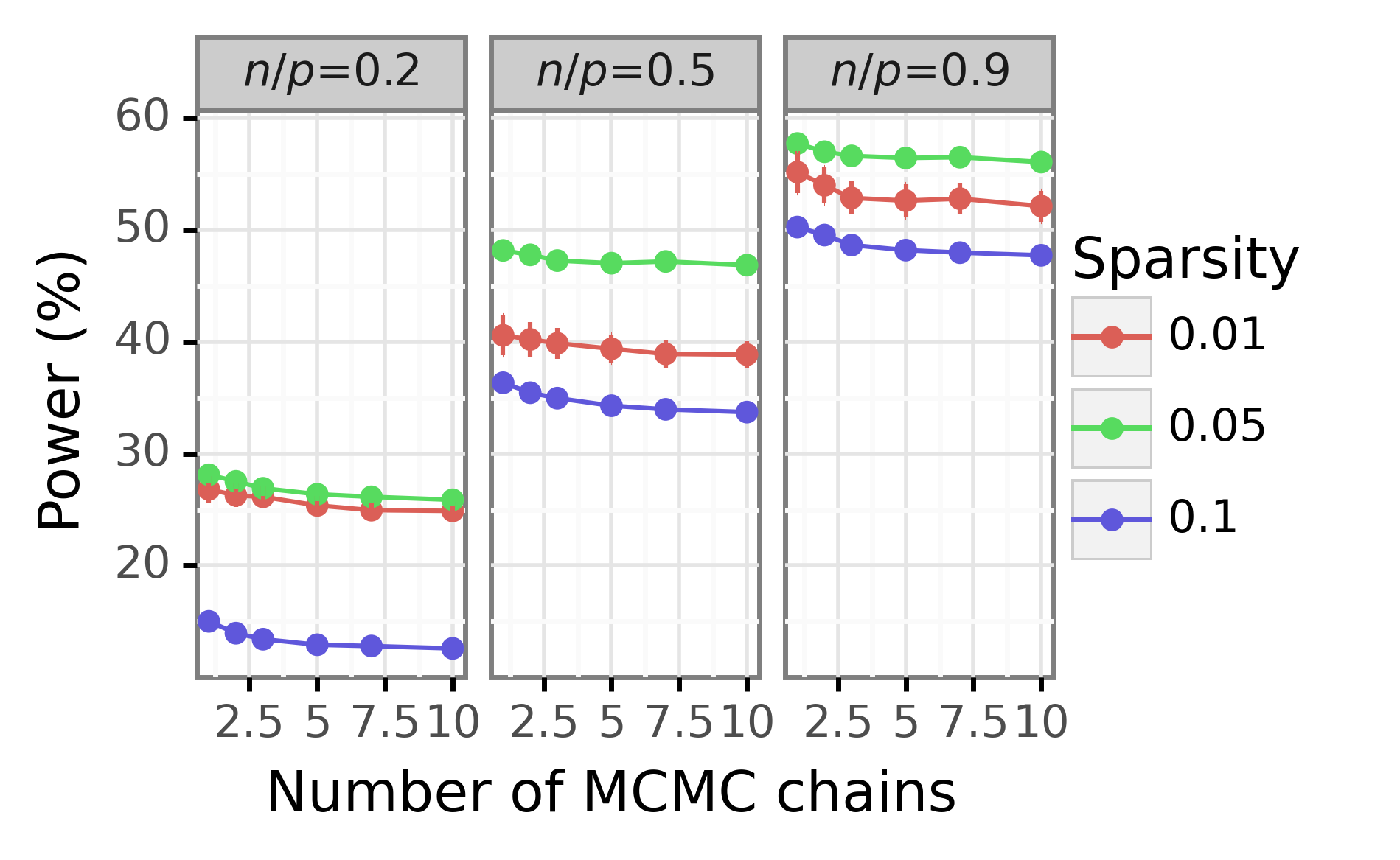}}}\label{subfig::manychainspower}
    \caption{This figure replicates the ``LSS + \blips (misspec.)" method from Figure \ref{fig::linear_hdim} but uses only $200$ samples per MCMC chain. The individual chains clearly do not converge, as shown by the left-hand plot, where the realized FDR is up to three times the nominal level when using just one chain. Despite this, aggregating results from $5$-$10$ chains leads to FDR control without substantially losing power (as shown in the right-hand figure). The simulation details are otherwise identical to Figure \ref{fig::linear_hdim} and are detailed in Appendix \ref{subsec::glmdetails}}.
    \label{fig::nchains}
\end{figure}

To empirically demonstrate the effect of using multiple MCMC chains, in Figure \ref{fig::nchains}, we rerun simulations in the setting of Figure \ref{fig::linear_hdim} but with only $200$ MCMC samples per chain and a burn-in of $20$ samples. (In contrast, in Figure \ref{fig::linear_hdim}, we used $5000$ samples per chain with a burn-in of $500$ samples.) This is a very high-dimensional setting where \blips searches over tens of thousands of candidate groups over $p=1000$ covariates, so we should not expect the first $220$ MCMC samples to converge. Indeed, when using only one chain, LSS + \blips dramatically violates FDR control. When using $10$ chains, however, LSS + \blips successfully controls the FDR while achieving nearly the same power.

\subsection{Details on the linear spike-and-slab sampler}\label{subsec::lss}

In this section, we review details on the linear spike and slab (LSS) Gibbs sampler we use in Sections \ref{sec::sims}-\ref{sec::gwas}. We assume $\by$ follows a Gaussian linear model, so $\by \mid \bX \sim \mcN(\bX \beta,\sigma^2 \ident_n)$. The prior specification is as follows:
\begin{equation*}
    \sigma^2 \sim \invGamma(a_\sigma,b_\sigma)
\end{equation*}
\begin{equation*}
    \tau^2 \sim \invGamma(a_\tau, b_\tau)
\end{equation*}
\begin{equation*}
    p_0 \sim \truncBeta(a_0, b_0, \minp)
\end{equation*}
where $\truncBeta(a_0, b_0, \minp)$ denotes a $\Beta(a_0, b_0)$ distribution truncated to the interval $[\minp, 1]$. We assume that $p_0, \sigma^2, \tau^2$ are jointly independent a priori. The exception to this prior scheme is that in the ``well-specified" case, we assume that $p_0, \sigma^2, \tau^2$ are known and use their true values. Finally, we assume the following prior for $\beta$:
\begin{equation*}
    \beta_j \mid \tau^2, p_0 \iid \begin{cases} 0 & \text{ w.p. } p_0 \\ \mcN(0, \tau^2) & \text{ w.p. } 1 - p_0. \end{cases}
\end{equation*}
To sample from the posterior of the parameters, we employ Gibbs sampling. However, because $\bX$ may exhibit high correlations, instead of sampling individual parameters from their conditional distributions, at each iteration we partition the covariates into blocks $J_1, \dots, J_m$ of size $k$ or less and sample from the joint distribution of $\beta_{J_1}$, then $\beta_{J_2}$, and so on. In our implementation, we choose $J_1, \dots, J_m$ to be contiguous groups of size $k$, but in principle one could apply any clustering algorithm. Although setting $k = 1$ and sampling from the distribution of individual parameters works fairly well, setting (e.g.) $k \approx 3$ can yield a substantial boost in power by helping the chain mix faster. Algorithm \ref{alg::lss} gives an outline of the general sampling scheme.

\begin{algorithm}[h!]
\caption{Linear spike-and-slab Gibbs sampler.}\label{alg::lss}
\algorithmicensure\, $\by, \bX$, hyperpriors, and the maximum block-size $k$.
\begin{algorithmic}[1]
        \State Initialize $\beta^{(1)} = 0$ and $\mathbf{r} = \by - \bX \beta^{(1)}$.
        \For {$\ell=1,2,\dots,n_{\mathrm{iter}}$}
            \State Partition $[p]$ into blocks $J_1, \dots, J_m$, e.g., by clustering $\bX$.
            \For {$J=J_1, \dots, J_m$}
                \State Update $\mathbf{r} = \mathbf{r} + \bX_J \beta_J^{(\ell)}$.
        		\State Sample $\beta_J \mid \by, \beta_{\negc J} = \beta^{(\ell)}_{\negc J}, \tau^2, \sigma^2, p_0$ using $\mathbf{r}$ for efficiency, as in Equations (\ref{eq::sampleA}),(\ref{eq::betaAmidA}).
    			\State Set $\beta_{J}^{(\ell)} = \beta_J$ and update $\mathbf{r} = \mathbf{r} - \bX_J \beta_J^{(\ell)}$
    		\EndFor
    		\State Sample $\tau^2, \sigma^2 \mid \by, \beta = \beta^{(\ell)}$ using inverse-Gamma conjugacy as in Equations (\ref{eq::sigma2sample}), (\ref{eq::tau2sample}).
    		\State Sample $p_0 \mid \by, \beta = \beta^{(\ell)}$ using rejection sampling as in Equation (\ref{eq::p0sample}).
		\EndFor
		\State Return $\beta^{(1)}, \dots, \beta^{(n_{\mathrm{iter}})}$.
	\end{algorithmic} 
\end{algorithm}

The most computationally intensive step in this algorithm is to efficiently sample $\beta_J \mid \beta_{\negc J}, \sigma^2, p_0, \tau^2, \by$ for any $J \subset [p]$. We describe how to do this in two steps. First, define $A = \{j \in J : \beta_j \ne 0\}$ to be the active set (set of nonzero coefficients) of $J$. Note that $A$ is a random latent variable. The first step is to sample $A \mid \beta_{\negc J}, \sigma^2, p_0, \tau^2, \by$. To lighten notation, let $H = (\sigma^2, p_0, \tau^2)$ denote the hyper-parameters. Note that for any $J_0 \subset J$,
\begin{align*}
        \P(A = J_0 \mid \beta_{\negc J}, H, \by)
    &\propto
        \P(A = J_0 \mid H, \beta_{\negc J}) p(\by \mid A = J_0, \beta_{\negc J}, H)
\end{align*}
where we use $p(\cdot \mid \cdot)$ to denote an arbitrary conditional density. Now, the form of the prior yields that $\P(A = J_0 \mid H, \beta_{\negc J}) = p_0^{|J|-|J_0|} (1-p_0)^{|J_0|}$. To analyze the second term in the product, note that since $\beta_{J_0} \mid A = J_0$ is multivariate Gaussian (and $\beta_{J \setminus J_0} = 0$), $$\by \mid A = J_0, \beta_{\negc J}, H \sim \mcN(\bX_{\negc J} \beta_{\negc J}, \sigma^2 \ident_n + \tau^2 X_{J_0} X_{J_0}^T).$$ As a result, we conclude
\begin{align*}
    p(\by \mid A = J_0, \beta_{\negc J}, H) \propto \det(\sigma^2 \ident_n + \tau^2 X_{J_0} X_{J_0}^T)^{-1/2} \exp\left(-\frac{1}{2} r^T (\sigma^2 \ident_n + \tau^2 X_{J_0} X_{J_0}^T)^{-1} r \right).
\end{align*}
To further simplify this, define $\mathbf{r} = \by - X_{\negc J} \beta_{\negc J}$ to be the residuals without $X_{J}$ and let $Q_{J_0} = I_{|J_0|} + \frac{\tau^2}{\sigma^2} X_{J_0}^T X_{J_0}$. Using the Woodbury identity and the matrix determinant lemma, we conclude
\begin{equation}\label{eq::sampleA}
        \P(A = J_0 \mid \beta_{\negc J}, H, \by)
    \propto
        p_0^{|J| - |J_0|} (1-p_0)^{|J_0|} \det\left(Q_{J_0} \right)^{-1/2} \exp\left(\frac{\tau^2}{2 \sigma^4} \mathbf{r}^T X_{J_0} Q_{J_0}^{-1} X_{J_0}^T \mathbf{r} \right).
\end{equation}
Since $Q_{J_0} \in \R^{|J_0| \times |J_0|}$ and $|J_0| \le |J|$, we can compute this expression in $O(n|J| + |J|^3)$. We repeat this for each of the $2^{|J|}$ possible values for $A$, yielding a computational complexity of $O(2^k (k^3 + nk))$ to sample $A$. Since typically $k$ is a constant which does not scale with $n$ or $p$ (we suggest taking $k \le 5$), we will write this as $O(2^k n k)$. Repeating this for each of the $m$ blocks and and $n_{\mathrm{iter}}$ iterations yields a time complexity of $O(2^k n k m n_{\mathrm{niter}}) = O(2^k n_{\mathrm{niter}} n p)$ since $k m = O(p)$. Since $k$ is a constant, this reduces to $O(n_{\mathrm{niter}} np)$, although the constant $2^k$ naturally affects the performance of the algorithm.

Once we have sampled $A$, the second step is to sample $\beta_J \mid \beta_{\negc J}, A, \by, H$. Of course, $\beta_{J \setminus A} = 0$, so we need only sample $\beta_A$. Observe that conditional on $A, \beta_{\negc J}$, we have that
\begin{equation*}
    (\mathbf{r}, \beta_{A}) \sim \mcN\left(0, \begin{bmatrix} \sigma^2 \ident_n + \tau^2 X_A X_A^T & \tau^2 X_A \\ \tau^2 X_A^T & \tau^2 \ident_{|A|} \end{bmatrix} \right).
\end{equation*}
Using standard Gaussian update rules, we can sample from the desired conditional distribution. To do this efficiently, we also apply the Woodbury identity to $\sigma^2 \ident_n + \tau^2 X_A X_A^T$. Thus, the second step is to sample from
\begin{equation}\label{eq::betaAmidA}
    \beta_{A} \mid \beta_{\negc J}, A, H, \by \sim \mcN\left(\frac{\tau^2}{\sigma^2} X_A^T \mathbf{r} - \frac{\tau^4}{\sigma^4} X_A^T X_A Q_A^{-1} X_A^T \mathbf{r}, \tau^2 \ident_{|A|} - \frac{\tau^4}{\sigma^2} X_A^T X_A + \frac{\tau^6}{\sigma^4} X_A^T X_A Q_A^{-1} X_A^T X_A \right)
\end{equation}
which can be done in $O(n |A|^2 + |A|^3) = O(n k^2) = O(n)$. Lastly, for completeness, we now specify how to sample the hyper-parameters $H = (\sigma^2, p_0, \tau^2)$ given $\beta$. Let $\mathbf{r} = \mathbf{y} - \bX \beta$ and let $n_{\mathrm{active}} = \#\{j : \beta_j \ne 0\}$ denote the number of active (nonzero) features. Then
\begin{equation}\label{eq::tau2sample}
    \tau^2 \mid \by, \beta, \sigma^2, p_0 \sim \invGamma\left(n_{\mathrm{active}} / 2 + a_{\sigma}, \frac{1}{2} \sum_{j : \beta_j \ne 0} \beta_j^2 + b_{\sigma} \right),
\end{equation}
\begin{equation}\label{eq::sigma2sample}
    \sigma^2 \mid \by, \beta, \tau^2, p_0 \sim \invGamma\left(n/2 + a_{\sigma}, \mathbf{r}^T \mathbf{r} / 2 + b_{\sigma}\right).
\end{equation}
In the case where $p_{\mathrm{min}} = 0$, one can use beta-binomial conjugacy to sample
\begin{equation}\label{eq::p0sample}
    p_0 \mid \by, \beta, \sigma^2, \tau^2 \sim \Beta(a_0 + p - n_{\mathrm{active}}, b_0 + n_{\mathrm{active}}).
\end{equation}
If $p_{\mathrm{min}} \ne 0$, one can repeatedly rejection sample using (\ref{eq::p0sample}) until the result is greater than $p_{\mathrm{min}}$. Since our algorithm maintains a running buffer of $\mathbf{r}$, sampling the hyperparameters requires either constant time or $O(n)$ computations, which is not a bottleneck. The overall time complexity is thus $O(n_{\mathrm{iter}} np)$ as discussed earlier.

\subsection{Details on the probit spike-and-slab sampler}\label{subsec::pss}

In the case of probit regression (Section \ref{subsec::binregsim}), we assume the same model as in Section \ref{subsec::lss} except that we observe $\bz = \I(\by \ge 0) \in \{0,1\}^n$ instead of the continuous outcome $\by$. To sample from the posterior $\beta, \sigma^2, \tau^2, p_0 \mid \bz, \bX$, we follow \cite{albertchib1993} and employ a data-augmentation strategy. In particular, we perform exactly the same steps as Algorithm \ref{alg::lss} except we add an intermediate step where we sample the latent variables $\by$ from the posterior distribution $\by \mid \bX, \beta, \sigma^2, \tau^2, p_0, \bz$. This is simple, since if we let $\mu = \bX \beta$, then conditional on $\mu, \bz, \sigma^2$, the coordinates of $\by$ are independent truncated Gaussians. In particular,
\begin{equation*}
    Y_j \mid \bX, \beta, \sigma^2, \tau^2, p_0 \simind \begin{cases} \truncNorm(\mu_j, \sigma^2, 0, \infty) & \bz_j = 1 \\ \truncNorm(\mu_j, \sigma^2, -\infty, 0) & \bz_j = 0, \end{cases} 
\end{equation*}
where $\truncNorm(\mu, \sigma^2, a, b)$ represents the Gaussian distribution with mean $\mu$ and variance $\sigma^2$ truncated to $(a,b)$. Note that we resample the whole vector $\by$ using this formula every time we update a block of coordinates $\beta_J$ in Algorithm \ref{alg::lss}: otherwise, the algorithm for the probit case is identical to Algorithm \ref{alg::lss}. It requires $O(n)$ operations to resample $\by$, but it also requires $\Omega(n)$ operations to update a block $\beta_J$, so the computational complexity of the probit sampler is still $O(n_{\mathrm{iter}} n p)$. Practically speaking, resampling $\by$ is a bottleneck, so we implemented the algorithm from \cite{truncnorm1991} to do this as quickly as possible.

\subsection{Further discussion of BLiP and SuSiE}\label{appendix::susie}

In this section, we describe how to run \blips on top of the outputs from SuSiE \citep{susie2020} and also explain why SuSiE + \blips can have much higher power than SuSiE alone. To begin with, a short review of SuSiE is in order.

\subsubsection{Review of SuSiE}

The building block of SuSiE is the \textit{single effect regression} (SER) model, which is essentially a spike-and-slab Gaussian regression model with exactly one nonzero coefficient. Formally, this model assumes that $\by \mid \bX \sim \mcN(\bX \beta, \sigma^2 \ident_n)$ and decomposes $\beta = b \gamma$, with priors $b \sim \mcN(0, \tau^2) \in \R$ and $\gamma \sim \Multi(1, \pi) \in \{0,1\}^p$ for $\pi \in [0,1]^n$ satisfying $\sum_{i=1}^p \pi_i = 1$. Also, $b$ and $\gamma$ are assumed to be independent a priori. Since the random vector $\gamma \in \{0,1\}^p$ has exactly one nonzero entry, only one entry of $\beta$ is nonzero. As notation, let $\ell\opt \in [p]$ denote the (random) index such that $\gamma_{\ell\opt} = 1$. For simplicity, we will assume that $\sigma^2, \tau^2$ and $\pi$ are known and fixed in advance, although this need not be the case.

Notably, it is easy to compute the posterior distribution of $b$ and $\gamma$ as
\begin{equation*}
    \gamma \mid \by, \bX \sim \Multi(1, \alpha)
\end{equation*}
\begin{equation*}
    b \mid \by, \bX, \gamma_j = 1 \sim \mcN(\mu_j, v_j^2)
\end{equation*}
where there are analytical formulas for $\alpha \in [0,1]^n, \mu = (\mu_1, \dots, \mu_p) \in \R^n$ and $v^2 = (v_1^2, \dots, v_p^2) \in \R_+^n$ given in \cite{susie2020}. Here, $\alpha_1, \dots, \alpha_p$ are the PIPs for coefficients $\beta_1, \dots, \beta_p$. A key advantage of this model is that one can easily construct a minimal-width credible set for the location of the single nonzero coefficient. Indeed, let $\varphi : [p] \to [p]$ be the permutation which sorts $\alpha$ in decreasing order, so that $\alpha_{\varphi(1)} \ge \alpha_{\varphi(2)} \ge \dots \ge \alpha_{\varphi(p)}$. Then for each $k$, the set $\{\varphi(1), \dots, \varphi(k)\}$ contains $\ell\opt$ (the index of the nonzero coefficient) with probability $\sum_{j=1}^k \alpha_{\varphi(j)}$. The minimal-width $1-q$ credible set simply chooses $k$ to be as small as possible such that $\sum_{j=1}^k \alpha_{\varphi(j)} \ge 1 - q$. As notation, let $\mathrm{CS}(\alpha) = \{\varphi(1), \dots, \varphi(k)\}$ for this choice of $k$.

The Sum of Single Effects (SuSiE) model stacks $L$ SER models on top of each other by assuming that $\beta = \sum_{i=1}^{L} b^{(i)} \gamma^{(i)}$, where $b^{(i)} \simind \mcN(0, \tau^2)$ and $\gamma^{(i)} \simind \Multi(1, \pi)$. To fit a SuSiE model, \cite{susie2020} proposed sequentially fitting $L$ SER models, and this algorithm is known as Iterative Bayesian stepwise selection (IBSS). To be precise, at the first iteration, the IBSS algorithm fits an SER model to obtain outputs $\alpha^{(1)}, \mu^{(1)}, v^{(1)}$ as well as a credible set $\mathrm{CS}(\alpha^{(1)})$ for the first signal. Let $B^{(1)} = \alpha^{(1)} \odot \mu^{(1)}$ be the estimated coefficients, where $\odot$ denotes elementwise multiplication. At this point, one computes the residuals $\br = \by - \bX B^{(1)}$ and then fits a second SER model on $\bX$ and the residuals $\br$ to obtain outputs $\alpha^{(2)}, \mu^{(2)}, v^{(2)}$.  The algorithm repeats for $L$ steps, at each step fitting an SER model and then recomputing the residuals based on the estimated coefficients. For our purposes, the main outputs of SuSiE are the PIPs $\alpha^{(1)}, \dots, \alpha^{(L)}$ and the SuSiE credible sets $\CS(\alpha^{(1)}), \dots, \CS(\alpha^{(L)})$. (It is worth noting that the original SuSiE algorithm only returns $\CS(\alpha^{(i)})$ as a discovery if $\CS(\alpha^{(i)})$ passes a heuristic post-processing check ensuring that the minimum pairwise correlation among variables in $\CS(\alpha^{(i)})$ is greater than $50\%$.)

\subsubsection{Computing the inputs to BLiP from the SuSiE output}

Given these outputs, it is straightforward to compute the inputs for \blip, namely the PIPs for a set of candidate groups $\mcG$. Recall that in a SER model with PIPs $\alpha$, for any group $G \subset [p]$, the PIP $p_G = \P(I_G = 1) = \sum_{j \in G} \alpha_j$ by definition of $\alpha$. The only question is how to combine the PIPs $\alpha^{(1)}, \dots, \alpha^{(L)}$ which correspond to fitting $L$ SER models. However, \cite{susie2020} showed that the IBSS algorithm is a variational approximation to the posterior which assumes that the posterior of the SuSiE model factors into $L$ independent SER models. Under this variational approximation, by independence, we conclude that
\begin{equation}\label{eq::susiepips}
    p_G = \P(I_G = 1) = 1 - \prod_{i=1}^L \left(1 - \sum_{j \in G} \alpha_j^{(i)} \right),
\end{equation}
which allows us to compute PIPs for arbitrary candidate groups based on the SuSiE outputs $\alpha^{(1)}, \dots, \alpha^{(L)}$. Note that this equation is a trivial extension of the formula for PIPs of single variables in the original paper. However, it has a useful consequence: if SuSiE discovers disjoint groups $G_1, \dots, G_m$, then for each $G_k$ there must exist some $i' \in \{1, \dots, L\}$ such that $\sum_{j \in G_k} \alpha^{(i')}_j \ge 1-q$. This plus the previous equation implies $p_{G_k} \ge 1 - q$, so the set of discoveries made by SuSiE are feasible for \blips when controlling the FDR. Since \blips optimizes to maximize the expected resolution-adjusted power, we should expect SuSiE + \blips to make discoveries with higher expected power than SuSiE for any input.

\subsubsection{Why SuSiE + BLiP outperforms SuSiE alone}

The previous paragraph only suggests that SuSiE + \blips weakly outperforms SuSiE. Now, we discuss why SuSiE + \blips can outperform SuSiE by wide margins. At first, this might seem a bit surprising, since at each iteration $i$ SuSiE obtains a minimal credible interval based on $\alpha^{(i)}$. However, SuSiE is not able to combine information across iterations, which can dramatically reduce power when there are multiple non-nulls. To see this, consider a simple example where $\sigma^2 = 1, \pi = \left(\frac{1}{4}, \frac{1}{4}, \frac{1}{4}, \frac{1}{4}\right), p = L = 4, \bX^T \bX = \ident_4$ and $\bX^T \by = (100, 100, 0, 0)$. In this case, many methods (such as a frequentist t-test) can detect that $\beta_1 \ne 0$ and $\beta_2 \ne 0$. However, by symmetry, the SuSiE outputs will be $\alpha^{(1)} = \alpha^{(2)} = \alpha^{(3)} = \alpha^{(4)} \approx (0.5, 0.5, 0, 0)$, suggesting that at each iteration, there is a $50\%$ chance that either $X_1$ or $X_2$ is a signal variable. As a result, when providing level $q=0.1$ error control, SuSiE will output the same discovery set $\{1, 2\}$ four times. In contrast, \blips can use PIPs which combine information across iterations, allowing it to discover the groups $\{1\}, \{2\}$, since $p_{\{1\}} = p_{\{2\}} = 1 - \left(\frac{1}{2}\right)^4 = 0.9375 \ge 1 - q$. 

Of course, this example is a bit contrived, since in this case anyone inspecting the marginal PIPs from SuSiE can tell that $X_1, X_2$ are signal variables, and furthermore it is unlikely the equality $\bX_1^T \by = \bX_2^T \by$ would hold exactly in real data. That said, this general phenomenon is not uncommon: if at any iteration of SuSiE $\bX_j^T \br \approx \bX_k^T \br$, SuSiE will assign PIPs satisfying $\alpha^{(i)}_j \approx \alpha^{(i)}_k \le 0.5$, even if it is obvious that both $X_j$ and $X_k$ are signal variables. The consequence is that SuSiE will only partially regress out the effects of $X_j$ and $X_k$, so it is also likely that $\alpha^{(i+1)}_j, \alpha^{(i+1)}_k$ will be fairly large in the next iteration as well. Admittedly, this does not always happen when there are a small number of signal variables, but when there are a large number of signal variables, the maximum signal size may concentrate around some value. As a result, $\bX_j^T \br \approx \bX_k^T \br$ may occur fairly frequently, and the SuSiE algorithm will have artificially low power. In our view, the best way to recover this power is by applying \blip, since \blips can leverage information from $\alpha^{(1)}, \dots, \alpha^{(L)}$ to compute more accurate PIPs. As exemplified by Figure \ref{fig::linear_hdim}, this can simultaneously lead to higher power and lower FDR. 

\begin{figure}[!htb]
    \centering
    \subfloat{{\includegraphics[width=0.46\textwidth]{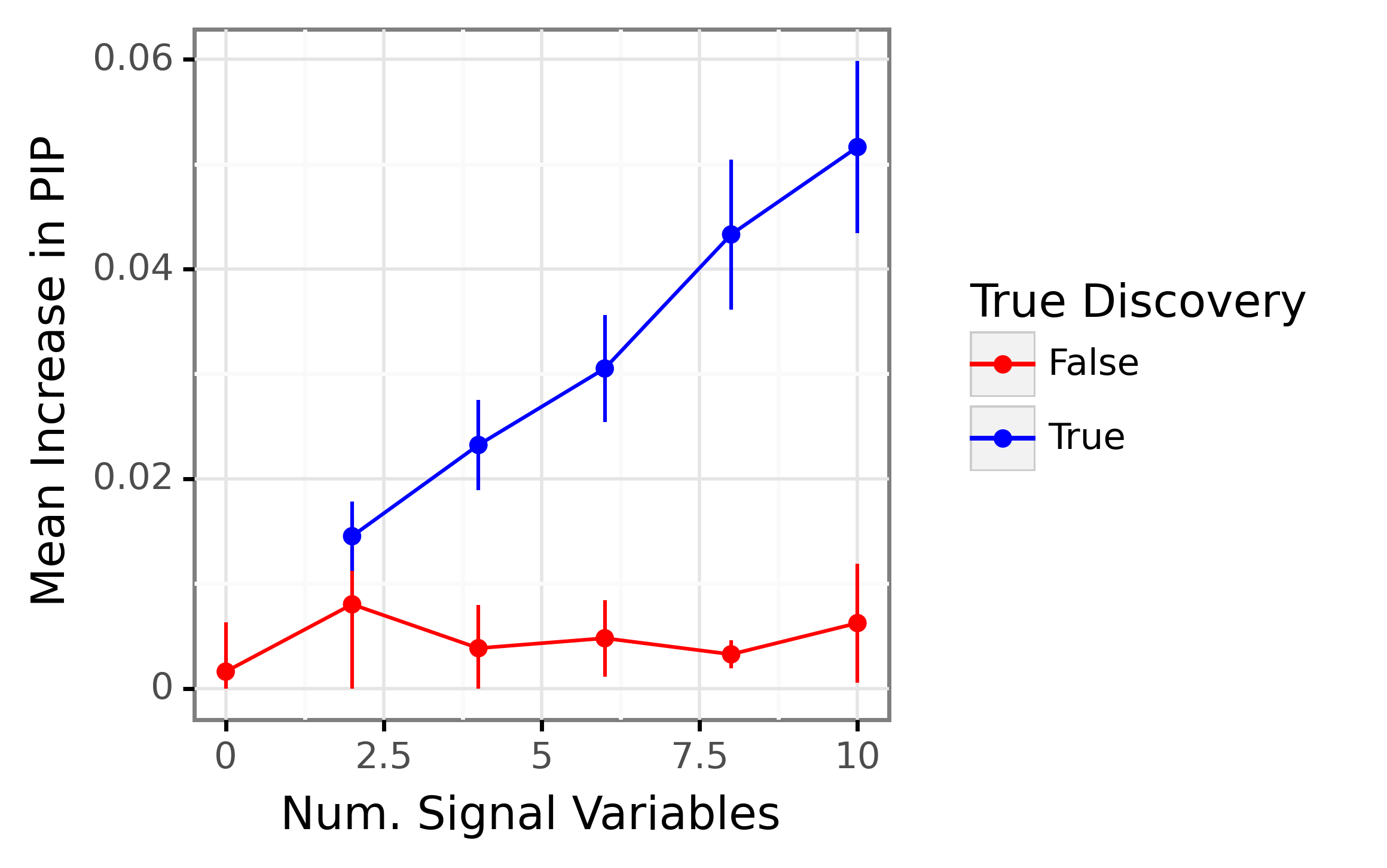}}}\label{subfig::spipsincrease}
    \subfloat{{\includegraphics[width=0.46\textwidth]{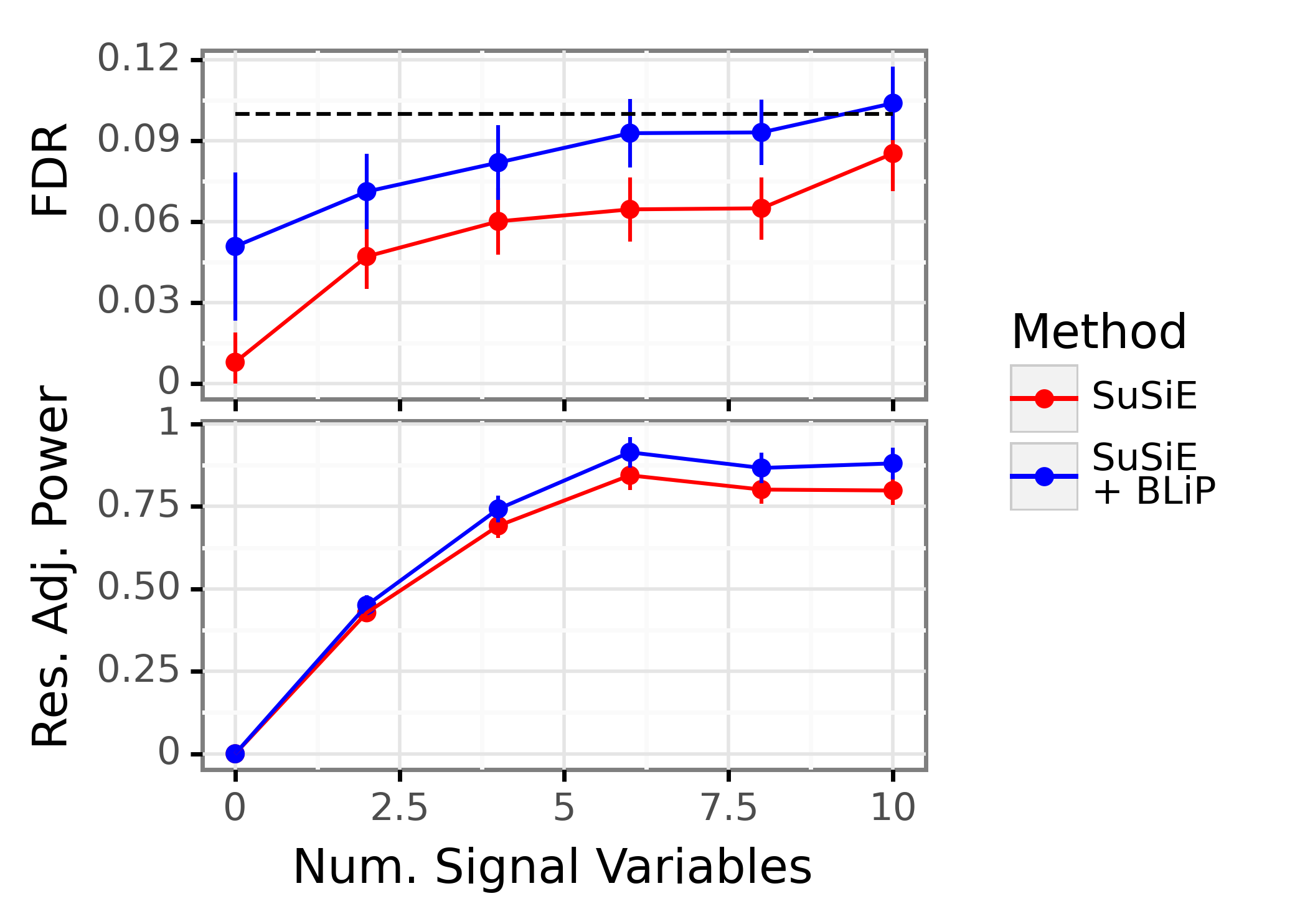}}}\label{subfig::spipsfdr}
     \caption{This figure illustrates why SuSiE + \blips outperforms SuSiE alone in a high-dimensional linear regression setting where $p=200, k=1, n / p = 0.2$, and we run SuSiE with $L=10$ iterations. For each group $G$ detected by SuSiE + \blip, the left plot shows the mean difference between the full PIPs $p_G$, which are used by SuSiE + \blip, and the iteration-specific PIPs $p_G^{\mathrm{max}}$ used by SuSiE. It shows that for true discoveries, $p_G$ is substantially larger than $p_G^{\mathrm{max}}$, whereas for false discoveries, $p_G$ and $p_G^{\mathrm{max}}$ are quite similar. This confirms the intuition in the preceding section and explains why SuSiE + \blips can have higher power than SuSiE alone but good FDR control even when $L$ is larger than the true number of signals. The right plot shows that SuSiE + \blips has higher (unnormalized) power than SuSiE and controls the FDR. See Appendix \ref{appendix::simdetails} for details.}
    \label{fig::susie_pips}
\end{figure}

To illustrate this further, we consider a simple setting where $p = 200$ and there are between $0$ and $10$ signal variables, although in each case we run SuSiE with $L=10$ iterations. For each of the detected groups $G$ returned by SuSiE + \blip, the left plot in Figure \ref{fig::susie_pips} shows the mean difference between $p_G = 1 - \prod_{i=1}^L \left(1 - \sum_{j \in G} \alpha_j^{(i)}\right)$, the aggregate PIP for group $G$, and $p_G^{\mathrm{max}} = \max_{1 \le i \le L} \sum_{j \in G} \alpha_j^{(i)}$, the maximum PIP for group $G$ at any iteration $i$ of SuSiE. This difference is relevant because SuSiE + \blips makes detections based on $p_G$ and SuSiE (roughly speaking) makes detections based on $p_G^{\mathrm{max}}$. Figure \ref{fig::susie_pips} shows that for true discoveries, the aggregate PIPs $p_G$ are often as much as five percentage points larger than $p_G^{\mathrm{max}}$ on average, which is a substantial difference since we are controlling the FDR at level $q = 0.1$. In contrast, $p_G$ is usually only half a percentage point larger than $p_G^{\mathrm{max}}$ for null variables. This suggests that aggregating over the iterations of SuSiE yields more accurate PIPs, which is remarkable since this appears to be true even when SuSiE runs for $L=10$ iterations and there are only $2$ signal variables. Indeed, the right plot in Figure \ref{fig::susie_pips} shows that SuSiE + \blips has higher power than SuSiE and successfully controls the FDR. See Appendix \ref{appendix::simdetails} for more details.

That said, we caution that if $L$ is \emph{much} larger than the number of signal variables, for example if $L = 100$ and there are only one or two signals, we expect that SuSiE + \blips as currently defined will violate FDR control. This motivates our discussion in the next subsection. That said, $L=10$ is the default in genetic fine-mapping, and we emphasize that when $L=10$ we have seen no evidence that SuSiE + \blips violates FDR control even when there are as few as zero signals.

\subsubsection{Further possibilities for SuSiE + BLiP}

We conclude this section by discussing two additional ways to apply \blips on top of SuSiE. First, one might worry that aggregating evidence across all $L$ SuSiE iterations in Equation (\ref{eq::susiepips}) could artificially increase the PIPs, since $L$ is often larger than the true number of signals. This did not appear to be a problem in any of our simulations, e.g., in Figures \ref{fig::susie_pips} and \ref{fig::gwassimpower}, where $L = 10$ and SuSiE + \blips controls the FDR even when there are zero signals. However, one additional way to address this concern would be to only aggregate evidence across iterations of SuSiE which pass the heuristic post-processing steps from \cite{susie2020}. In other words, before computing the PIPs in Equation (\ref{eq::susiepips}), we could completely discard $\alpha^{(i)}$ if $\alpha^{(i)}$ does not yield a sufficiently ``pure" credible set (see \cite{susie2020} for details). Alternatively, one could compute the PIPs as normal but discard all candidate groups which are not sufficiently ``pure." The main point is that any post-processing heuristic which can be applied to SuSiE can likely be applied to SuSiE + \blips as well, although in this paper, we never found it necessary to use postprocessing heuristics when applying \blips on top of SuSiE.

Second, \cite{zou2021} recently introduced a refinement of the IBSS algorithm which is guaranteed to produce a better model fit. Unfortunately, we could not use this new algorithm in our simulations, since it is computationally prohibitive when SuSiE makes many discoveries. For example, in Figure \ref{fig::linear_hdim}, the refined SuSiE algorithm would involve refitting the IBSS procedure hundreds or thousands of times with different initializations. That said, in principle, \blips can apply on top of this refined SuSiE procedure in exactly the same manner as described above. Note that although the refined procedure produces a better fit, its confidence sets still do not aggregate evidence across multiple iterations. Thus, we expect that \blips will improve the power of the refined IBSS procedure for the same reasons outlined in this section.

\section{Applying BLiP when the locations are continuous}\label{appendix::infloc}

In this section, we discuss how to apply \blips when $\mcL$ is a ``continuous" set of locations. For example, in our application to astronomical point source detection in Section \ref{sec::astro}, $\mcL = [0,1]^2$. The continuous case can be challenging when the candidate groups $\mcG$ and locations $\mcL$ are infinite sets. This poses issues both for solving the relaxed LP in the \blips algorithm and for computing the PIPs $\{p_G\}_{G \in \mcG}$ efficiently. The following two sections address these problems separately. First, Appendix \ref{subsec::pipsinfinite} introduces efficient algorithms to compute $\{p_G\}_{G \in \mcG}$ even when $\mcG$ is extremely large or infinite. Second, Appendix \ref{subsec::edgecover} shows that even when $|\mcL| = \infty$ and applying \blips seems to require solving an infinite-dimensional LP, there exist efficient algorithms to reduce the infinite-dimensional LP to an equivalent finite-dimensional LP of (approximately) minimal dimension. Overall, as demonstrated by our astronomical application in Section \ref{sec::astro}, \blips can be very efficient and powerful even when $\mcL$ is continuous and infinite.

Throughout this section, we will assume for simplicity that $\mcL \subset \R^d$, although the ideas in this section extend beyond $\R^d$. We make three further assumptions. First, we assume that $M$ is finite, where $M$ denotes the expected number of signals under the posterior. This assumption is satisfied in any setting where the analyst does not expect to make an infinite number of discoveries. Second, we assume there exists some constant $R \in \N$ such that for each $x \in \R^d$, there are at most $R$ sets $G \in \mcG$ containing $x$. This intuitively means that $\mcG$ contains groups at roughly $R$ resolutions. For example, in our astronomical application, the candidate groups are a collection of circles of radius $r \in \{r_1, \dots, r_m\}$ centered at lattice points. It turns out that each point $x$ is contained by up to $5$ circles of radius $r$ for each $r$, so therefore each point $x$ is contained by at most $R = 5 m$ candidate groups. Overall, this is not a particularly restrictive assumption because $R$ can be arbitrarily large, and this assumption still permits $\mcG$ to be infinite, e.g., if $\mcL = \R^d$ and $\mcG$ covers $\mcL$ with finite rectangles. Lastly, we assume that given any $x \in \R^d$, we can find all groups $G_1, \dots, G_{m} \in \mcG$ containing $x$ in a computational complexity of $O(\varphi(d, R))$, where $m \le R$ and $\varphi$ is some function not depending on $x$. Throughout this paper, $\varphi(R,d) = Rd$, because $\mcG$ is a collection of shapes (spheres, cubes) on a collection of grids of lattice points (see Appendix \ref{appendix::astro} for a precise description). Thus, to find the groups containing $x$, we just round $x$ to the $R$ nearest lattice points, which requires $O(dR)$ time. However, we use the notation for $\varphi(d,R)$ to allow for greater generality.

As discussed in Section \ref{sec::discussion}, there are possible extensions to \blips which apply when $M$ is infinite, although we do not consider them here.

\subsection{Computing PIPs for a continuous set of locations}\label{subsec::pipsinfinite}

Suppose we have access to $N$ samples $\{\gamma^{(1)}, \dots, \gamma^{(N)}\}$ from the posterior of the locations of the signals, which means that the $i$th posterior sample is a set of the form $\gamma^{(i)} = \{\ell_1^{(i)}, \dots, \ell_{m_i}^{(i)}\} \subset \mcL \subset \R^d$. Note that the $i$th posterior sample asserts that there are $m_i$ signals at locations $\ell_1^{(i)}, \dots, \ell_{m_i}^{(i)}$, which implies that the expected number of signals is approximately $M \approx \frac{1}{N} \sum_{i=1}^N m_i$. Given these inputs and a set of candidate groups $\mcG$, in this section we consider the following question: how can we compute the PIPs $\{p_G\}_{G \in \mcG}$ efficiently? 

Recall that for any candidate group $G \in \mcG$, the PIP $p_G$ is simply the proportion of posterior samples $\{\gamma^{(i)}\}$ which include a signal inside $G$. Thus, a naive algorithm would loop through all candidate groups $G \in \mcG$ and compute this proportion. Unfortunately, the number of candidate groups is potentially infinite, and even when it is finite, $|\mcG|$ generally grows exponentially in the dimension $d$, so the naive algorithm is too slow. However, we assume the number of signals is finite, and each signal $\ell_i^{(j)}$ in our posterior samples is contained by at most $R$ candidate groups. Therefore, at most $R N M$ candidate groups contain a signal, and thus when $\mcG$ is large, the vast majority of candidate groups $G \in \mcG$ have PIPs of zero and are irrelevant. This suggests the following algorithm: instead of looping through the candidate groups, we can loop through the signals in the posterior samples $\{\ell_i^{(j)}\}$. By the assumptions at the start of Appendix \ref{appendix::infloc}, for each $1 \le i \le N$, we can compute the groups $G_1, \dots, G_{M_i} \in \mcG$ which contain a signal in $\gamma^{(i)}$, and we are guaranteed $M_i \le R m_i$. After eliminating duplicates in $G_1, \dots, G_{M_i}$, we can then increment the PIPs corresponding to these candidate groups by $\frac{1}{N}$ and proceed to the next posterior sample. This is formally stated in Algorithm \ref{alg::subpartitions}.

\begin{algorithm}[h!]
\caption{Computing PIPs when $\mcL \subset \R^d$.}\label{alg::subpartitions}
\algorithmicensure\, Candidate groups $\mcG$ and posterior samples $\{\gamma^{(i)}\}_{i=1}^N$ for $\gamma^{(i)} \subset \mcL$.
\begin{algorithmic}[1]
        \State Initialize a hash map \texttt{PIP}, which will map candidate groups to PIPs.
        \For {$i \in \{1,\dots,N\}$}
            \State Initialize $\mcA = \{\}$, the set of candidate groups which contain signals according to $\gamma^{(i)}$.
            \For {$\ell \in \gamma^{(i)}$}
                \State Let $C_1, \dots, C_R$ denote the candidate groups in $\mcG$ containing $\ell$.
                \State Set $A = A \cup \{C_1, \dots, C_R\}$.
            \EndFor
            \For {$C \in \mcA$}
                \State If \texttt{PIP}$(C)$ is uninitialized, initialize \texttt{PIP}$(C) = 1/N$. Else, set \texttt{PIP}$(C) = $\texttt{PIP}$(C) + 1/N$.
            \EndFor
        \EndFor
		\State Return \texttt{PIP}, whose keys are the set of candidate groups with nonzero PIPs.
	\end{algorithmic} 
\end{algorithm}

As an illustrative example, suppose $d = 2$ and consider the problem of computing PIPs for $\mcG = \{S_r(r z) : z \in \Z^d, r \in \{r_1, \dots, r_m\}\}$, a collection of circles of varying radii centered at lattice points. As noted previously, for any $\ell \in \R^2$, there are at most $5 m$ circles in $\mcG$ containing $\ell$, and we can find these $R=5m$ circles in $O(R)$ by checking the distances between $\ell$ and its surrounding lattice points. Furthermore, by assumption, the cost of finding the candidate groups $G_1, \dots, G_{M_i}$ containing $\ell$ is $\varphi(d, R)$, where $d$ is the dimension. Thus, Algorithm \ref{alg::subpartitions} runs in $O(NM \varphi(R,d))$. In this paper, $\varphi(d,R)= O(Rd)$, so Algorithm \ref{alg::subpartitions} runs in $O(NMRd)$ for our choices of $\mcG$.

Algorithm \ref{alg::subpartitions} is implemented in \pyblips and \bliprs when $\mcG$ is composed of rectangles and/or spheres, although it applies (in principle) to arbitrary shapes.

\subsection{Reducing the locations to a finite set}\label{subsec::edgecover}

Given a set of candidate groups $\mcG$ and PIPs $\{p_G\}_{G \in \mcG}$, it is not immediately obvious how to run \blips when $\mcL$ has infinite cardinality. To start, note that the size of $\mcG$ is not a problem, per se, for two reasons. First, when the estimated PIPs $\{\hat p_G\}_{G \in \mcG}$ are estimated via sample means from a finite number of samples from the posterior, then in Appendix \ref{subsec::pipsinfinite} we showed that $\{G \in \mcG: \hat p_G > 0\}$ has finite cardinality. Second, as discussed in Appendix \ref{subsec::prefilter}, we recommend adaptively preprocessing $\mcG$ and only considering the sets $\{G \in \mcG : p_G > \epsilon\}$ for some $\epsilon > 0$. In practice, in sparse problems, choosing a reasonable $\epsilon$  (e.g., $\epsilon = 0.5$ when controlling the FDR at $q=0.1$) ensures $\mcG$ is not too large. And formally, the assumptions at the beginning of Appendix \ref{appendix::infloc} imply that $\{G \in \mcG : p_G > \epsilon\}$ has finite cardinality for any $\epsilon > 0$. To see this, let $\nu$ be the measure which maps $G \subset \mcL$ to the expected number of signals in $G$ under the posterior, so for example $M = \nu(\mcL)$ is the total number of expected signals. Let $f$ be the density of $\nu$ with respect to a base measure $\mu$ on $\R^d$. Then note that
\begin{equation*}
    \sum_{G \in \mcG} p_G \le \sum_{G \in \mcG} \int_G f d\mu \le R \int_{\mcL} f d\mu = R M < \infty,
\end{equation*}
where the second inequality follows because each point $x \in \mcL$ is contained in at most $R$ groups in $\mcG$. This shows that $\{G \in \mcG : p_G > \epsilon\}$ has finite cardinality. Thus, under our assumptions, $\mcG$ can be infinite without causing too many problems.

However, when $|\mcL| = \infty$, it is impossible to apply \blips as written. To see this and to ease readability, we reproduce the \blips integer LP below:
\begin{align}
    \max_{\{x_G\}_{G \in \mcG} : x_G \in \{0,1\}} \,\,\,\,\,\,\,\,\,\,\,\, & \sum_{G \in \mcG} p_G \weight(G) x_G \tag{\ref{eq::power_reduction}} \\
    \mathrm{s.t.}\,\,\,\,\,\,\,\,\,\,\,\,\,\,\,\,\,\,\,\, & \sum_{G \in \mcG} (1-p_G-q) x_G \le 0, \tag{\ref{eq::fdr_red}} \\
    & \sum_{G \in \mcG : \ell \in G} x_G \le 1 \,\,\,\,\,\, \forall \ell \in \locations, \tag{\ref{eq::disj_red}} \\
    & x_G \in \{0,1\} \,\,\,\,\,\, \forall G \in \mcG. \tag{\ref{eq::intconstraint}}
\end{align}
The penultimate line represents an infinite number of linear constraints when $|\mcL| = \infty$. Since there are only a finite number of optimization variables, clearly all but finitely many of these linear constraints are redundant! The question is how to compute a (minimal) subset of the constraints in (\ref{eq::disj_red}) which are equivalent to the original constraints. Since each constraint is associated with a location $\ell \in \mcL$, this is equivalent to finding a finite subset $\mcLz \subset \mcL$. As the following proposition makes clear, it turns out this can be reduced to the problem of finding a (minimal) clique cover of the edges of a graph. Before proving this, we first formally define an edge clique cover.

\begin{definition}[Edge clique cover] Let $(V, E)$ be a simple undirected graph with vertices $V = \{v_1, \dots, v_n\}$ and edges $E \subset V \times V$. A subset of the vertices $C \subset V$ is a \textit{clique} if it is fully connected, i.e., for all $v_1 \ne v_2 \in C$, $(v_1, v_2) \in E$. A set of cliques $C_1, \dots, C_k$ is an \textit{edge clique cover} (ECC) of the edges if for all edges $(v_1, v_2) \in E$, there exists some $C_i$ such that $v_1, v_2 \in C_i$.
\end{definition}

The next proposition tells us the following: form the graph with $\mcG$ as the vertices and an edge between $G_1, G_2 \in \mcG$ iff $G_1$ and $G_2$ intersect. Given an edge clique cover $C_1, \dots, C_k$ of this graph, we can reduce the integer LP (\ref{eq::power_reduction})-(\ref{eq::intconstraint}) to an equivalent integer LP with only $k$ constraints. Proposition \ref{prop::ecc} makes this claim precise.

\begin{proposition}\label{prop::ecc} Consider the integer linear program (\ref{eq::power_reduction})-(\ref{eq::intconstraint}). Consider the graph $(V,E)$ where the vertices are the candidate groups, so $V = \mcG$, and $(G_1, G_2) \in E$ if and only if $G_1 \ne G_2$ and $G_1 \cap G_2 \ne \emptyset$. Let $C_1, \dots, C_k$ be an edge clique cover of $E$. Then the integer linear program (\ref{eq::power_reduction})-(\ref{eq::intconstraint}) is equivalent to the same integer LP but with Equation (\ref{eq::disj_red}) replaced by the following constraints:
\begin{equation}
    \sum_{G \in C_i} x_G \le 1 \,\,\,\,\,\, \forall i \in [k]. \label{eq::clique_disj_red}
\end{equation}
\begin{proof} It suffices to show that the constraints in Equation (\ref{eq::disj_red}) are equivalent to the constraints in Equation (\ref{eq::clique_disj_red}).  As notation, for any $\ell \in \mcL$, let $\mcG_{\ell} = \{G \in \mcG : \ell \in G\}$.

For the first direction, suppose that $\{x_G\}_{G \in \mcG}$ are feasible for Equation (\ref{eq::disj_red}): under this assumption, we will show $\{x_G\}_{G \in \mcG}$ satisfy Equation (\ref{eq::clique_disj_red}). In particular, fix any clique $C_i$. It suffices to show that $\sum_{G \in C_i} x_G \le 1$. There are two cases here. First, if $x_G = 0$ for all $G \in C_i$, this condition is satisfied trivially. Second, suppose there exists at least one $G \in C_i$ such that $x_{G} = 1$. We must then show that for every other $G' \in C_i$, $x_{G'} = 0$. This is true because if $G, G' \in C_i$, there is an edge between $G$ and $G'$, and thus there exists some common $\ell \in G \cap G'$. Therefore, the constraint corresponding to $\ell$ in (\ref{eq::disj_red}) guarantees that $x_G + x_{G'} \le 1$, so therefore $x_{G'} = 0$. This completes the proof of the first direction.

For the second direction, suppose that $\{x_G\}_{G \in \mcG}$ are feasible for Equation (\ref{eq::clique_disj_red}). To show $\{x_G\}_{G \in \mcG}$ are also feasible for Equation (\ref{eq::disj_red}), it suffices to show that for any $\ell \in \mcL$, $\sum_{G \in \mcG : \ell \in G} x_G = \sum_{G \in \mcG_{\ell}} x_G \le 1$. (The equality is by definition of $\mcG_\ell$.) To see this, once again, there are two cases. The first case is that $x_G = 0$ for all $G \in \mcG_\ell$, in which case the inequality holds trivially. The second case is that there exists at least one $G \in \mcG_\ell$ such that $x_{G} = 1$. In this case, it suffices to show that for any other $G' \in \mcG_\ell$, $x_{G'} = 0$. However, because $G, G'$ both contain $\ell$, they form an edge in the graph $(V,E)$. Therefore, there exists a clique $C_i$ such that $G, G' \in C_i$. This guarantees that $x_G + x_{G'} \le 1$, and since $x_G = 1$, this implies $x_{G'} = 0$. This completes the proof.
\end{proof}
\end{proposition}

This proposition motivates the following strategy for applying \blip. First, we form the graph $(V, E)$ by checking whether each pair of candidate groups $G_1, G_2 \in \mcG$ intersect. A naive algorithm which loops over all pairs $G_1, G_2$ already runs in polynomial time of the size of $\mcG$, specifically $O(|\mcG|^2)$. The naive algorithm proved to run quickly in our applications, but in large-scale applications, it is likely possible to use spatial tree-structures such as (for example) R-trees \citep{rtree1993} to compute $(V,E)$ substantially more efficiently, although we did not explore that possibility. Second, given the graph $(V,E)$, we find an edge clique cover of (approximately) minimal size and use it to form a finite-dimensional version of \blip, following Proposition \ref{prop::ecc}.

Unfortunately, for any $k_0$, the decision problem asking whether or not there exists an ECC $C_1, \dots, C_k$ with $k \le k_0$ is known to be NP complete \citep{orlin1977}. This means that there is no known polynomial-time algorithm which can find a \textit{minimal} ECC. However, we do not need to find an optimal solution: we only need to find a ``good" solution where $k$ is small. This is certainly possible in polynomial time. At worst, a trivial edge clique cover of $E$ is simply the set of cliques corresponding to all the edges in $E$, e.g., $\{(v_1, v_2) : (v_1, v_2) \in E\}$, since these are cliques of size $2$. This shows that it is always possible to find \textit{an} edge clique cover of size $|E|$ in polynomial time, although in many problems, we can do much better. For example, in our astronomical application, after preprocessing, $|\mcG| \approx 26,000$ and the graph formed by $\mcG$ has over $1,100,000$ edges. Despite this, we are able to find an edge clique cover of size $\approx 12,000$, a $100$x improvement.

There are known polynomial time heuristics for the minimal ECC problem \citep{kou1978, gramm2006}, but we found it easier to implement a simple iterative algorithm (Algorithm \ref{alg::ecc}), and we found it worked well. Algorithm \ref{alg::ecc} does the following. Given $(V, E)$,  we iterate through the edges $e \in E$ and maintain a running list of cliques $C_1, \dots, C_k$ as well as a set of ``uncovered" edges $E'$. For each $e \in E$, if $e$ is uncovered $(e \in E'$), then we pick a maximal clique $C_{k+1}$ containing $e$ and add it to our list $C_1, \dots, C_k$ and remove all edges in $C_{k+1}$ from $E'$. Note that we try to pick $C_{k+1}$ to contain as many uncovered edges as possible using a heuristic based on the degrees of the graph $(V, E')$. Otherwise, if $e$ has already been covered, we move on to the next edge.

\begin{algorithm}[h!]
\caption{Polynomial-time heuristic to find minimal edge clique covers.}\label{alg::ecc}
\algorithmicensure\, A graph $(V, E)$ with $V = \{1, \dots, n\}$.
\begin{algorithmic}[1]
        \State Check that $(V,E)$ has one connected component. Else, apply the algorithm separately to each connected component.
        \State Initialize $E' = E$, $\mcC = \{\}$.
        \State Initialize $D[1], \dots, D[n]$ to be the degrees of vertices $1, \dots, n$.
        \For {$e = (v_1, v_2) \in E$}
            \State If $e \not \in E'$, proceed to the next edge $e$.
            \State Else, initialize a new clique $C = \{v_1, v_2\}$.
            \State Let $N_C = \{v \in V : D[v] > 0, (v_1, v) \in E, (v_2, v) \in E\}$ denote the common neighbors of $C$ with nonzero degree in the uncovered graph $(V, E')$.
            \While {$|N_C| > 0$}
                \State Let $v\opt = \arg\max_{v \in N_c} D[v]$.
                \State Set $C = C \cup \{v\opt\}$ and update $N_C = N_C \cap \{v : (v, v\opt) \in E, D[v] > 0\}$.
            \EndWhile
            \For {$(v, v') \in C$}
                \State Set $E' = E' \setminus (v, v')$
                \State Set $D[v] = D[v]-1$, $D[v'] = D[v'] - 1$.
            \EndFor
            \State Set $\mcC = \mcC \cup \{C\}$.
        \EndFor
		\State Return $\mcC$, an edge clique cover of $(V,E)$.
	\end{algorithmic} 
\end{algorithm}

The worst-case complexity of Algorithm \ref{alg::ecc} is $O(|V|^2 |E|)$, since the cost of adding any clique is at most $O(|V|^2)$, and we add at most one clique per edge. Of course, when the degree-based heuristic performs well, adding a clique of size $k$ requires $O(|V|k)$ operations and will cover $k-1$ uncovered edges, yielding a complexity of $O(|V|^2)$. In our astronomical application, Algorithm \ref{alg::ecc} ran in only a few seconds when $|V| = 26,000, |E|=1,100,000$. 

\section{Further simulations and simulation details}\label{appendix::simdetails}

\subsection{Our choice of candidate groups}\label{subsec::simcandgroups}

In this section, we describe our method for choosing the candidate groups $\mcG$, which we use in all simulations in the paper except those in Appendix \ref{appendix::gwas}. To introduce notation, recall that all of our simulation settings can be represented as a sparse (Bayesian) GLM. Let $\by, \bX$ be the data and let $\beta \in \R^p$ be the linear coefficients, where nonzero coefficients correspond to signals. The set of locations is $\mcL = \{1,\dots,p\}$. Furthermore, in each setting, we use various methods to compute samples $\tilde{\beta}^{(1)}, \dots, \tilde{\beta}^{(N)} \in \R^p$ approximating the posterior of $\beta$. (The exception to this is when we run \blips on top of SuSiE, which we will discuss at the end of this section.) Given these samples, let $\epsilon_{ij} = \I(\tilde{\beta}^{(i)}_j \ne 0)$ be the indicator of whether there is a signal at location $j$ in posterior sample $i$, forming the binary matrix $\epsilon \in \{0,1\}^{N \times p}$. As a reminder, given any group $G \subset \mcL$, the estimated PIP $p_G$ is
\begin{equation*}
    p_G \approx \frac{1}{N} \sum_{i=1}^N \I(\exists \, j \in G : \epsilon_{ij} = 1). 
\end{equation*}
Finally, let $\hat \Sigma = \frac{1}{n} \tilde{\bX}^T \tilde{\bX}$ be the empirical correlation matrix of $\bX$, where $\tilde{\bX}$ is $\bX$ but shifted and scaled so the columns of $\tilde{\bX}$ sum to zero and have unit variance. Similarly, let $\hat \Sigma_{\epsilon}$ denote the empirical correlation matrix of $\epsilon$.

Below, we specify how we choose the candidate groups.
\begin{enumerate}
    \item First, for some choice of $\kappa > 0$ (we will specify $\kappa$ in step 4), we adaptively pre-filter the locations to obtain $\mcL_{\kappa} = \{\ell \in \mcL : p_{\{\ell\}} \ge \kappa\} \subset \mcL$. Note that we can represent $\mcL_{\kappa} = \{\ell_1, \dots, \ell_k\}$ for some $k$ such that $\ell_1 < \ell_2 < \dots < \ell_k$. 
    \item Second, we consider all \textit{contiguous} sub-groups of $\mcL_{\kappa}$ of maximum size less than $m$: recall that a contiguous group is a group of the form $\{\ell_{i}, \ell_{i + 2}, \ell_{i+3}, \dots, \ell_{i+m_0}\}$ where we restrict $m_0 \le m$, so there are $O(m k)$ contiguous subgroups. By default, we pick $m = 25$ unless specified otherwise. Let $\mcG_{\kappa}^{\mathrm{seq}}$ denote these subgroups.
    \item Third, we hierarchically cluster $\mcL_{\kappa}$ based on a dissimilarity metric $d : \mcL_{\kappa} \times \mcL_{\kappa} \to \R_+$. For each dissimilarity metric, we apply single-linkage, average-linkage, and complete-linkage clustering, yielding three tree-structured sets of candidate groups $\mcG_{\kappa,d}^{\mathrm{single}}, \mcG_{\kappa,d}^{\mathrm{average}}, \mcG_{\kappa,d}^{\mathrm{complete}}$. We also restrict these sets to have cardinalities of $m$ or less. We repeat this for two dissimilarity metrics, one based on $\bX$ where $d_1(i,j) = |1 - \hat \Sigma_{i,j}|$ and a second based on $\epsilon$ where $d_2(i,j) = 1 + (\hat \Sigma_{\epsilon})_{i,j}$.
    \item We repeat this process for $\kappa \in \mcK$ where we pick $\mcK = \{0,0.01, 0.02, 0.03, 0.05, 0.1, 0.2\}$. Then, the final set of candidate groups is
    \begin{equation*}
        \mcG = \bigcup_{\kappa \in \mcK} \mcG_{\kappa}^{\mathrm{seq}} \cup \left(\bigcup_{d \in \{d_1, d_2\}} \mcG_{\kappa,d}^{\mathrm{single}} \cup \mcG_{\kappa,d}^{\mathrm{average}} \cup \mcG_{\kappa,d}^{\mathrm{complete}}\right).
    \end{equation*}
\end{enumerate}

After creating $\mcG$, we adaptively preprocess $\mcG$ and only include candidate groups $G$ such that $p_G \ge 0.5$. Since we control the FDR at level $q = 0.1$, it is unlikely that including groups with $p_G < 0.5$ would lead to higher power (Appendix \ref{subsec::prefilter}). 

Lastly, if we are applying \blips on top of SuSiE, we use the same procedure, with two exceptions. First, we add the groups discovered by the default SuSiE algorithm to the candidate groups. Second, naturally, we do not include the hierarchical groups generated using the covariance matrix $\hat \Sigma_{\epsilon}$, since SuSiE does not produce posterior samples $\epsilon$. See Appendix \ref{appendix::susie} for more discussion of how to compute PIPs when working with outputs from SuSiE.

\subsection{Details for variable selection simulations}\label{subsec::glmdetails}

\begin{figure}
    \centering
    \subfloat[]{{\includegraphics[width=0.49\textwidth]{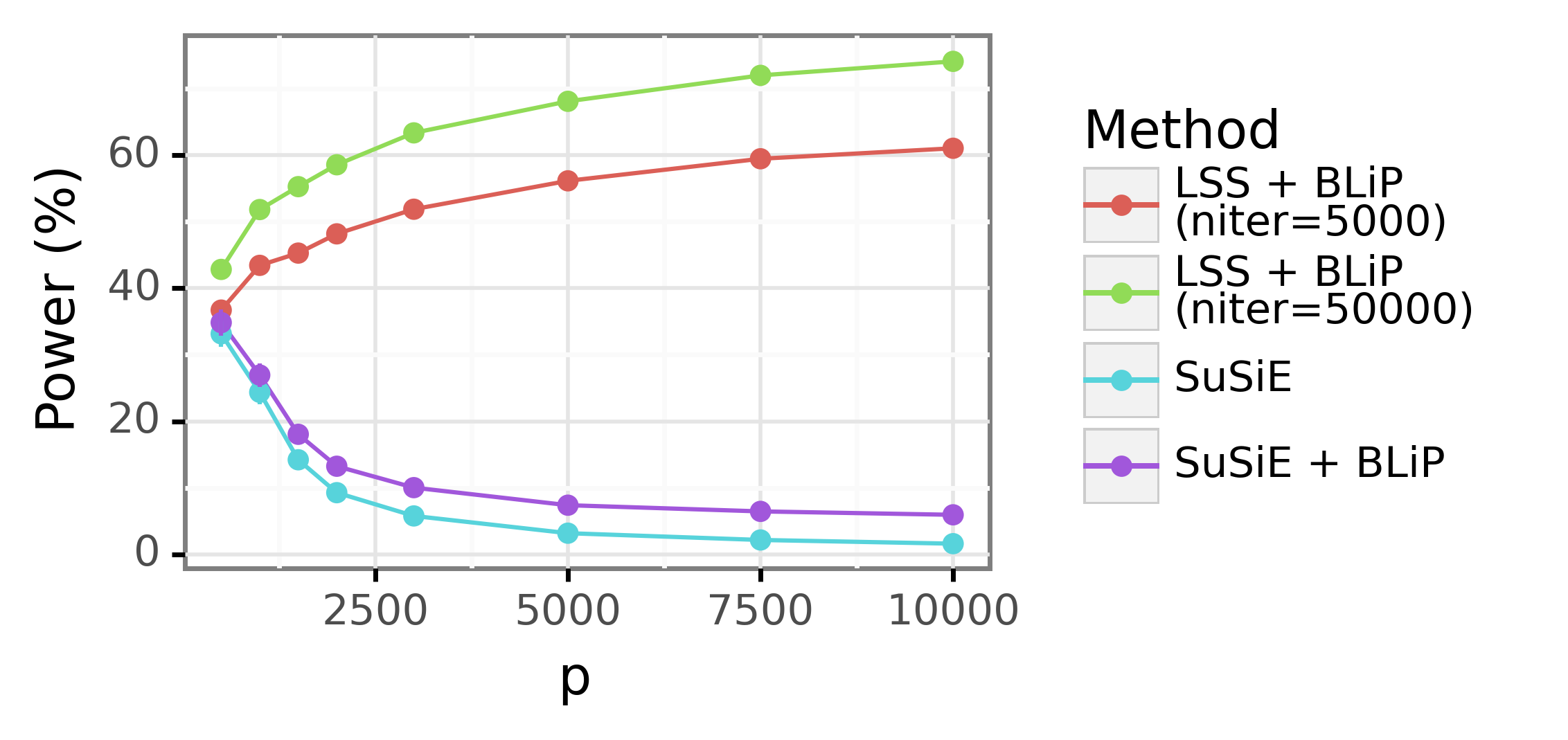}}}
    \subfloat[]{{\includegraphics[width=0.49\textwidth]{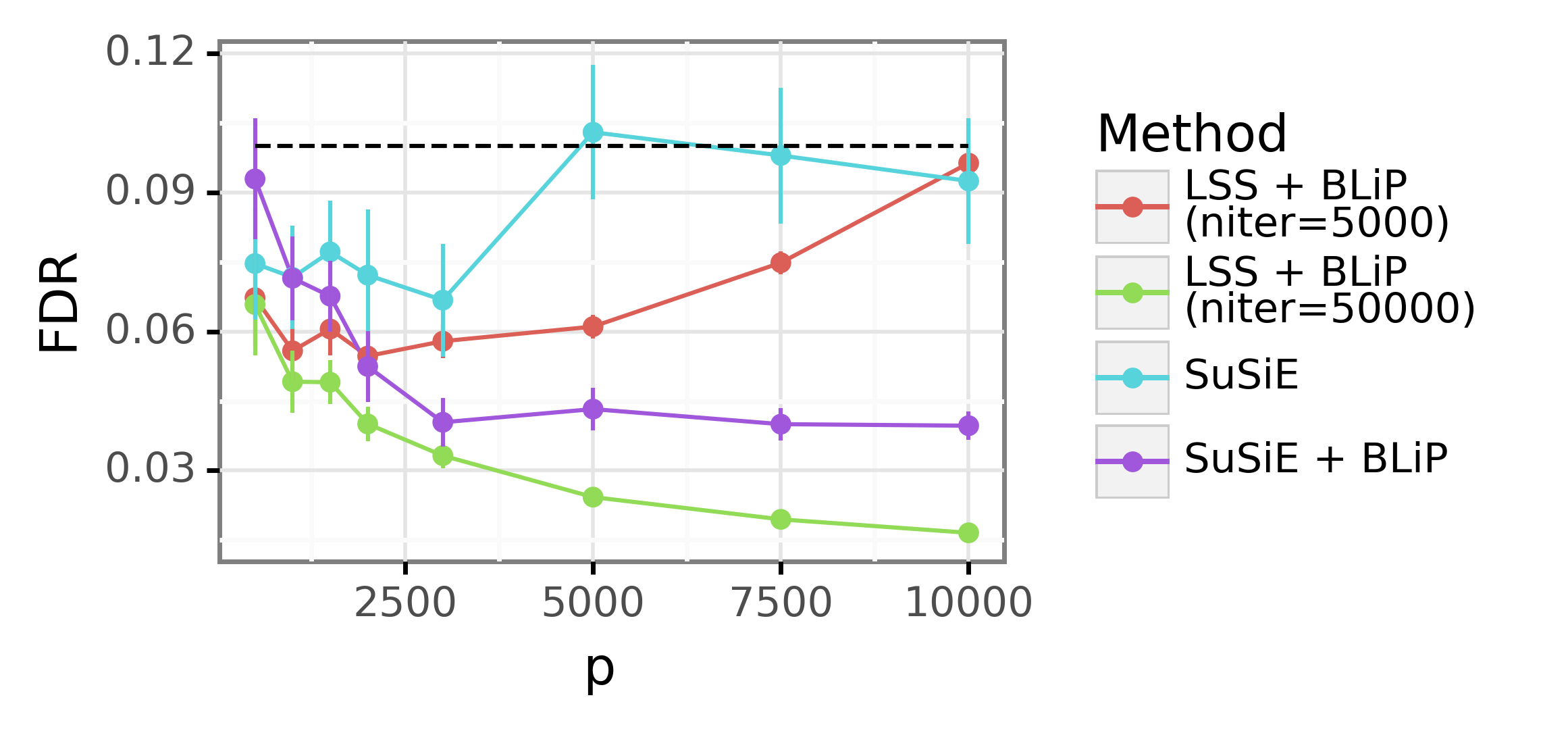}}}
    \caption{The corresponding power and FDR plots of Figure \ref{fig::linear_vp}. Panel (a) shows SuSiE loses power compared to the other methods as $p$ grows, even as $n / p$ and the signal size are held constant. Panel (b) shows that all methods controlled the FDR up to Monte Carlo error.}
    \label{fig::linear_vp_fdr}
\end{figure}

In this section, we describe the data-generating process and the methods used for all variable selection simulations in the paper. To be clear, all of the descriptions in this section apply to Figures \ref{fig::blipintsol}, \ref{fig::linear_hdim}, \ref{fig::linear_vp}, \ref{fig::randcounts}, \ref{fig::fwerintsol}, \ref{fig::nchains}, \ref{fig::susie_pips} and \ref{fig::linear_vp_fdr}. We also include the corresponding power and FDR plots for Figure \ref{fig::linear_vp}, which show that the power of SuSiE decreases compared to other methods as the absolute number of signals grows. All code for these simulations is available at \url{https://github.com/amspector100/blip_sims}.

\begin{figure}
    \centering
    \includegraphics[width=0.5\linewidth]{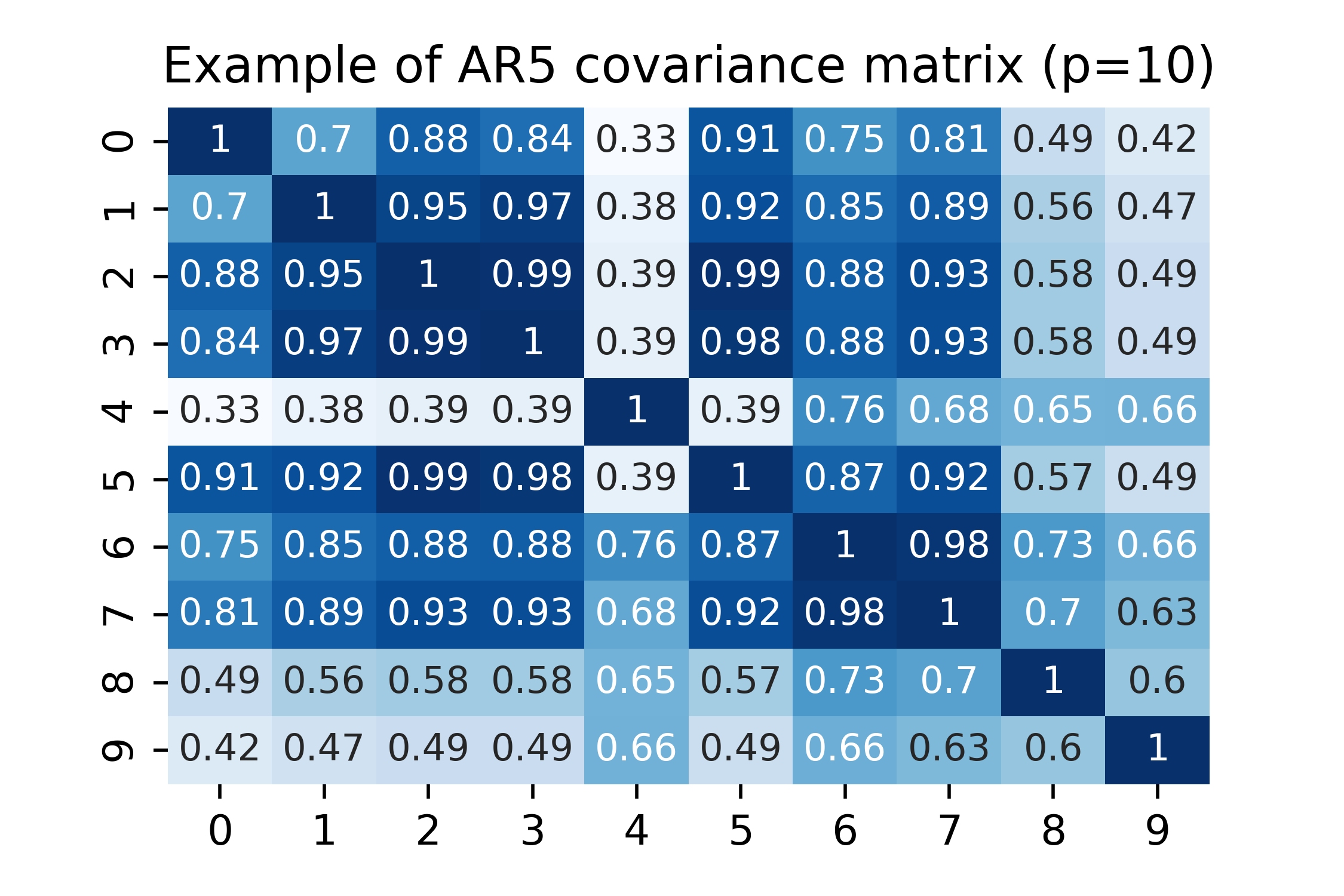}
    \caption{An example of the AR5 covariance matrix we used in variable selection simulations.
    }
    \label{fig::ar5ex}
\end{figure}

To start, we describe the data-generating processes for all figures involving variable selection.
\begin{enumerate}
    \item Throughout, we sample $X \in \R^p$ to follow a non-stationary AR(k) process. In particular, to sample $X$, we sample $Z_1, \dots, Z_p \iid \mcN(0,1)$ and initialize $X_1 = Z_1$. We then iteratively set 
    \begin{equation*}
        X_j = \rho^{(j)}_{0} Z_j + \sum_{\ell=1}^{\min(j-1,k)} \rho^{(j)}_{\ell} X_{j-\ell}.
    \end{equation*}
    For each $j$, we sample $(\rho_0^{(j)}, \dots, \rho_{k}^{(j)}) \iid \mathrm{Dirichlet}\left(\left(0.2, \frac{0.8}{k-1}, \dots, \frac{0.8}{k-1} \right)\right)$, and then we rescale $(\rho_0^{(j)}, \dots, \rho_{k}^{(j)})$ so that $\var(X_j) = 1$. This sampling structure was motivated by fine-mapping studies, since it creates very strong local dependencies. Figure \ref{fig::ar5ex} gives an example of what the covariance matrix looks like with $k = 5$ and $p = 10$. In all simulations, we choose $k = 5$ unless otherwise specified in the plot.
    \item In all regression settings, for simulations with sparsity $s \in [0,1]$, we set $\beta \in \R^p$ to equal all zeroes except for $\lceil s p \rceil$ randomly chosen coefficients. The nonzero coefficients are sampled as i.i.d. $\mcN(0, \tau^2)$ random variables, except that we ensure all nonzero coefficients have absolute values greater than $0.1 \tau$. We do this so that the resolution-adjusted power of each method approaches $\lceil s p \rceil$ with a reasonable sample size (otherwise, there will be a few coefficients which are extremely close to zero, and no method will detect them without $n >> p$). We sample $\by \mid \bX, \beta \sim \mcN(\bX \beta, \sigma^2 \ident_n)$ with $\sigma^2 = 1$. In the linear regression simulations, we observe $(\bX, \by)$: in the probit simulations, we only observe $(\bX, \bz)$ where $\bz = \I(\by \ge 0)$. The only parameter which changes between figures is $\tau^2$. We picked $\tau^2 = 0.5$ in Figures \ref{fig::blipintsol}, \ref{fig::binreg_power_fdr}, and \ref{fig::fwerintsol}, and $\tau^2 = 1$ otherwise.
\end{enumerate}

Next, we describe the methods we compare between in more detail.
\begin{enumerate}
    \item \textbf{LSS + \blip}: For the method LSS + \blip, we use the LSS sampler described in Section \ref{subsec::lss} with the true values of $\sigma^2, \tau^2, p_0$. This means that we do not use hyperpriors. The LSS + \blips (misspec.) method is identical except that we use the hyperpriors specified in Section \ref{subsec::lss} with $a_{\sigma} = 2, b_{\sigma} = 1, a_{\tau} = 2, b_{\tau} = 1$, $a_0 = 1, b_0 = 1$ and $\minp = 0.9$. Note that these are fairly uninformative priors. For the Gibbs update, we use a block size of $5$ for all simulations except Figures \ref{fig::blipintsol} and \ref{fig::linear_vp}, where we use a block size of $1$. We run $N = 10000$ samples per chain with $10$ chains, unless otherwise specified. After fitting these samplers, we apply \blips to the candidate groups described in Appendix \ref{subsec::simcandgroups}.
    
    \item \textbf{PSS + \blip}: The methods PSS + \blips and PSS + \blips (misspec.) are identical to the corresponding LSS-based methods except that we use the probit sampler with the extra data-augmentation step described in Appendix \ref{subsec::pss}.
    \item \textbf{SuSiE-based methods}: In all simulations, we apply SuSiE with the default parameters in the susieR package \citep{susie2020} except that we provide SuSiE the oracle value for the number of signals. For SuSiE + \blip, we use the same SuSiE model and extract PIPs from SuSiE as described in Appendix \ref{appendix::susie}. 
    
    \item \textbf{DAP-G}: In all instances, we apply DAP-G to Z-statistics based on $\bX, \by$. We run DAP-G with the default parameters in the package from \cite{dapg2018} except that we give DAP-G oracle knowledge of the sparsity $s$. For the sake of computation, we tell DAP-G the exact number of non-nulls, so that it does not search over models with more than $\lceil s p \rceil$ parameters.
    \item \textbf{dCRT+FBH and dCRT+Yekutieli}: For both of these methods, we apply the $d_0$CRT based on a cross-validated lasso with analytical p-values as described in \cite{dcrt2020}. The FBH and Yekutieli methods apply to tree-structured p-values, so to create the tree, we hierarchically cluster the covariates with average linkage based on the empirical correlation matrix of $\bX$. We cut the correlation tree at $10$ evenly spaced values between $0$ (which corresponds to individual covariates) and the first correlation at which the maximum group size exceeds $m=25$, the same maximum groupsize that we apply with \blip. We cut the correlation tree at $10$ levels for two reasons. First, we chose the number of levels to approximately balance the benefit of more levels (high adaptivity) against the drawback, namely that it makes the FBH more conservative by increasing the multiplicity correction. Note we were only able to do this because we could simulate from the true data-generating process, and in general it is not clear how one would choose the number of levels. Second, computing many $d$CRT p-values is very expensive, and cutting the tree at (e.g.) $100$ different levels would have made these procedures computationally intractable. As is, the dCRT-based methods were $100$x more expensive than the Bayesian samplers in Figure \ref{fig::linear_hdim}.

    It may also be helpful to review exactly how we extended the $d_0$CRT of \cite{dcrt2020} to test group hypotheses. As a short review, the dCRT is used to test the conditional independence hypotheses $H_j : X_j \Perp Y \mid X_{-j}$, where in a generalized linear model this is equivalent to testing $H_j : \beta_j = 0$ under mild conditions on the distribution of $X$. The dCRT yields valid p-values under the model-X assumption \citep{mxknockoffs2018}, which assumes knowledge of the distribution of $X$. (In contrast, the Bayesian procedures, including \blip, assume a full Bayesian model of $Y \mid X$ but no knowledge of $X$.) For fair comparison, we give the dCRT oracle knowledge of the distribution of $X$.
    
    Consider the problem of obtaining a $d_0$CRT-based p-value for the group null hypothesis $H_G : \bigcap_{j \in G} H_j$, for any $G \subset [p]$. We will describe this in the linear regression case first and then generalize to the probit case. The first step is identical to the normal $d_0$CRT, which is to compute $\bd_y$, the \textit{distillation} of $\by$. Intuitively, $\bd_y$ is a proxy of $\E[\by \mid \bX_{-G}]$, although it can be any function of $\bX_{-G}, \by$ while maintaining Type I error control. We choose $\bd_y$ to be predictions from a cross-validated lasso fit on $\by, \bX_{-G}$. The next step is to compute analytical p-values based on $\by, \bd_y, \bX_{G}$. In particular, in our setting, $X \sim \mcN(0, \Sigma)$ is multivariate Gaussian. As a result, $X_{G} \mid X_{-G} \sim \mcN(A_{-G} X_{-G}, \Sigma_G) $ where $A_{-G} \in \R^{|G| \times (p-|G|)}$ and $\Sigma_{G} \in \R^{|G| \times |G|}$ are known quantities depending on $\Sigma$. Under the null $H_G$, this is also true conditional on $\by, \bd_y$. Let $\bx_G = A_{-G} \bX_{-G}$. Then, under the null, we have that
    \begin{equation*}
        \left(\bX_{G} - \bx_G\right)^T (\by - \bd_y) \mid \by, \bX_{-G} \sim \mcN\left(0, ||\by - \bd_y||_2^2 \Sigma_G \right)
    \end{equation*}
    and therefore if we apply the whitening transformation $\Sigma_{G}^{-1/2}$, we can use the test statistic
    \begin{equation*}
        T_G \defeq ||((\bX_{G} - \bx_G) \Sigma_{G}^{-1/2})^T (\by - \bd_y)||_2^2
    \end{equation*}
    where $T_G \mid \bX_{-G}, \by \sim \chi_{|G|}^2$ under the null. Therefore, the p-value for group $G$ is simply $1 - F_{\chi_{|G|}^2}(T_G)$, where $F_{\chi_{|G|}^2}$ is the CDF of a $\chi_{|G|}^2$ random variable. The p-value for the probit case is identical except we replace $\by$ with $\bz$ and apply a cross-validated logistic lasso instead of a lasso during the distillation.
    
    Note that we do not use the pre-screening procedure from \cite{dcrt2020} in either the linear or probit case. We made this choice because pre-screening reduced power empirically, perhaps because the very high correlations in $\bX$ make it very easy to accidentally screen out signal variables.
\end{enumerate}

Lastly, Figure \ref{fig::linear_vp_fdr} is the corresponding power and FDR plot for Figure \ref{fig::linear_vp}.

\subsection{Sensitivity to different weight functions}\label{appendix::weightfn}

In this section, we show empirically that the output of \blips is not too sensitive to the choice of weight function. 
As a simple example, we compare the weight functions $w_1(\mcG) = \frac{1}{|\mcG|}$ and $w_2(\mcG) = \frac{1}{1 + \log_2(|\mcG|)}$. Note the latter weight function is a monotone transformation of the former; it still encodes the principle that it is preferable to discover smaller groups, but it quantifies this principle rather differently. 

\begin{figure}[!h]
    \centering
    \includegraphics[width=\linewidth]{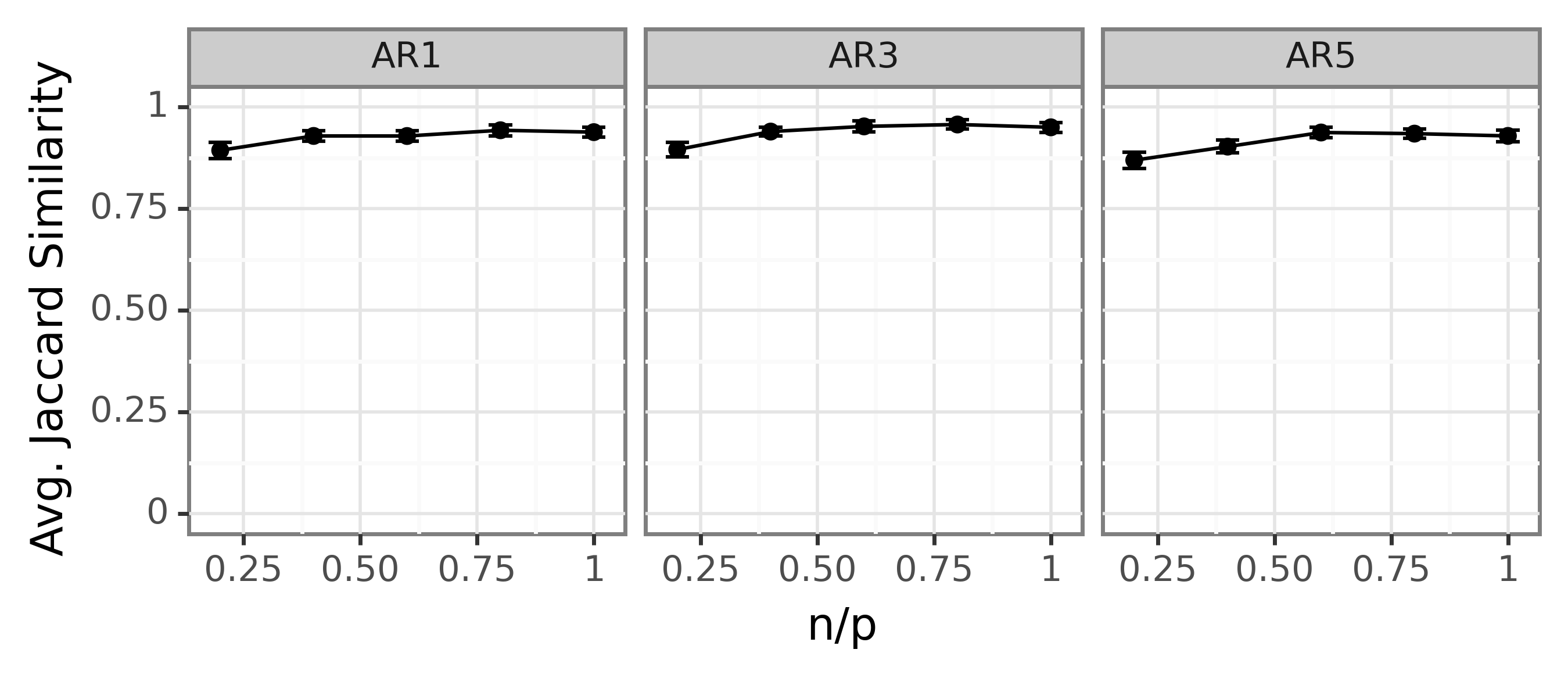}  
    \caption{This plot shows the average Jaccard similarity (defined above) of the rejection sets from \blips with two weight functions in exactly the same simulation setting as Figure \ref{fig::blipintsol}. It shows that the results from \blips are not too sensitive to the choice of weight function.}
    \label{fig::weightfnsim}
\end{figure}

In particular, we exactly replicate the simulation setting of Figure \ref{fig::blipintsol}, but we run \blips with both weight functions and obtain two rejection sets $\mcR = \{G_1, \dots, G_R\}$ and $\mcR' = \{G_1', \dots, G_{R'}'\}$. To measure how similar these rejections are, recall that the \textit{Jaccard similarity} of two sets $A, B$ is defined as $J(A,B) = \frac{|A \cap B|}{|A \cup B|}$. For each rejected region $G \in \mcR$ outputted by the first weight function, we compute its maximum Jaccard similarity with any rejection from the second weight function. Formally, we define
\begin{equation*}
    J(G, \mcR') = \max_{G' \in \mcR'} \frac{|G \cap G'|}{|G \cup G'|}.
\end{equation*}
Intuitively, when $J(G, \mcR')$ is high, this means there exists a rejected region $G' \in \mcR'$ that is highly similar to $G$. To measure how similar $\mcR$ and $\mcR'$ are, we average this quantity over all $G \in \mcR, G' \in \mcR'$, and we call the result the \textit{average Jaccard similarity} $J(\mcR, \mcR')$, as defined below: 
\begin{equation}
    J(\mcR, \mcR') = \frac{1}{|\mcR| + |\mcR'|} \left(\sum_{G \in \mcR} J(G, \mcR') + \sum_{G' \in \mcR'} J(G', \mcR) \right).
\end{equation}
Figure \ref{fig::weightfnsim} shows the results, namely that the average Jaccard similarity is above $85\%$ for the two weight functions (up to MCMC error). This suggests that using either weight function should give reasonably similar scientific results.

\subsection{Simulations for change point detection}\label{subsec::changepointsim}



In this section, we show that \blips has potential as a method for detecting change points in a time series. As notation, consider a time series $Y_1, \dots, Y_T$ which is stationary except at a few times, called change points, where the process changes: here, the locations $\locations = \{1, \dots, T\}$ represent $T$ discrete times, there is a signal at time $t$ if there is a change point at time $t$. Resolution-adaptivity is important because often, we cannot identify exactly \textit{when} a time series has changed (perhaps because each observation $Y_t$ is noisy), even if it is clear that $\{Y_t\}$ has change points. Thus, we opt to make statements like ``there was at least one change point between time $t_0$ and $t_1$."

\begin{figure}[!h]
    \includegraphics[width=\linewidth]{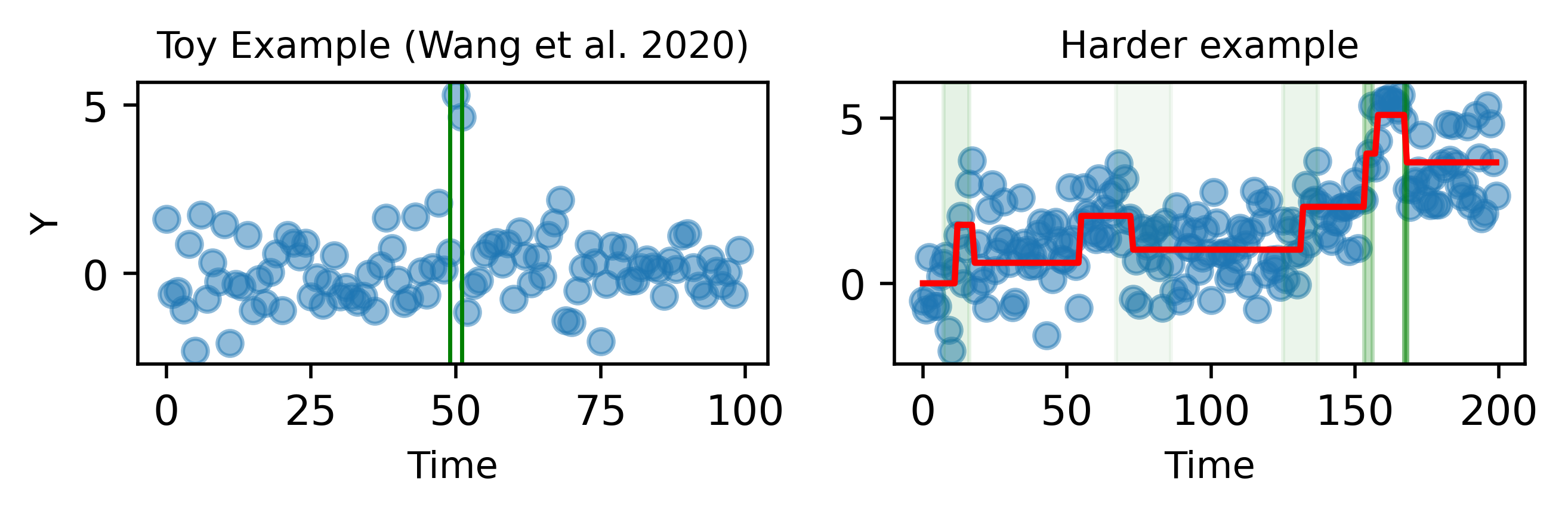}  
    \caption{Two examples of LSS + \blips applied to change point detection. In both cases, the blue points represent $\{Y_t\}$ and the green regions represent detections from LSS + \blip. The left example is a setting from \cite{susie2020} where SuSiE cannot escape a local minimum and makes no discoveries, but LSS + \blips discovers both change points. The right example shows \blip's output in a more challenging setting. Note in the right example, the discontinuities in the red line denote true change points.}
    \label{fig::cp_s3_ex}
\end{figure}

We consider a model where $Y_t \simind \mcN(\mu_t, \sigma^2)$ and $\mu_t = \mu_{t-1} + \beta_t$ where $\beta_t$ is a mixture of a point mass at zero and a $\mcN(0, \tau^2)$ distribution. We chose this model for sake of simplicity and because it has been frequently discussed in prior work, such as in \cite{susie2020, fang2020, cappello2021}, although \blips could easily wrap on top more complex models, for example those which account for temporal dependence. As noted by \cite{susie2020}, this problem can be described as a sparse regression problem where $\bX \in \R^{T \times T}$ is defined so that $\bX_j = (\mathbf{1}_{T-j+1}, \mathbf{0}_{T-1}) \in \R^T$. As illustrated by Figure \ref{fig::cp_s3_ex}, \blips will output regions where it detects at least one change point, and it makes these regions as narrow as possible while controlling the FDR. As discussed in Section \ref{subsec::literature}, there are several frequentist methods which can accomplish this while controlling the FWER \citep{frick2014, fang2016, fang2020, fryzlewicz2020}. Similarly, \cite{limunk2016} control a slightly modified FDR, although they do not aim to precisely localize signals. Since these methods are not directly comparable to \blips while controlling the FDR, our simulations only compare against SuSiE.

To illustrate one possible advantage of \blip, we first consider a toy example from \cite{susie2020}, where there is one very large change point at $T = 50$ which immediately is cancelled out by another change point at $T = 52$. Since SuSiE is a greedy method and adding a single change point never improves the model fit, SuSiE discovers no change points. In contrast, LSS + \blips overcomes this optimization hurdle and correctly identifies the two change points, as in Figure \ref{fig::cp_s3_ex}. We emphasize that this toy example is motivated by real applications, where (for example) the mean of an economic indicator might temporarily jump before reverting to its previous value \citep{fang2020}. Of course, as discussed in \cite{susie2020}, SuSiE can perform very well as a change point detection algorithm in other settings, and in these cases, SuSiE + \blips can slightly improve SuSiE's power.

For more comprehensive simulations, we set $T = 300$ with $\lceil s T \rceil$ change points and vary the signal size $\tau^2$ and the sparsity $s$. To capture this ``cancellation" effect, we chose the change points so that the distances between consecutive change points are distributed as i.i.d. $\Unif([2d])$ random variables. Intuitively, $d$ represents the ``spacing" or average distance between change points. We consider six methods. To start, as discussed in Section \ref{subsec::changepointsim}, this problem can be represented as a sparse linear regression problem, so we apply the methods SuSiE, SuSiE + \blip, LSS + \blip, and LSS + \blips (misspec.) from the linear regression setting in Section \ref{subsec::linregsim}. We use the exact same parameters as described in Appendix \ref{subsec::glmdetails} for these methods. Second, we apply \blips on top of posterior samples from the BCP method from \cite{bcp2015}. We use the default samples from the BCP package except in the well-specified setting, where we use oracle values for the parameters governing the sparsity level and signal size. For all methods involving \blip, we use the candidate groups described in Appendix  \ref{subsec::simcandgroups}.

\begin{figure}[!h]
    \includegraphics[width=\linewidth]{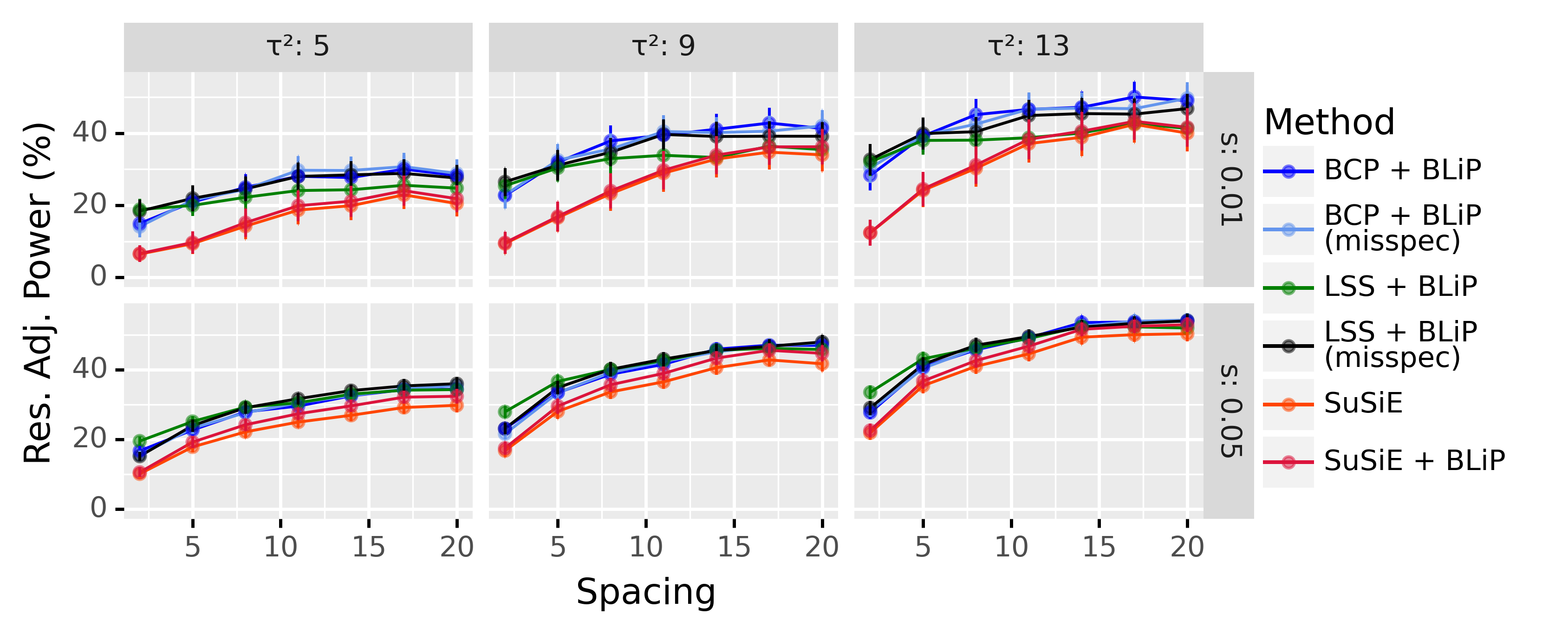}  
    \caption{We apply LSS + \blip, BCP + \blip, SusiE, and SuSiE + \blips to perform change point detection in a Gaussian spike and slab model, as described in Section \ref{subsec::changepointsim}. The x-axis denotes the average spacing between consecutive change points.
    }
    \label{fig::cp_power}
\end{figure}

\begin{figure}
    \includegraphics[width=\linewidth]{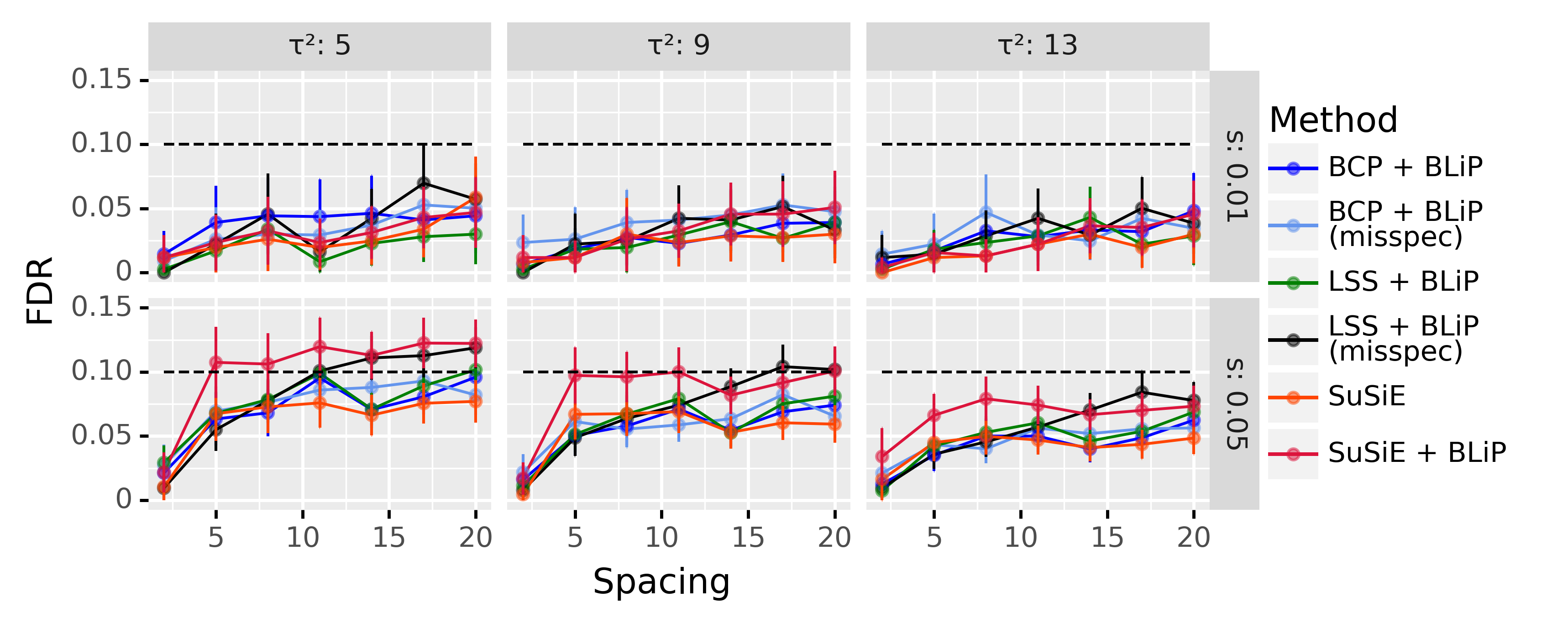}  
    \caption{The corresponding FDR plot of Figure \ref{fig::cp_power}.}
    \label{fig::cp_fdr}
\end{figure}

As shown in Figure \ref{fig::cp_power}, for change points which are close together, BCP + \blips and LSS + \blips are often much more powerful than the SuSiE-based methods. 
That said, when the change points are fairly spaced out, SuSiE performs quite well, although the \blip-based methods do give a slight boost in power. All methods controlled the FDR up to Monte Carlo error, as shown by Figure \ref{fig::cp_fdr}.

\begin{figure}
    \centering
    \includegraphics[width=\textwidth]{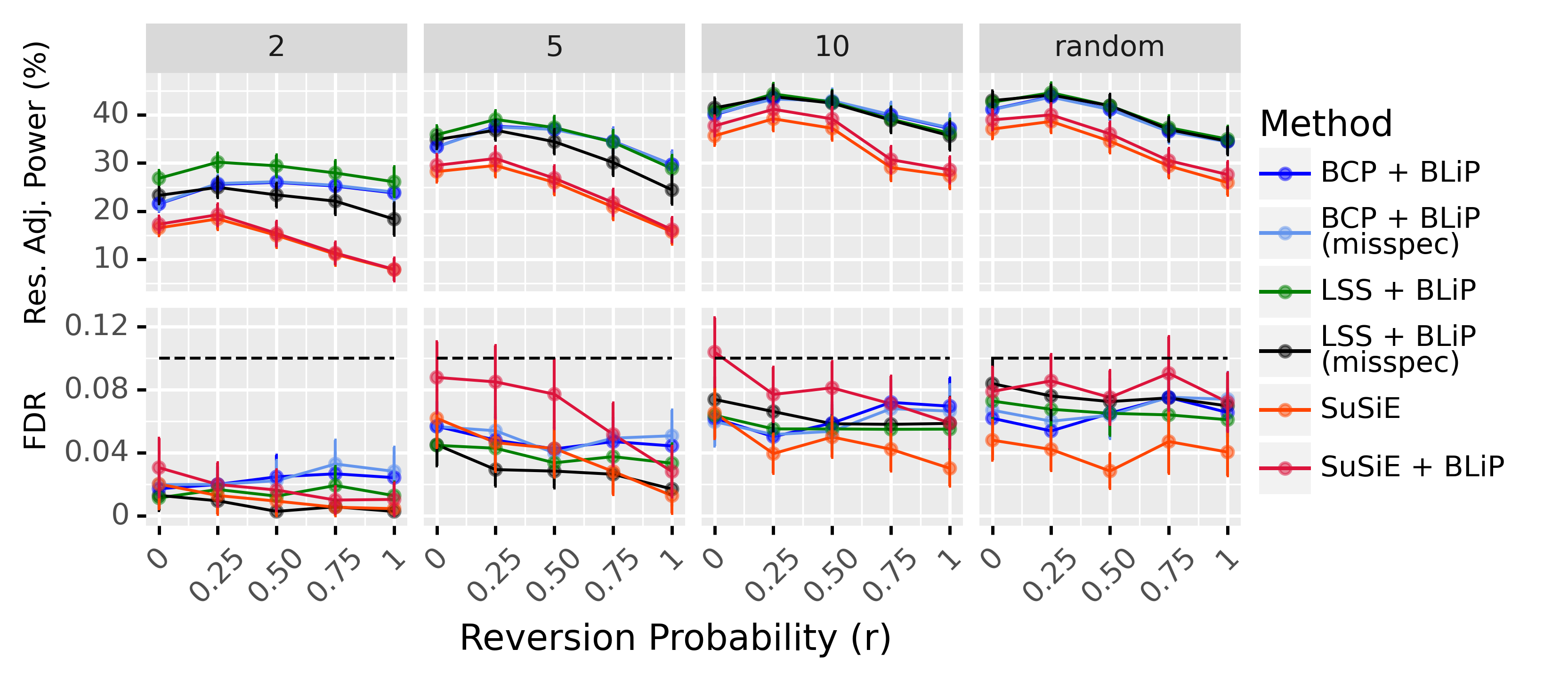}
    \caption{This plot shows the same setting as Figure \ref{fig::cp_power} with $\tau^2 = 9$ and $s = 0.05$, with one exception. At each change point, with probability $r$, the time series reverts to a global mean of zero. When $r$ is high, this causes many successive change points to cancel each other out, which can trap change point detection algorithms in local maxima. In particular, SuSiE has comparatively lower power than the other methods in this setting. The y-panel shows the average spacing between change points (``random" spacing means the locations of the change points are chosen uniformly at random).}
    \label{fig::rprob}
\end{figure}

Next, we consider the same setting as before, but with an additional twist: instead of sampling the nonzero coefficients as i.i.d. $\mcN(0, \tau^2)$ random variables, we specify that at each change point $t$ where $\mu_t \ne 0$, with probability $r$ we set $\mu_{t+1} = 0$, i.e., the process reverts to having mean zero. This is a challenging setting because consecutive change points may cancel out previous change points, causing change point detection algorithms to get stuck in local maxima. However, Figure \ref{fig::rprob} shows that \blip-based methods perform comparatively well in this setting.

\section{Further details on Section \ref{sec::gwas}}\label{appendix::gwas}

\subsection{Methodological details}\label{subsec::gwasmethods}

In this section, we describe our fine-mapping analysis in detail. As a review, \cite{polyfun2019} split the genome into $m=2763$ loci spanning $3$ mega-base-pairs (Mb) each. For each of these loci, \cite{polyfun2019} fit SuSiE models with $L = 10$ causal SNPs, yielding $m=2763$ SuSiE outputs $\alpha^{(1)} \in \R^{p_1 \times 10}, \dots, \alpha^{(m)} \in \R^{p_{m} \times 10}$, where $p_i$ denotes the number of SNPs in locus $i$ and $\alpha^{(i)}$ denote the matrix of SuSiE PIPs as described in Appendix \ref{appendix::susie}. Please see \cite{polyfun2019} for further details on these SuSiE models and the generation of the loci. Note that \cite{polyfun2019} fit SuSiE models both with uninformative priors and with (novel) priors based on functional annotation data. For simplicity, we use the SuSiE models with uninformative priors, although more sophisticated priors could improve the power of SuSiE and SuSiE + \blip.

For each $\alpha^{(i)}$, we generated the candidate groups as follows. To speed up computation, we pre-filter the SNPs to exclude SNPs from consideration unless they have a posterior inclusion probability of at least $1\%$, since it is quite unlikely we would discover any groups containing those SNPs. Formally, if $\mcL^{(i)}$ denotes the set of SNPs in locus $i$, let $\mcL_{\kappa}^{(i)} = \{\ell \in \mcL^{(i)} : p_{\{\ell\}} > 0.01\}$. Then, we generate the set of all contiguous subgroups of $\mcL_{\kappa}^{(i)}$ of maximum size less than $25$, as described in Appendix \ref{subsec::simcandgroups}. For good measure, we also include the groups discovered by the default SuSiE model as candidate groups. Let $\mcG^{(i)}$ denote these candidate groups.

Given the candidate groups $\mcG^{(1)}, \dots, \mcG^{(m)}$ for each locus, let $\mcG = \bigcup_{1 \le i \le m} \mcG^{(i)}$ denote the set of candidate groups for the whole genome. As discussed in Appendix \ref{subsec::prefilter}, we pre-filter $\mcG$ to only include groups $G$ where the PIP is greater than $75\%$, so $G \in \mcG \implies p_G \ge 0.75$. We then apply \blips to $\mcG$ to control the FDR at level $q = 0.05$. This yields the results described in Section \ref{sec::gwas}.

\subsection{Simulations on real genotype data}\label{subsec::gwassims}

In this section, we describe the simulations we performed using the real LD matrix from the fine-mapping analysis.

We create simulated data using exactly the same set-up as \cite{polyfun2019}. As notation, let $\bX \in \R^{n \times p}$ denote the genetic covariate data for a fixed locus, so $n \approx 337,000$ and $p$ ranges between $5,000$ and $50,000$. Suppose that $\bX$ has been scaled so its columns have mean zero and unit variance, let $\by \in \R^n$ be the phenotype data scaled so that $\E[\by^T \by] \approx n$, and let $R \approx \frac{1}{n} \bX^T \bX$ be the reported LD matrix from \cite{polyfun2019}. As in \cite{polyfun2019}, we assume a Gaussian linear model where $\by \sim \mcN(\bX \beta, \sigma^2 \ident_n)$ and $\sigma^2 = 1 - \beta^T \bX^T \bX \beta \defeq 1 - h_g^2$, so $h_g^2$ is the locus-specific heritability. Note that $R$ is not exactly equal to $\frac{1}{n} \bX^T \bX$, but in the fine-mapping literature, it is common to assume that $\frac{1}{n} \bX^T \by \sim \mcN(R \beta, \sigma^2 R / n)$, and \cite{polyfun2019} make this assumption. Initially, the raw LD matrix $R$ may not be positive semidefinite (PSD), but we use the same procedure as \cite{polyfun2019} to perturb $R$ to be positive definite while changing $R$ as little as possible. 

\begin{figure}
    \centering
    \includegraphics[width=0.85\linewidth]{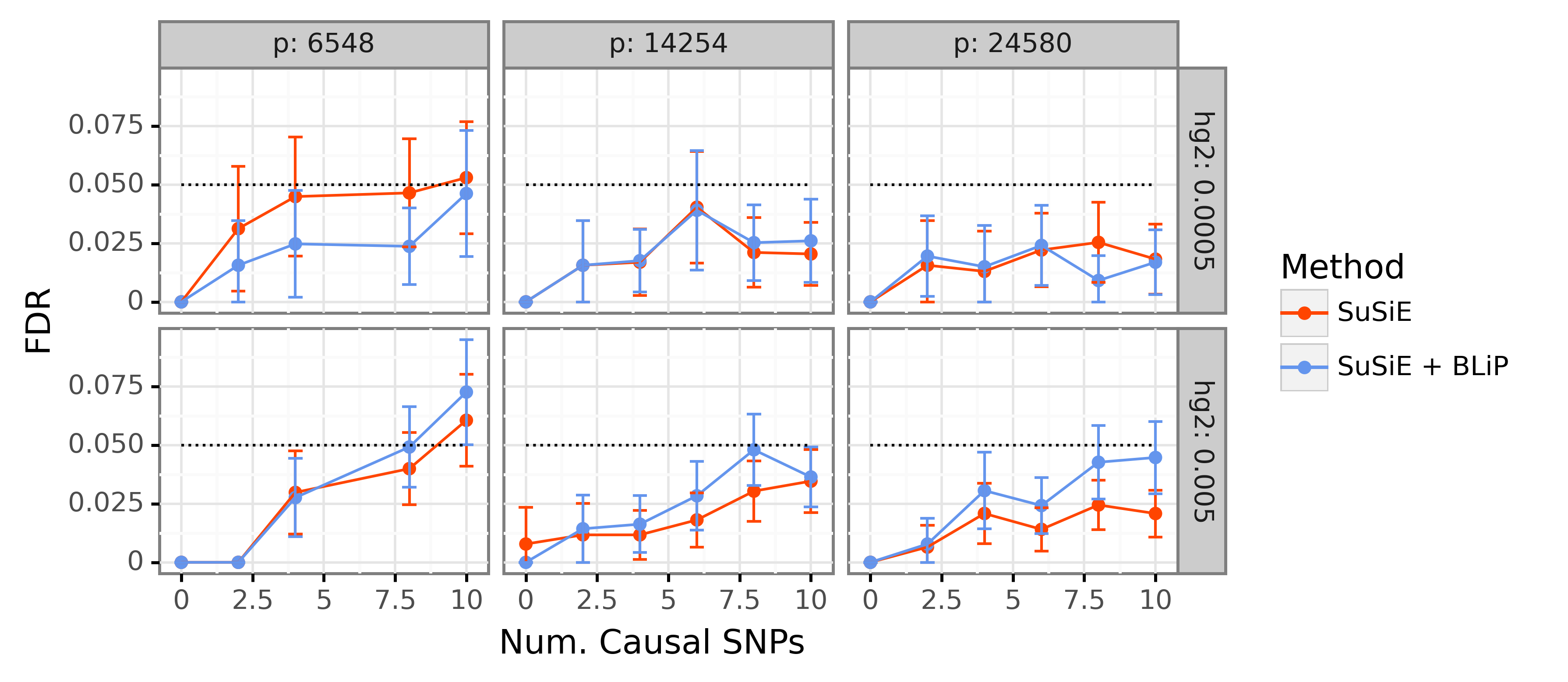}
    \caption{This figure shows the FDR of SuSiE and SuSiE + \blips in the fine-mapping simulations described in Appendix \ref{subsec::gwassims}. It shows that both methods control the FDR (up to Monte Carlo error) in a variety of settings, even when the number of causal SNPs is smaller than $10$.}
    \label{fig::gwassimfdr}
\end{figure}

\begin{figure}
    \centering
    \includegraphics[width=0.85\linewidth]{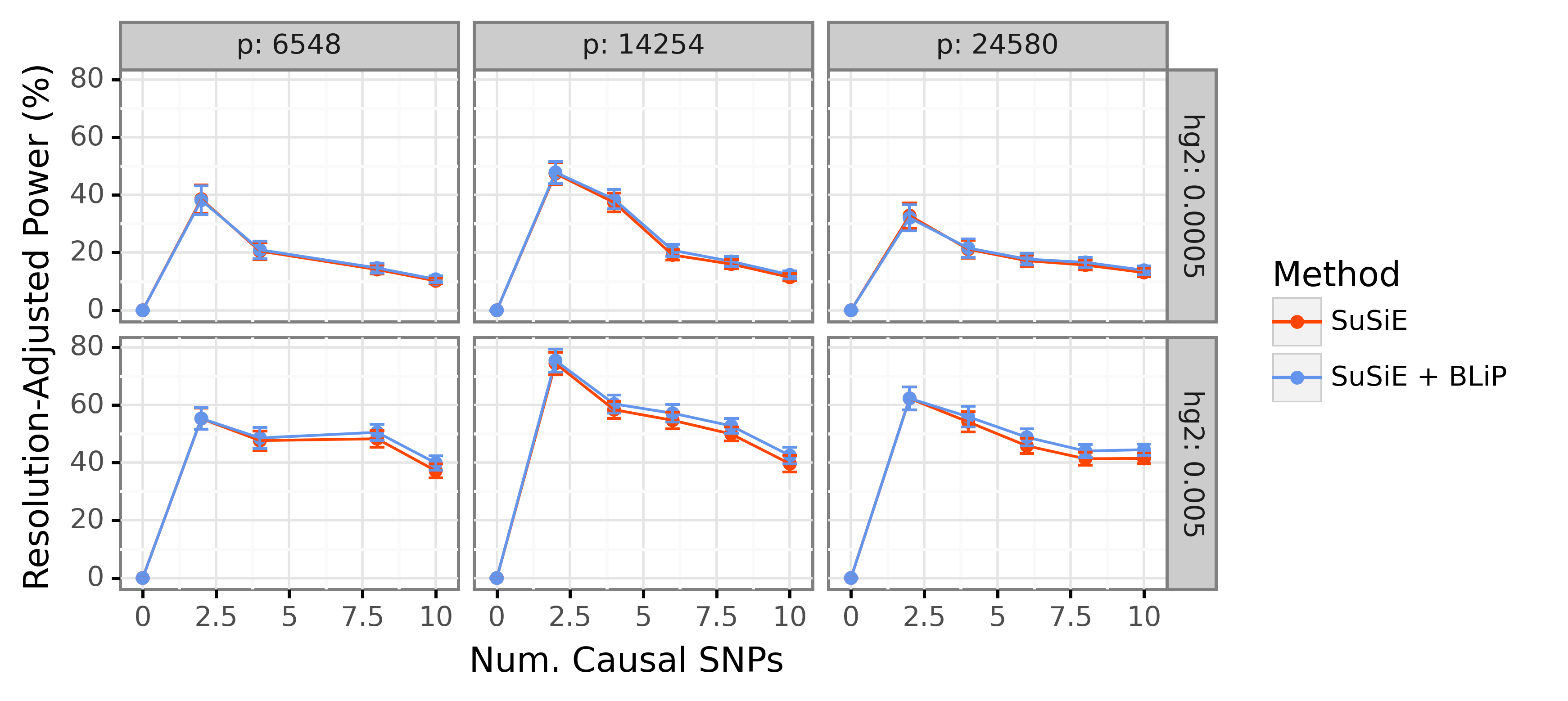}
    \caption{The corresponding power plot for Figure \ref{fig::gwassimfdr}---see Appendix \ref{subsec::gwassims} for a discussion.}
    \label{fig::gwassimpower}
\end{figure}

For a fixed value of $h_g^2$ and a fixed number of causal SNPs $L$, we run simulations as follows. First, we randomly choose $L$ nonzero coefficients of $\beta$ and sample their values as $\mcN(0,1)$ random variables. Then, we scale $\beta$ such that $\beta R \beta = h_g^2$. Second, for this $\beta$, we sample the sufficient statistics $\frac{1}{n} \bX^T \by \sim \mcN(R \beta, \sigma^2 R / n)$. Third, we fit a SuSiE model based on the sufficient statistics $(\frac{1}{n} \bX^T \by, R)$ using the algorithm from \cite{zou2021}, and we run \blips on top of SuSiE as specified in Appendix \ref{subsec::gwasmethods}. Note that when fitting the SuSiE model, we follow \cite{polyfun2019} and use a modified HESS estimator to estimate the prior variance of the non-null coefficients---see \cite{polyfun2019} for details. Then, we evaluate the resolution-adjusted power and FDR of SuSiE and SuSiE + \blip. We repeated this process for $128$ replications on three loci from chromosomes $1, 10$, and $12$, chosen because they represent various values of $p$. The results are shown in Figures \ref{fig::gwassimfdr} and \ref{fig::gwassimpower} for various values of $h_g^2$ and $L$, where we adapted our values of $h_g^2$ and $L$ from \cite{polyfun2019}. Note that in all cases, we fit the SuSiE model assuming there are $L=10$ causal variants, even when in truth there are fewer. As expected, the figures show that \blips improves the power of SuSiE and that both methods successfully control the FDR. 

That said, the power boost from \blips in these simulations is smaller than in the real analysis. We suspect this is because these simulations are locus-specific, whereas in the real analysis, we apply \blips to the whole genome. Since there are vastly more candidate groups when working on the whole genome, the optimization problem of which groups to discover is much harder, and it is much more likely that \blips can substantially improve upon the heuristic solution from SuSiE. In contrast, when working with only one 3Mb locus, it seems that SuSiE comes close to finding the optimal solution. Admittedly, this difference suggests that these simulations are not perfectly representative of the real analysis, and we would prefer to have done simulations on the entire genome. Unfortunately, these locus-specific simulations were already quite computationally expensive, so a genome-wide simulation was not possible. (Note \cite{polyfun2019} did not perform a genome-wide simulation either.)

\subsection{Comparison to previous studies}\label{subsec::gwascomp}

In this section, we compare our findings to those of \cite{farh2015} and \cite{polyfun2019}, and we also check which of our discoveries have previously been reported by a separate study in the NHGRI-EBI GWAS Catalog \citep{gwascatalog2019}. First, as shown in Table \ref{tab::farh2015comp}, we replicate four of the eight findings from \cite{farh2015}, who analyzed the HDL cholesterol and LDL direct traits using a separate dataset. This replication rate roughly agrees with that of \cite{polyfun2019}, who also compared their findings to those from \cite{farh2015}. Note SuSiE + \blips makes far more than eight discoveries for these two traits, likely due to the fact that the SuSiE model we use from \cite{polyfun2019} was fit on a much larger dataset than the one used in \cite{farh2015}. 

\begin{table}[]
    \centering
    \begin{tabular}{|c|c|c|c|c|}
        \hline
        Trait &         SNP &  PIP (Farh et al. (2015)) & Replicated by SuSiE + BLiP? \\
        \hline
        HDL &   rs581080 &                     0.978 &                          No \\
        \hline
        HDL &    rs964184 &                     1.000 &                         Yes \\
        \hline
        HDL &  rs12229654 &                     1.000 &                          No \\
        \hline
        HDL &   rs2074356 &                     1.000 &                          No \\
        \hline
        HDL &   rs1532085 &                     0.986 &                          No \\
        \hline
        HDL &   rs1800961 &                     1.000 &                         Yes \\
        \hline
        LDL &  rs11591147 &                     1.000 &                         Yes \\
        \hline
        LDL &    rs964184 &                     1.000 &                         Yes \\
        \hline
    \end{tabular}
    \caption{This table shows the eight significant PIPs from \cite{farh2015} for the traits HDL cholesterol and LDL direct. It also shows whether SuSiE + \blips replicated these findings. Note that in each case where SuSiE + \blips replicated one of these findings, SuSiE + \blips discovered the individual SNP (by coincidence SuSiE + \blips did not discover groups of size larger than $1$ containing any of these SNPs).}
    \label{tab::farh2015comp}
\end{table}

Second, we compare the results from SuSiE + \blips to those of SuSiE alone, using the SuSiE model from \cite{polyfun2019}. Note that this is certainly \emph{not} a replication analysis, since the dataset is the same for both analyses---rather, it is a sanity check to ensure that SuSiE + \blips does not completely change the findings from \cite{polyfun2019}. For each of the four traits we analyze, Table \ref{tab::polyfuncomp} shows the degree of overlap between SuSiE and SuSiE + \blip. To be precise, we say that the proportion of overlap for SuSiE is the proportion of discoveries $G$ made by SuSiE which overlap with at least one other group $G'$ discovered by SuSiE + \blip, and analogously for SuSiE + \blip. For example, for the cardiovascular trait, the first row of Table \ref{tab::polyfuncomp} shows that $100\%$ of the discoveries made by SuSiE overlapped with at least one discovery made by SuSiE + \blip, and $81.4\%$ of the discoveries made by SuSiE + \blips overlapped with at least one discovery made by SuSiE. Holistically, Table \ref{tab::polyfuncomp} shows that SuSiE + \blips confirms every discovery that SuSiE makes but also that SuSiE + \blips makes roughly $15$--$20\%$ more discoveries that do not overlap with any previous discoveries from SuSiE. Since \blips increased SuSiE's resolution-adjusted power by $30$--$50\%$, this suggests that the power gain comes both from more precisely localizing existing discoveries \textit{and} from making some completely new discoveries.

\begin{table}[]
    \centering
    \begin{tabular}{|c|c|c|}
        \hline
        Trait & Prop. Overlap for SuSiE & Prop. Overlap for SuSiE + \blips \\
        \hline
        Height & $100.0\%$ & $87.5\%$ \\
        \hline
        HDL & $100.0\%$ & $84.8\%$ \\
        \hline
        LDL & $100.0\%$ & $86.2\%$ \\
        \hline
        Cardiovascular & $100.0\%$ & $81.4\%$ \\
        \hline
    \end{tabular}
    \caption{For the four traits analyzed in Section \ref{sec::gwas}, this table shows the proportion of overlap (as defined in Appendix \ref{subsec::gwascomp}) for SuSiE and SuSiE + \blip. For example, the first row shows that for the cardiovascular trait, $100\%$ of the discoveries made by SuSiE overlapped with at least one discovery made by SuSiE + \blip, whereas $81.4\%$ of the discoveries made by SuSiE + \blips overlapped with at least one discovery made by SuSiE.}
    \label{tab::polyfuncomp}
\end{table}

Lastly, for the new discoveries made by SuSiE + \blip, we check how many were also reported in a separate study in the NHGRI-EBI GWAS Catalog \citep{gwascatalog2019} using SNPnexus \citep{snpnexus2008}. In particular, we say that a discovery $G$ has been \textit{corroborated} if at least one SNP in $G$ has been previously reported by a separate study. The results are shown in Table \ref{tab::gwascatalog}, which also shows the corroboration rate for the original SuSiE model from \cite{polyfun2019}. Notably, the corroboration rate for the new discoveries is comparable to that of the original model, although it is slightly smaller for the new discoveries. We interpret this as reasonably strong evidence that SuSiE + \blips is largely discovering real causal variants, especially since a priori, one would expect that novel discoveries are harder to corroborate since they may be (informally) harder to discover, as the initial model did not discover them. Nonetheless, the \textit{additional} discoveries made by SuSiE + \blips were corroborated at roughly the same rate as those from the original model.

\begin{table}[]
    \centering
    \begin{tabular}{|c|c|c|}
        \hline
        Trait & Prop. Corroborated (SuSiE) & Prop. Corroborated (additional new discoveries) \\
        \hline
        Height & $53.5\%$ & $45.0\%$ \\
        \hline
        HDL & $57.0\%$ & $50.0\%$ \\
        \hline
        LDL & $67.3\%$ & $60.0\%$ \\
        \hline
        Cardiovascular & $82.2\%$ & $65.2\%$ \\
        \hline
    \end{tabular}
    \caption{For the analysis in Section \ref{sec::gwas}, this table shows the proportion of discoveries which can be corroborated by a separate study in the NHGRI-EBI GWAS Catalog \citep{gwascatalog2019}. The first column shows the results for the SuSiE model from \cite{polyfun2019}, and the second column shows the corroboration rate for the new discoveries made by SuSiE + \blip. Notably, the corroboration rate for the \textit{new} discoveries is similar to, if slightly smaller than, that of the original model.}
    \label{tab::gwascatalog}
\end{table}

\subsection{Discussion on making discoveries at a single-variant resolution}\label{subsec::singletons}

One notable result from Section \ref{sec::gwas} is that in addition to having higher resolution-adjusted power, SuSiE + \blips discovers more individual variants than SuSiE alone. This result raises the question: does \blips help make more discoveries at a single-variant resolution? The answer to this question is subtle, so we pause to discuss it here.

First, if one aims to perform \textit{resolution-adaptive inference}, then yes, we expect SuSiE + \blips to make more single-variant discoveries than SuSiE alone. The reason for this, as explained in Section \ref{appendix::susie}, is that SuSiE + \blips can simultaneously use full PIPs from the SuSiE model and also make resolution-adaptive discoveries. In contrast, prior to \blip, analysts using SuSiE faced a trade-off: SuSiE can construct credible sets, but to do so, SuSiE must use PIPs which are provably more conservative than the full PIPs. This may cause SuSiE to make fewer single-variant discoveries, as seen in Section \ref{sec::gwas}. Alternatively, one could use SuSiE's full PIPs to make many discoveries at the single-variant resolution, but in this case it is not clear how to make any discoveries at coarser resolutions. Thus, SuSiE + \blips gets the best of both worlds, since it can both use the full SuSiE PIPs \textit{and} make resolution-adaptive discoveries.
 
Of course, if one's \textit{exclusive} goal is to discover single variants without discovering any larger groups, then \blips will reduce to the following trivial (known) procedure: calculate PIPs $\{p_{\{\ell\}}\}$ for each variant $\ell$, sort them in decreasing order, and sequentially make rejections until the average PIP of the rejections drops below $1-q$. In short, \blips is not a meaningful innovation in settings where one does not wish to take a resolution-adaptive approach. That said, there is basically no cost to being resolution-adaptive, since \blips will discover single variants whenever this is possible, and when this is not possible, it will (likely) discover other, larger groups which may still be of scientific interest. Note that this same argument applies when applying \blips on top of any Bayesian model.

A last note is that studies which primarily aim to make discoveries at a single-variant resolution may wish to consider a different notion of the FDR than the (standard) one we consider in this paper. In particular, controlling the FDR as defined in Equation (\ref{eq::rasdfdr}) does not guarantee that one will control the FDR when only considering singleton discoveries---for example, the FDR of the singleton discoveries in our application in Section \ref{sec::gwas} is $5.99\% > 5\%$. Of course, this is easy to fix, since \blips can control a variety of error rates. For example, one might require that for each $k \in \N$, the FDR for each set of discoveries of size $k$ is controlled at level $q$. Similarly, one could also define a weighted or ``resolution-adjusted" FDR constraint. In general, we suspect that the right notion of the error rate depends on the study in question: the good news is that \blips can be easily adapted to control any of these error rates.

\section{Further details on Section \ref{sec::astro}}\label{appendix::astro}

In this section, we describe our method from Appendix \ref{sec::astro}. Note that this section follows the notation of Appendix \ref{appendix::infloc}. We create the inputs to \blips as follows:

\begin{enumerate}
    \item First, we load the prefit StarNet ``wake-sleep" encoder published by \cite{starnet2021}, and we sample $N=1000$ samples from the posterior of this StarNet model. Let $\gamma^{(1)}, \dots, \gamma^{(N)} \subset \mcL= [0,1]^2$ denote these posterior samples. Furthermore, let $\mcM = \{\ell^{(1)}, \dots, \ell^{(m)}\} \subset \mcL$ denote the MAP estimators of point source locations from the StarNet model.
    \item As candidate groups, for a fixed radius $r$, let $\{S_r(rz) : z \in \Z^2\}$ denote the set of circles of radius $r$ on the lattice points $rz$ for $z \in \Z^2$. Furthermore, let $\{S_r(\ell) : \ell \in \mcM\}$ denote the circles of radius $r$ around the MAP estimators. We set
    \begin{equation*}
        \mcG = \bigcup_{r \in \mcR} \{S_r(rz) : z \in \Z^2\} \cup \{S_r(\ell) : \ell \in \mcM\},
    \end{equation*}
    where we take $\mcR$ to be $25$ evenly spaced values on the log scale between $0.01$ and $0.0005$. For intuition, recall that a radius of $0.01$ corresponds to one pixel.
    \item We compute PIPs $\{p_{\mcG}\}_{G \in \mcG}$ for these candidate groups using Algorithm \ref{alg::subpartitions} from Appendix \ref{appendix::infloc}. We then pre-filter $\mcG$ to only include groups $G$ such that $p_G \ge 0.25$ (so each candidate group must have at least a $25\%$ chance of containing a signal). This takes roughly eight minutes of serial computation on a laptop. Note that this step is the bottleneck, but it is quite parallelizable, so in large-scale applications it should not pose a barrier to running \blip.
    \item In this case, there are an infinite number of locations, so we compute a finite-dimensional reduction of \blips using the algorithms from Appendix \ref{subsec::edgecover} (notably Algorithm \ref{alg::ecc}). This takes only a few seconds.
    \item Finally, we apply \blips to control the FDR at levels $q \in \{0.05,0.1,0.15, \dots, 0.75\}$, which takes 5-10 seconds per value of $q$. We use the weight functions described in Section \ref{sec::astro}.
\end{enumerate}

To evaluate both \blips and the baseline, we use the Hubble locations as the true sources and calculate resolution-adjusted power using the weight functions from Section \ref{sec::astro}. We give all methods $0.01$ pixels of slack when calculating false discoveries to account for measurement differences between the Hubble and SDSS surveys.

\end{document}